\documentclass[12pt]{article}
\usepackage{hyperref}
\usepackage{times,amssymb,euscript,epsfig,latexsym,amsmath}
\usepackage[normalem]{ulem}
\usepackage{epsfig,graphicx,subfigure}
\usepackage{color}
\usepackage{backref}
\begin{document}
\newcommand{\todo}[1]{\vspace{5 mm}\par \noindent
\framebox{\begin{minipage}[c]{0.95 \textwidth}
\tt #1 \end{minipage}}\vspace{5 mm}\par}
\newtheorem{thm}{Theorem}
\newtheorem{prop}[thm]{Proposition}
\date{}
\title{ \Large{Viscous vortex-pair-cylinder interactions near inviscid translating equilibria: a numerical study}}
\author{
Banavara N. Shashikanth\footnote{Mechanical and Aerospace Engineering
Department, MSC 3450, PO Box 30001,
New Mexico State University, Las Cruces, NM 88003, USA.
E-mail:shashi@nmsu.edu} \\ \\
Yanxing Wang \footnote{Mechanical and Aerospace Engineering
Department, MSC 3450, PO Box 30001,
New Mexico State University, Las Cruces, NM 88003, USA.
E-mail:yxwang@nmsu.edu}}
\maketitle
\abstract{Numerical simulations using the Lattice Boltzmann Method are presented of the following two-dimensional incompressible flow problem. Starting from configurations corresponding to translating inviscid equilibria, namely, the translating F\"{o}ppl equilibria (counter-rotating point vortex pair, either fore or aft of a translating circular cylinder) and the translating Hill equilibria (counter-rotating point vortex pair, either fore or aft of an elliptic cylinder), viscosity is turned on for $t >0$ and the subsequent viscous interaction is simulated. The interaction is in a dynamically coupled setting where the neutrally buoyant cylinder is free to move along the symmetry axis under the action of the instantaneous fluid stresses on its surface. It is observed that for starting configurations in which the vortex pair trails the cylinder, the viscous evolution stays close to the inviscid equilibrium. However, for starting configurations in which the vortex pair leads the cylinder, there is significant deviation from the inviscid equilibrium. In such cases, the vortices either accelerate and leave the cylinder behind or, more interestingly,  leave their leading positions and are attracted towards the trailing positions. In other words, the cylinder in such cases threads through the leading vortices, overtakes them and the vortices are observed to trail the cylinder again. The symmetry of the inviscid dynamics about the perpendicular axis is thus broken. }

\newpage
\tableofcontents
\section{Introduction.} 
      The topic of vortices interacting with bodies with prescribed motion is not a new subject in fluid mechanics. Examples of such interactions abound in aerospace engineering, some well-researched ones being the  wake hazard problem due to wing tip vortices shed from large aircraft and the topic of blade-vortex interactions in turbines \cite{Ro1998}. There are also well-studied canonical examples involving simple geometries such as the famous Karman vortex shedding problem for the streaming flow over a fixed circular cylinder. With oscillatory motions of the cylinder included, this topic leads to the subject of vortex induced vibrations (VIV).  The case of freely falling heavy objects in a fluid, and their interaction with the vortices that they shed, has also been studied though perhaps to a lesser extent \cite{HoWi2010a, HoWi2010b}.  Vortex-body interactions have also been examined in another area of relevance, namely, that of vortices on collision or near-collision 
courses with fixed solid objects in their path \cite{Pe1992, Or1993, OrVe1993, ChChLiChLe1995, VeFlVaOr1995, PePo2013}. An interesting effect observed in some of these papers is that of rebound and reversal of the vortex trajectories.

     However, the topic of vortices interacting with neutrally buoyant bodies that are free to move has been far less researched than any of these topics. In such a dynamically coupled setting, the body has no prescribed motion (or lack of motion) and is free to move under the instantaneous action of the fluid stress field on its surface. There are environmental applications of this topic, though with significantly more complex features than in the simple models of this paper. Passive objects being transported on or beneath the surface of water bodies may have vortices present in their vicinity due to self-shedding or shedding by other solid objects in proximity or by the free surface. For example, a growing environmental concern of the past decade has been the transport of rigid plastic waste by rivers into oceans and their subsequent pollution, ex. `The Great Pacific Garbage Patch'. One may also mention the environmental engineering example of floating offshore wind turbines, where the supporting column of the turbine assembly is only partially constrained  by the sea-bed. With the same caveat, there are also potential applications to the subjects of biomechanical and biomimetic locomotion. For example, the problem of an underwater swimming creature or vehicle that may have vortices in its proximity shed from  itself or from other sources. 

    But the motivation for this paper comes more from a fundamental point of view. Though Hamiltonian models in an inviscid setting with a free-slip boundary condition have been developed in the the last three decades or so \cite{Ko1987, ShMaBuKe2002, Ra2002, Sh2005, Sh2006, BoMaRa2007, ShShKeMa2008}, there is a lack of basic research  done on vortices interacting with a neutrally buoyant body free to move in a viscous Navier-Stokes setting. Basic features and principles of the coupled dynamical interactions  and momentum and energy exchanges have yet to be elucidated. This situation may be contrasted with the amount of papers written on and textbooks that describe the Karman vortex street problem. A study of models in such settings could also potentially shed light on the coupled fluid-solid dynamics in which the body is partially or fully constrained or with prescribed motions as in the more traditional settings. Some previous studies of such interactions, which however focused on the collision scenarios mentioned above, may be found in \cite{AlJoSh2007, HaScScSh2010, HaScSh2012}.

    A dynamically coupled model of simple geometry, which derives from a classical work due to F\"{o}ppl \cite{Fo1913}, is numerically investigated in this paper. F\"{o}ppl obtained  the equilibrium positions of a symmetrically located counter-rotating pair of point vortices in the wake region\footnote{There is also an equilibrium configuration in which the point vortex pair is located upstream of the cylinder but as pointed out in \cite{VaMoSc2011} this has been relatively less studied.} of a stationary circular cylinder in the streaming flow of an incompressible, inviscid fluid. The stability of this equilibrium was studied by F\"{o}ppl himself and since then the model has been further studied from both dynamics and control perspectives \cite{TaAu1997, CaLiLu2003, Pr2004, VaMoSc2011}. There is an analogous translating equilibrium which has been far less studied, termed the moving F\"{o}ppl equilibrium, in the dynamically coupled problem of a circular cylinder and a counter-rotating point vortex pair \cite{ShMaBuKe2002, Sh2006}. In this case, the entire configuration of cylinder plus symmetric vortex pair translates steadily.  There are also moving vertical line  equilibria in this configuration in which the vortices in the pair are located vertically above and below the translating cylinder. Much later, Hill \cite{Hi1998} showed using conformal maps that analogs of F\"{o}ppl equilibria exist for streaming flows over stationary elliptical cylinders as well. Like the classical F\"{o}ppl equilibria these equilibria also carry over to the dynamically coupled setting in a straightforward manner and result in what will be referred to as moving Hill equilibria; see Appendix for a review of results pertaining to these moving equilibria.  

   There are few or no investigations on the effects of adding viscosity to these translating equilibria. Clearly, the equilibrium is lost due to viscous effects like drag and diffusion. However, there seems to be no systematic study of how much the viscous evolution deviates from the equilibria. This paper numerically examines dynamically coupled evolutions in an incompressible Navier-Stokes framework for initial configurations that correspond to that of the above equilibria. The (pre-exisiting) vortices have cores in these simulations and therefore the configuration is defined by the centers of the vortices.  The viscosity is turned on for $t>0$ and the dynamically coupled motion of the cylinder and the vortices is tracked. Several such simulations for various choices of vortex starting positions, on and near the equilibrium curves, are presented in this paper.  In addition, some off-equilibrium and some asymmetric  initial configurations are also considered. 
  
     Two particularly interesting phenomenon are observed in the  time frames of the simulation. Firstly, in cases corresponding to both the moving F\"{o}ppl and moving Hill  equilibria, the trailing vortex configuration seems stable to viscous effects, but the leading vortex configuration displays unstable behavior.  All cases corresponding to the vertical equilibria also display unstable behavior. Secondly, in several of the unstable configurations the vortices are attracted to the trailing vortex configuration. The cylinder in such cases threads through the vortex pair which in turn moves towards the more stable trailing configuration. Boundary layer development, separation and entrainment of boundary layer vorticity by the primary vortices is observed in cases where the vortices start close to the cylinder. But as such no secondary vortex formation is observed. The results obtained are discussed and analyzed. 


\section{Set-up and method.}

The numerical simulations are based on the low Mach number Lattice Boltzmann Method (LBM). Compared with the other numerical methods, the LBM has particular advantages in its ability to handle, with relative ease, complex moving boundaries with scalar and other complexities. 


\subsection{Basic Lattice Boltzmann Algorithm with Single and Multiple Relaxation Times.}
The LBM can be viewed as an explicit finite difference representation of the continuous Boltzmann equation, with the dependent variable being a particle distribution function $f_\alpha (\mathbf{x},t)$ that quantifies the probability of finding an ensemble of molecules at position $\mathbf{x}$ with velocity $\mathbf{e}_\alpha$ at time $t$ \cite{Qi1992, ChDo2002, WaBrBaWeAiNe2010}. In this work, a D2Q9 scheme with nine velocity elements was developed and implemented. To overcome the numerical instability issue of single-relaxation-time (SRT) Bhatnagar-Gross-Krook (BGK) model, a multi-relaxation-time (MRT) model was developed and implemented, which nevertheless retains the simplicity and computational efficiency of the SRT-BGK model \cite{LaLu2000, Dh2002, LaLu2003}. The MRT lattice Boltzmann equation reads: 
\begin{align}
\mathbf{f}(\mathbf{x}+\mathbf{e}\delta_t,t+\delta_t )-\mathbf{f}(\mathbf{x},t)=-\mathbf{M}^{-1} \mathbf{\hat{S}}(\mathbf{\hat{f}}(\mathbf{x},t)-\mathbf{\hat{f}}^{eq} (\mathbf{x},t))				
\end{align}
where $\mathbf{\hat{f}}=\mathbf{M}\mathbf{f}$ and $\mathbf{f}=\mathbf{M}^{-1}\mathbf{\hat{f}}$. The collision matrix $\mathbf{\hat{S}}=\mathbf{M}\mathbf{S}\mathbf{M}^{-1}$ in moment space is a diagonal matrix, and $\mathbf{\hat{f}}^{eq} (\mathbf{x},t))$ is the equilibrium value of the moment $\mathbf{\hat{f}}(\mathbf{x},t)$.  The transformation matrix $\mathbf{M}$ can be constructed via the Gram-Schmidt orthogonalization procedure. Macroscopic variables such as density $\rho$ and velocity $\mathbf{u}$ are calculated from the moments of the distribution functions,
\begin{align}
\rho(\mathbf{x},t)=\Sigma_\alpha f_\alpha (\mathbf{x},t), \quad \rho(\mathbf{x},t)\mathbf{u}(\mathbf{x},t)=\Sigma_\alpha f_\alpha (\mathbf{x},t) \mathbf{e}_\alpha,	
\end{align}
It has been shown that the MRT model is of better numerical stability than the SRT model.
In the treatment of non-slip conditions at cylinder surface, the scheme with 2nd order of accuracy proposed by \cite{LaLu2003} was applied. This method is based on the simple bounce-back boundary scheme and interpolations.  If the distance fraction of the first fluid node from the solid boundary is less than half lattice space, interpolation of distribution functions is conducted before propagation and bounce-back collision. If the distance fraction is greater than half lattice space, interpolation is conducted after propagation and bounce-back collision. When the boundary is in motion, the additional momentum on the fluid generated by the movement of the boundary must be included in the streaming step of the lattice-Boltzmann algorithm \cite{La1994, BoFiLa2001}.
\paragraph{Motion of Neutrally Buoyant Cylinder.}
The solid cylinder is considered to be neutrally buoyant in the fluid. The motion of the particle is found by solving Newton’s equations of motion,
\begin{align*}
M d \mathbf{U}(t)⁄dt&=\mathbf{F}(t), \quad  \mathbf{I} \cdot d \mathbf{\Omega}(t)⁄dt+\mathbf{\Omega}(t)\times [\mathbf{I} \cdot \mathbf{\Omega}(t)]=\mathbf{T}(t),     			
\end{align*}
where $M$ is the mass of the cylinder, $\mathbf{I}$ is the inertial tensor, $\mathbf{U}$ is the translational velocity, $\mathbf{\Omega}$  is the angular velocity, and $\mathbf{F}$ and $\mathbf{T}$ are the total force and torque on the cylinder, respectively. At each fluid boundary node, the force is calculated by the exchange of distribution function between the flow and solid.
\begin{align}
\mathbf{F}^{(b)} (\mathbf{x}_b,t)=-\Sigma_\alpha\left[f_{\bar{\alpha}} (\mathbf{x}_1,t+\delta_t)-\tilde{f}_\alpha (\mathbf{x}_2,t_+)+f_{\bar{\alpha}}' \right] ,		   		
\end{align}
where $f_{\bar{\alpha}}$ is the distribution bounced back from the wall, $\tilde{f}_\alpha$ is the post-collision distribution transferred from the flow to the solid wall, and $f_{\bar{\alpha}}' $ is the additional momentum given by the wall due to boundary motion.

\subsection{Initialization.} The vortices at initial time are chosen to have circular cores of radius $a$ and uniform vorticity $\omega_0$, with their centers $(x_1,y_1)$ and $(x_1,-y_1)$, relative to a frame fixed to the center of the cylinder, and lying on the equilibrium locii given by (\ref{eq:fopplcurve}), the $y$-axis or (\ref{eq:hillcurve}), as the case may be. The vortex in the upper half has counter-clockwise circulation (assumed positive by convention). The corresponding point vortex strength is chosen as $\pm \Gamma=\pm \omega_0 \pi a^2$ and the translational speed $V$ of the corresponding equilibrium configuration is obtained from the  formulas (\ref{eq:velposition}), (\ref{eq:velpositionN}) or (\ref{eq:velpositionell}), as the case may be. 

The initial fluid velocity field outside the vortex cores is then chosen as the inviscid velocity field corresponding to point vortices of strength $\pm \Gamma=\pm \omega_0 \pi a^2$ located at these points in the presence of circular cylinder translating with speed $V_{c,0}=V$, where the latter is given by  formulas (\ref{eq:velposition}) or (\ref{eq:velpositionN}). Note that this initial velocity field is the sum of the velocity field of each point vortex, of each image point vortex and the Kirchhoff velocity field due to the translating cylinder. For the cases of the elliptic cylinder, the initial velocity field is taken as simply the sum of the velocity fields of the two point vortices. Inside the core of each vortex one replaces the point vortex velocity field due to that vortex with solid body rotation.  Since the no-slip boundary condition is not satisfied at time $t=0$, there is some initial transient behavior which however lasts for a very short time.

    The main parameter that was varied was the initial distance of the vortex center from the center of the cylinder, denoted by $L$ in the figures. A few runs changing the core size while keeping the strength were also done, but these showed no significant changes.

\section{Viscous simulations: circular cylinder.}

In all the following simulations, the parameters have the following values: 
\[\frac{a}{R_c}=0.2, \quad Re=\frac{V_{c,0} D_c}{\nu}, \]
where $R_c(=D_c/2)$ denotes the radius of the circular cylinder. Since $V_{c,0}$ depends on the distance of the vortex from the cylinder center, the values of $R_c$ varied in the range 10 to 150 (approximately).

\begin{figure}
\centering
\includegraphics[scale=0.45]{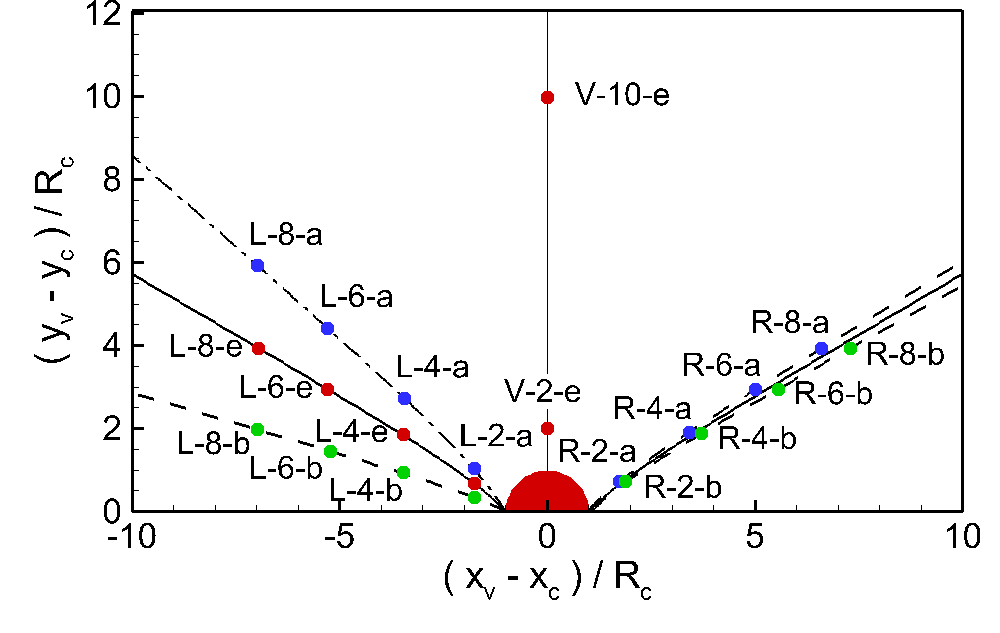}
\caption{The initial positions of the vortex centers on the F\"{o}ppl equilibrium curves, on two curves that are adjacent to each branch and on the vertical equilbrium line. }
\label{TrFo}
\end{figure}

 Figure \ref{TrFo} shows a variety of initial positions of the vortex center. The thick black curve represents the F\"{o}ppl equilibira curves, both the left and the right branches. Some of the initial positions were chosen above and below the equilibria curves, and lie on the dashed curves shown. These curves were chosen arbitrarily. On the left side, the $y$ coordinates on the lower dashed curve are half of those on the equilibrium curve, and the $y$ coordinates on the upper dashed curve are one and a half times those on the equilibrium curve. On the right side, the $y$ coordinates on the lower and upper dashed curves are 0.95 and 1.05 of those on the equilibrium curve, respectively. Due to the close proximity of these curves, not all starting positions are labeled  in the figure. A few positions were also chosen on the vertical equilibrium line.
\paragraph{Remark.} The time instants indicated in the snapshots that follow are approximate times and mainly provide a qualitative measure of time elapsed between different snapshots and events. 

\subsection{Starting configuration: moving F\"{o}ppl equilibria, trailing vortices.}

\begin{figure}
\centering
\begin{subfigure}
 \centering \includegraphics[scale=0.12]{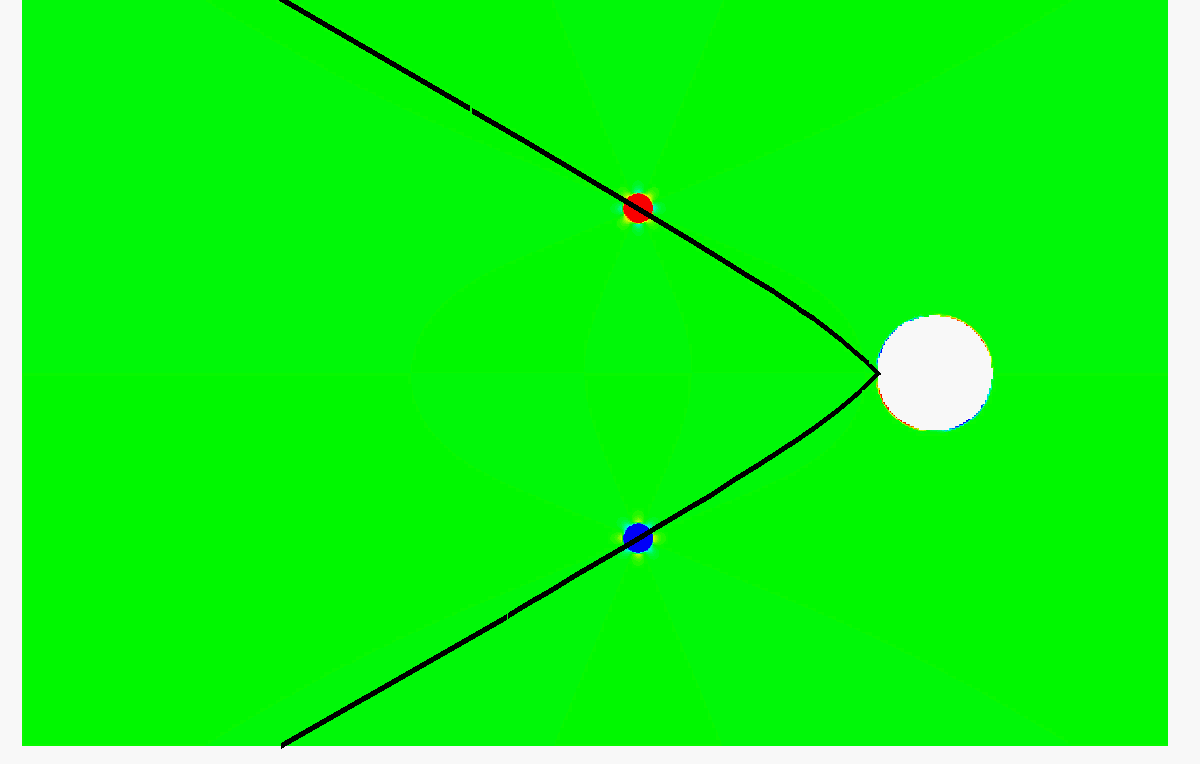} 
 {$t=0$}
\end{subfigure}
\begin{subfigure}
\centering \includegraphics[scale=0.12]{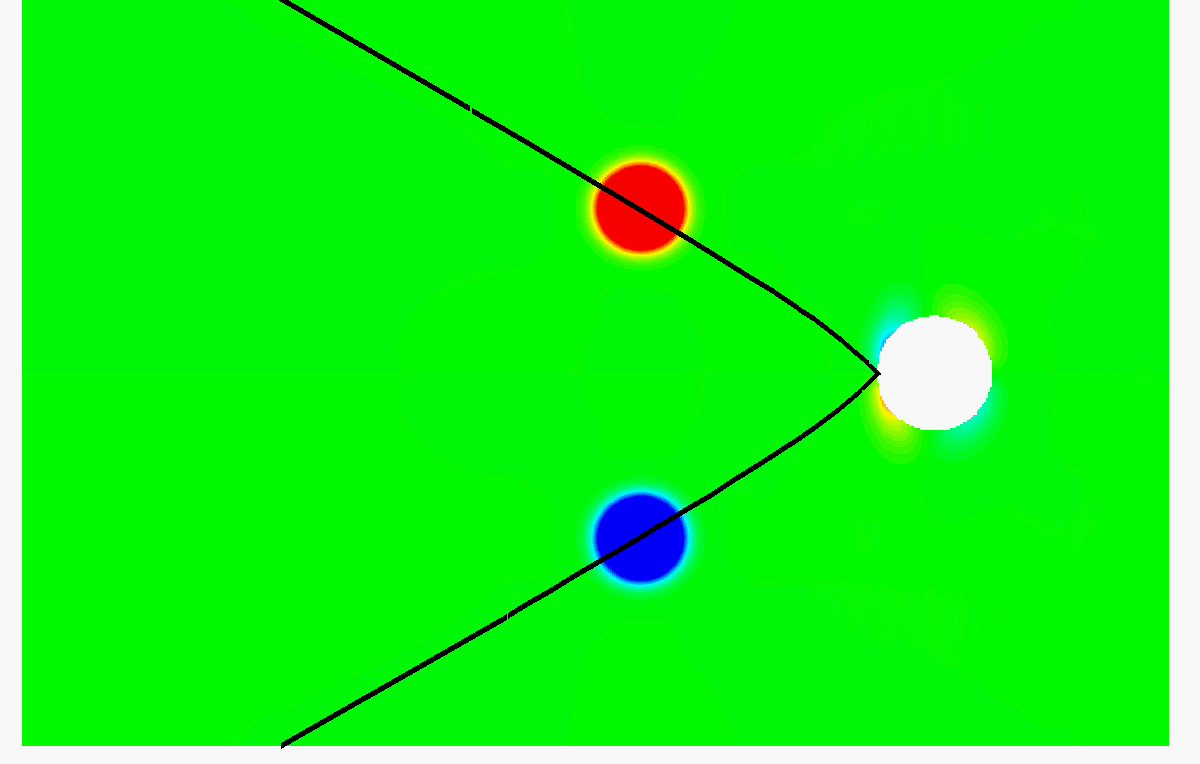}
 {$t=2$}
\end{subfigure}
\begin{subfigure}
\centering \includegraphics[scale=0.12]{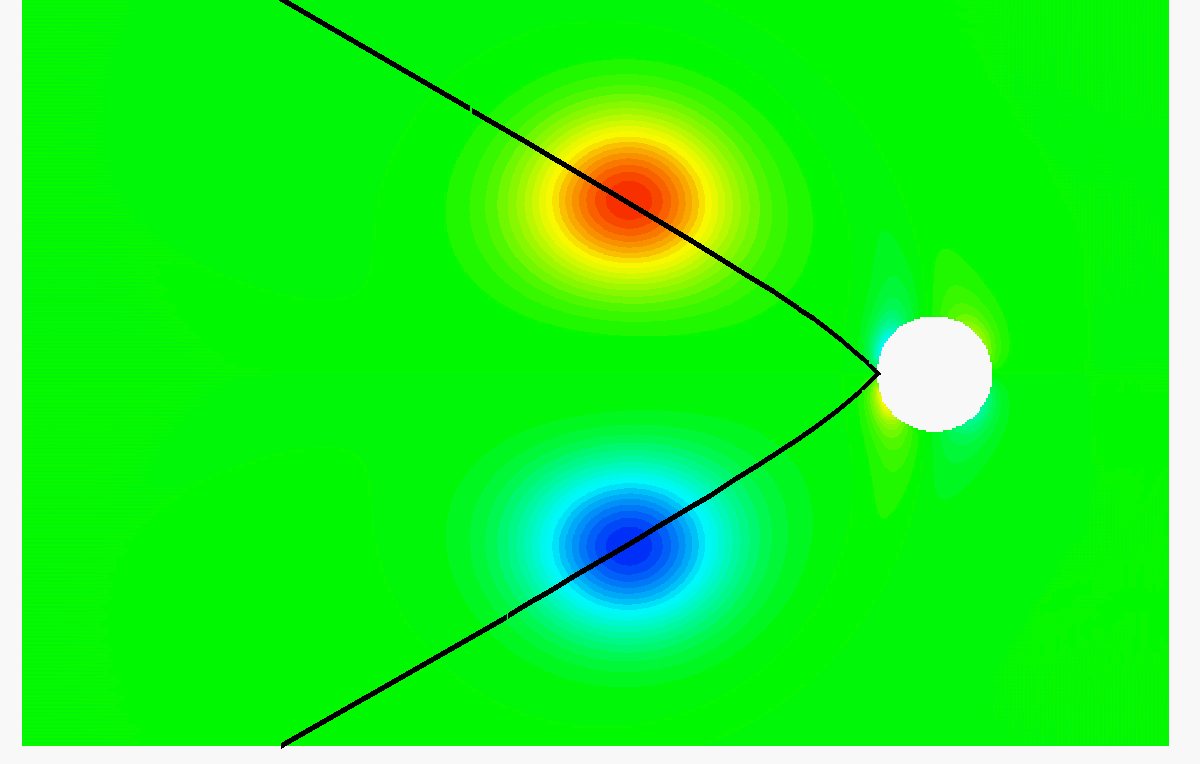}
 {$t=14$}
\end{subfigure}
\caption{Viscous interactions in the body-fixed frame corresponding to starting position L-6-e of Figure \ref{TrFo}. }
\label{L-6-e}
\end{figure}

\begin{figure}
\centering
\begin{subfigure}
 \centering \includegraphics[scale=0.12]{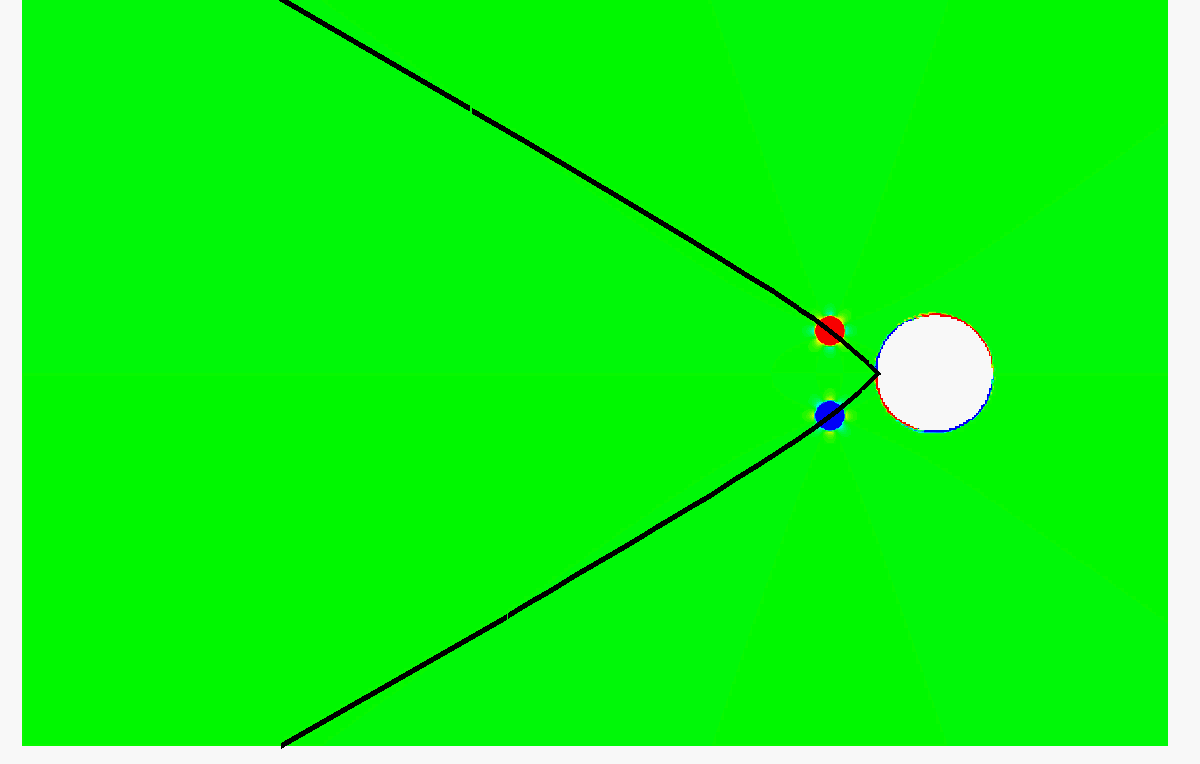} 
 {$t=0$}
\end{subfigure}
\begin{subfigure}
\centering \includegraphics[scale=0.12]{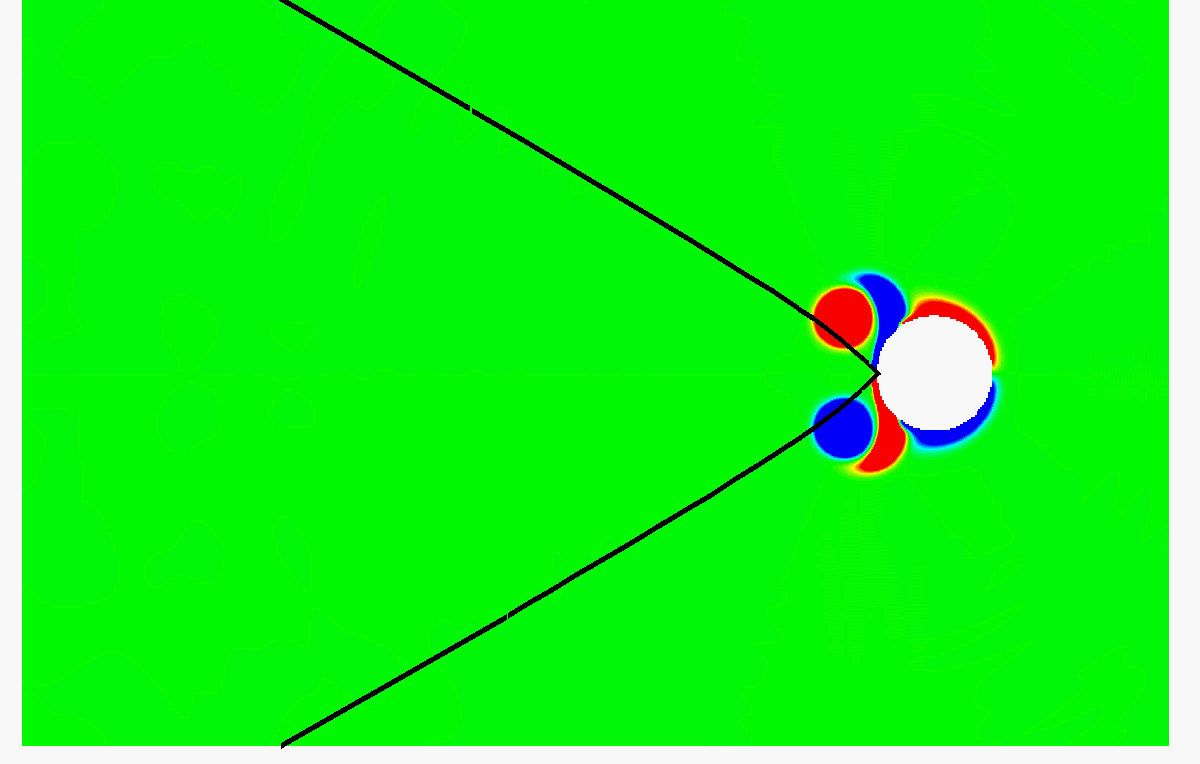}
 {$t=0.75$}
\end{subfigure}
\begin{subfigure}
\centering \includegraphics[scale=0.12]{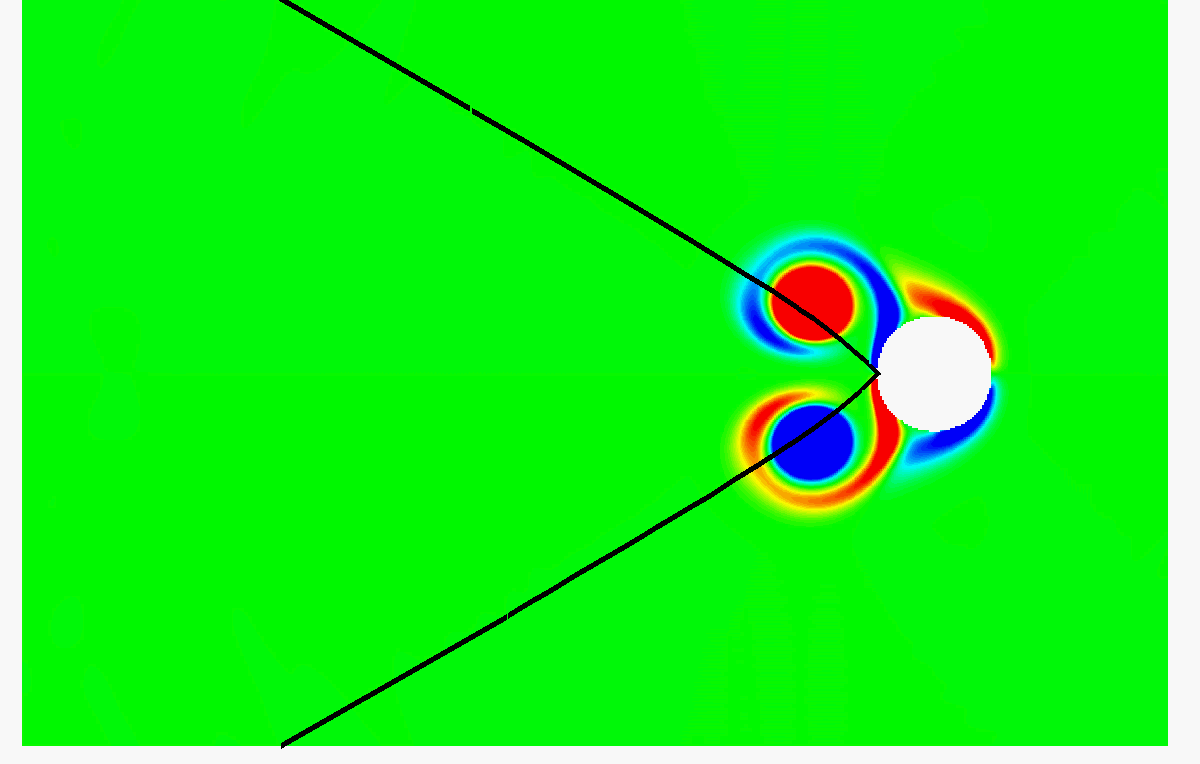}
 {$t=1.5$}
\end{subfigure}
\begin{subfigure}
\centering \includegraphics[scale=0.12]{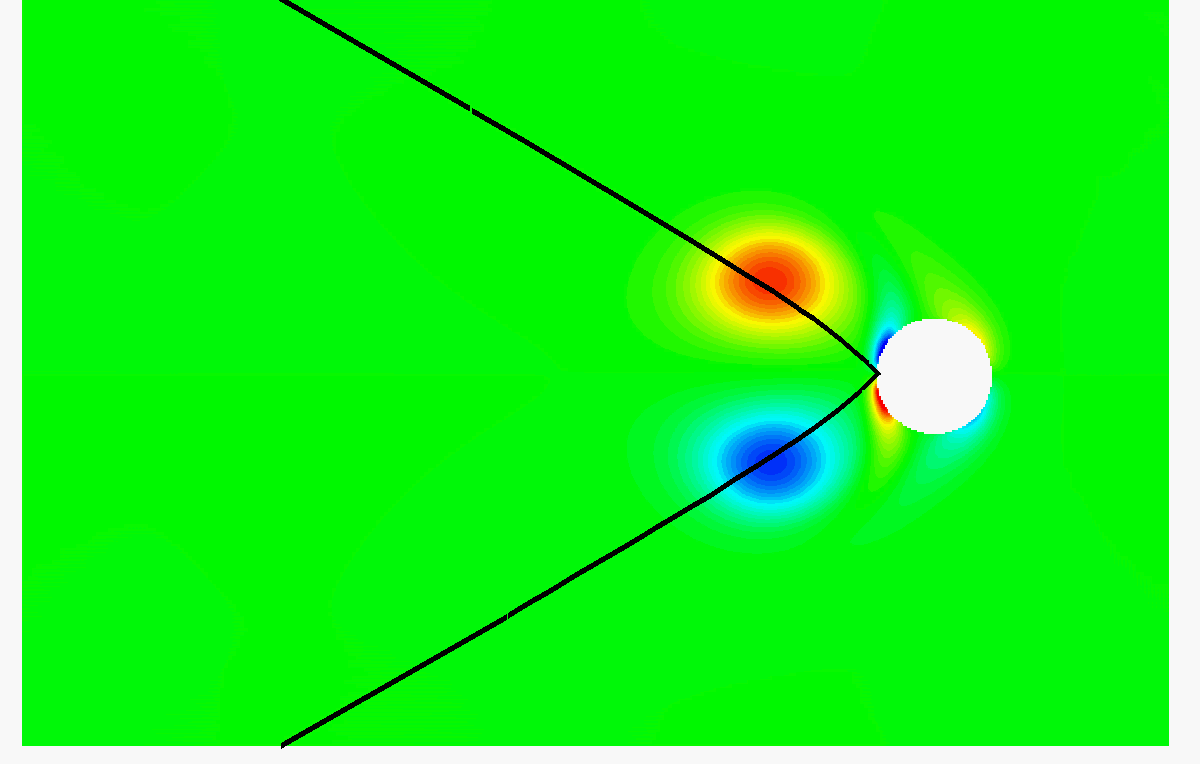}
 {$t=7$}
\end{subfigure}
\caption{Viscous interactions in the body-fixed frame corresponding to starting position L-2-e of Figure \ref{TrFo}. }
\label{L-2-e}
\end{figure}

Snapshots at different instants of time corresponding to the initial points L-6-e and L-2-e are shown in Figures \ref{L-6-e} and \ref{L-2-e}, respectively. Also, shown in these snapshots are the {\it instantaneous} F\"{o}ppl curves, i.e. the equilibrium curves drawn with respect to the instantaneous position of the cylinder. Note that, as expected,  in all evolutions the vortex cores diffuse due to viscosity.

Starting from position L-8-e, the vortex cores remain on the instantaneous F\"{o}ppl curves while the vorticity diffuses and, moreover, maintain their initial center positions on the curves. The cylinder is pushed by the vortices for a significant distance and no significant development of boundary layers or secondary vorticity  is seen. From starting position L-6-e, the qualitative behavior is mostly the same, and is shown in Figure \ref{L-6-e}. There is slightly more development of the viscous boundary layers. There is also a slight rearward movement of the vortex along the equilibrium curve towards the end of the sequence as the vorticity becomes weaker. Both these phenomena are more easily discerned in the movie.

    From starting position L-2-e, more development of the viscous boundary layers is observed and significantly more events are seen, as shown in Figure \ref{L-2-e}.
It may be noticed in each half-pane (upper or lower) that the boundary layer nearer to each vortex has the opposite-signed vorticity while the boundary layer on the farther side has the same-signed vorticity. This suggests the presence of a stagnation point somewhere near the top surface of the cylinder. Boundary layer separation occurs on the windward side (i.e. side closer to the vortex). The separated shear layer of opposite-signed vorticity wraps around the primary vortex. During this process the primary vortex deforms, rotates in a rigid body manner and also gets displaced from the instantaneous curves. With diffusion, the entrainment of the separated shear layer gets weaker and the primary vortex moves back to the equilibrium curves.The rearward movement of the vortex along the equilibrium curves towards the end of the sequence is more clearly seen in this case. 

\begin{figure}
\centering
\begin{subfigure}
 \centering \includegraphics[scale=0.12]{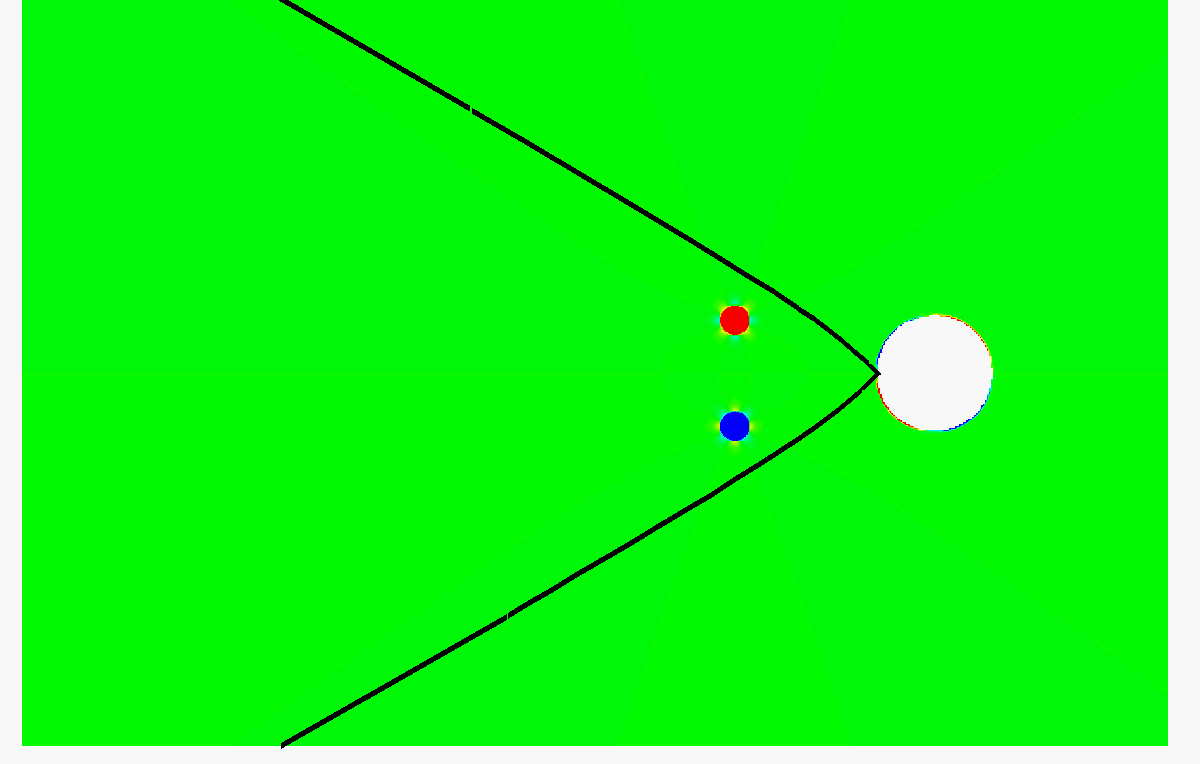} 
 {$t=0$}
\end{subfigure}
\begin{subfigure}
\centering \includegraphics[scale=0.12]{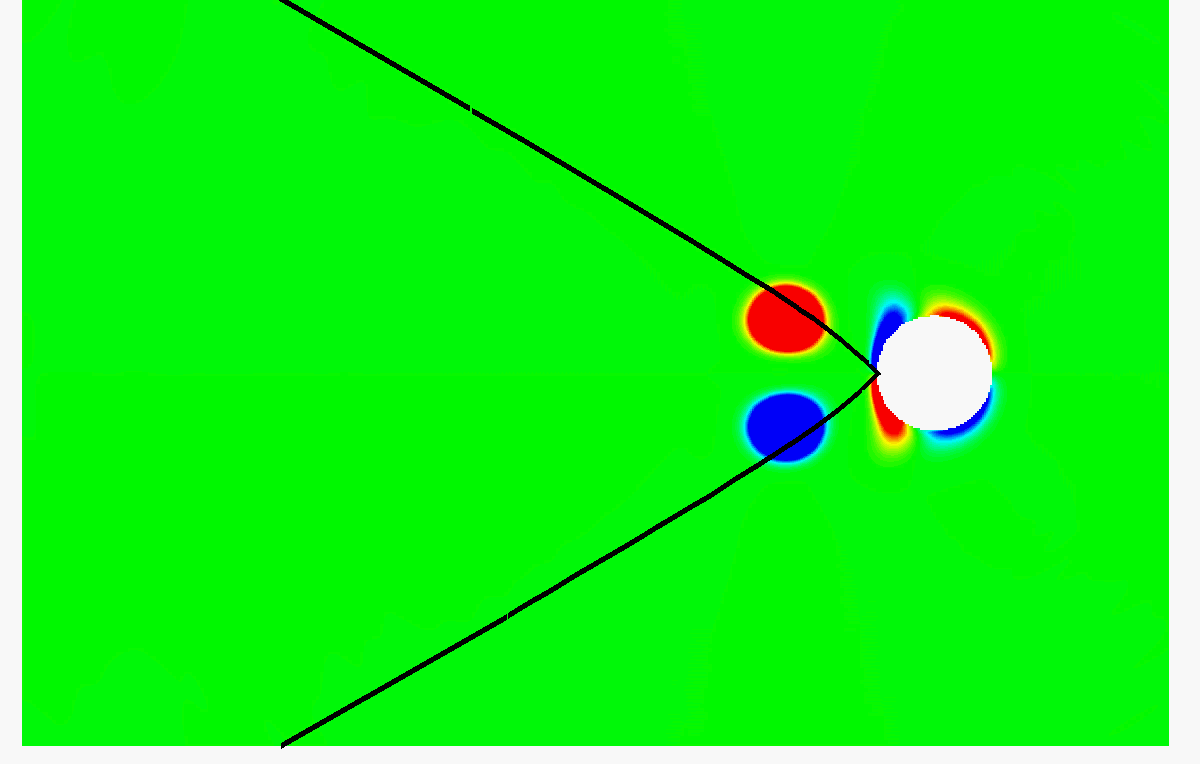}
 {$t=1$}
\end{subfigure}
\begin{subfigure}
\centering \includegraphics[scale=0.12]{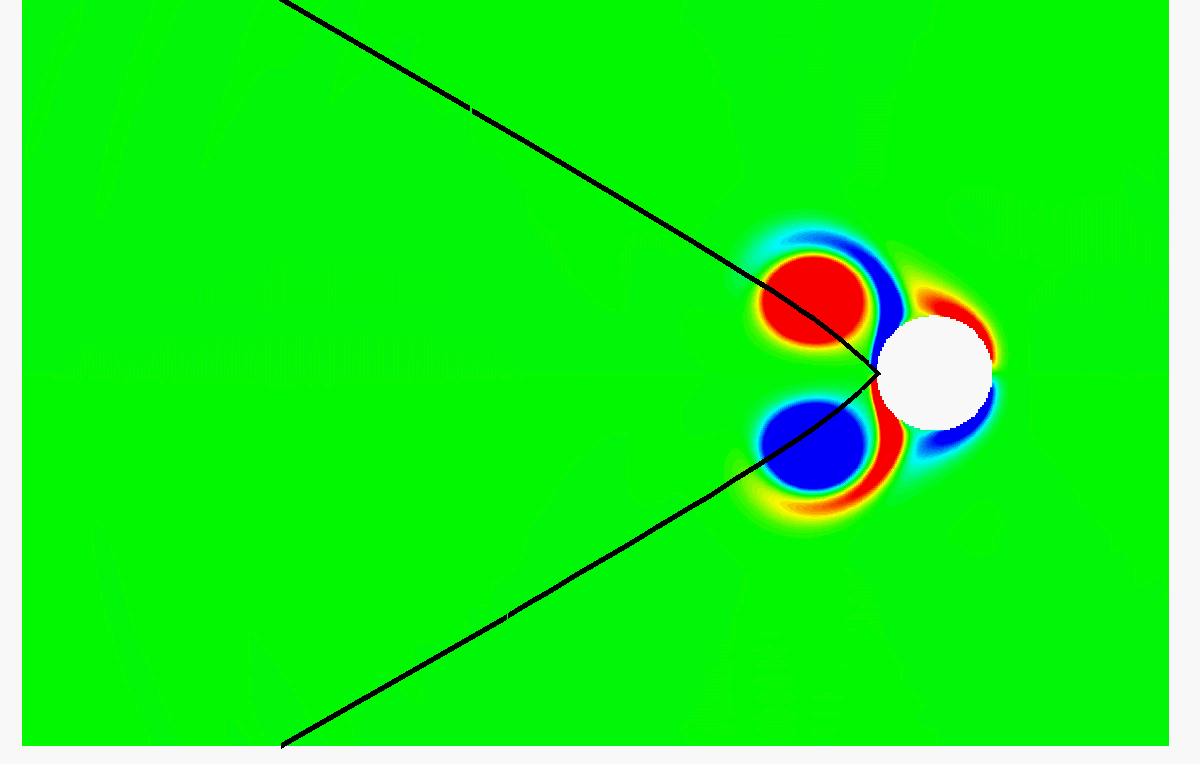}
 {$t=3$}
\end{subfigure}
\begin{subfigure}
\centering \includegraphics[scale=0.12]{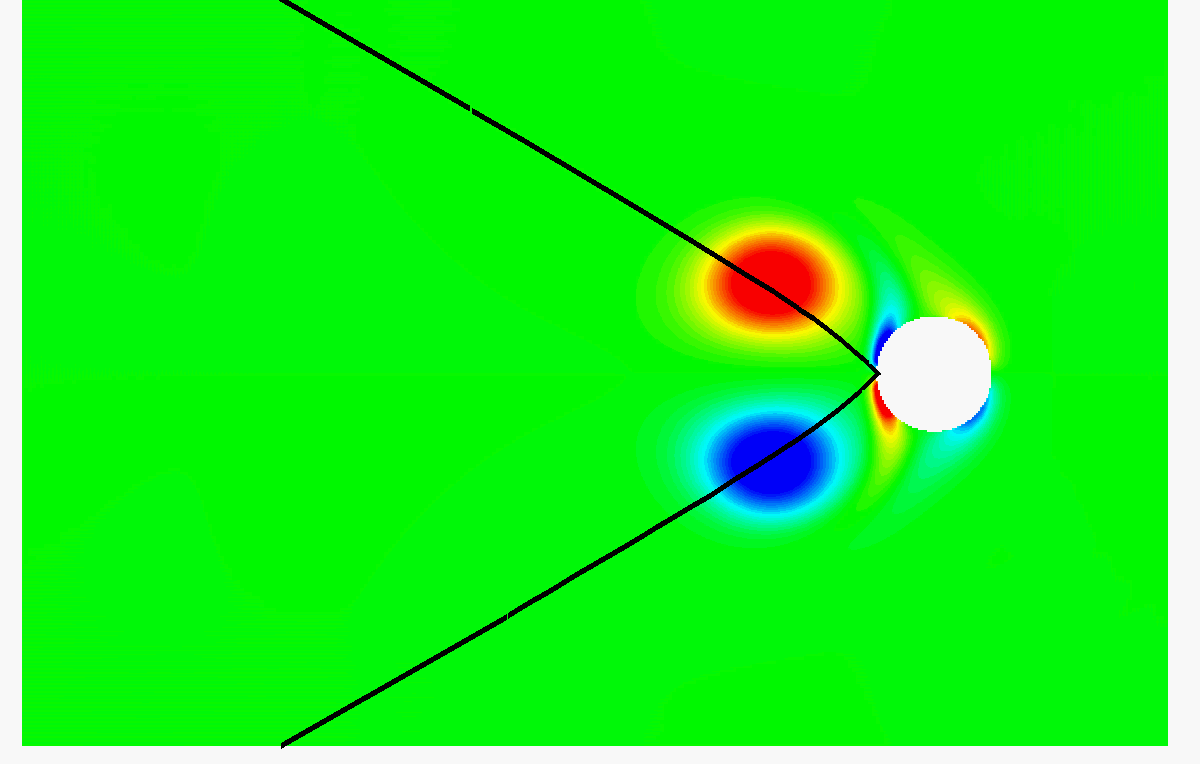}
 {$t=8$}
\end{subfigure}
\caption{Viscous interactions in the body-fixed frame corresponding to starting position L-4-b of Figure \ref{TrFo}. }
\label{L-4-b}
\end{figure}

    In another set of interactions, the starting positions of the vortices are chosen below the F\"{o}ppl equilibrium curves corresponding to the initial points L-2-b, L-4-b, L-6-b  and L-8-b  in Figure \ref{TrFo}. The qualitative trends from cases L-8-b to L-2-b, as far as development, separation and entrainment of the boundary layers and vorticity diffusion go, is similar as in the previous set of interactions. For near starting positions, like L-2-b and L-4-b there is more significant separation and entrainment of the boundary layers, compared to that for the far starting position L-8-b. However, another interesting feature is seen. The vortices in all cases all seem to approach the cylinder but are seemingly unable to  cross the instantaneous equilibrium curves and are then pushed back along them. A typical sequence starting from position L-4-b is shown in Figure \ref{L-4-b}. 
 
\begin{figure}
\centering
\begin{subfigure}
 \centering \includegraphics[scale=0.12]{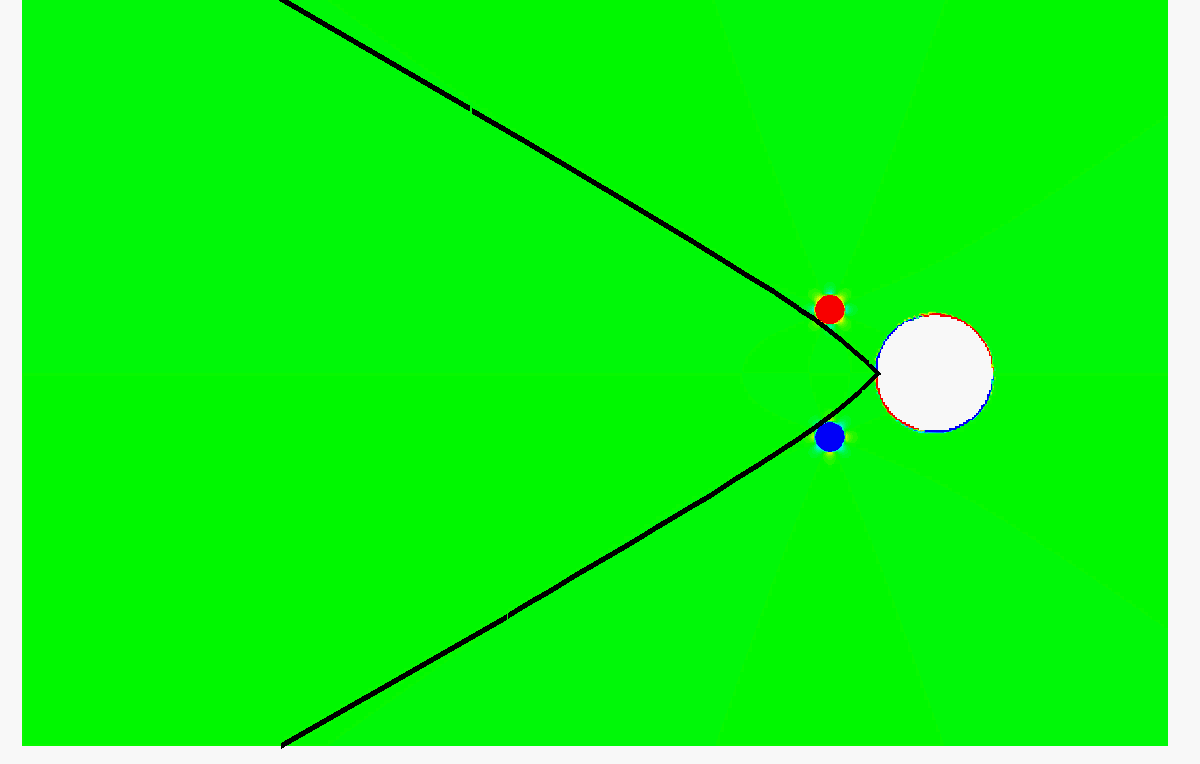} 
 {$t=0$}
\end{subfigure}
\begin{subfigure}
\centering \includegraphics[scale=0.12]{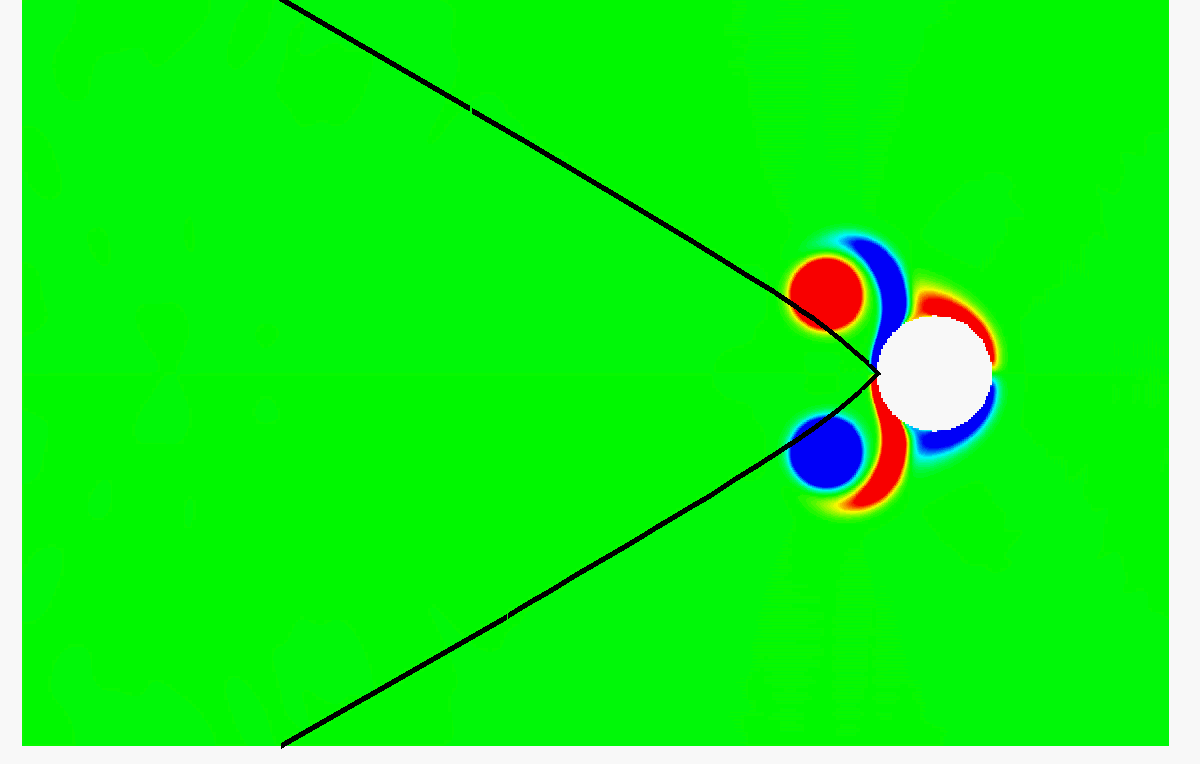}
{$t=1$}
\end{subfigure}
\begin{subfigure}
\centering \includegraphics[scale=0.12]{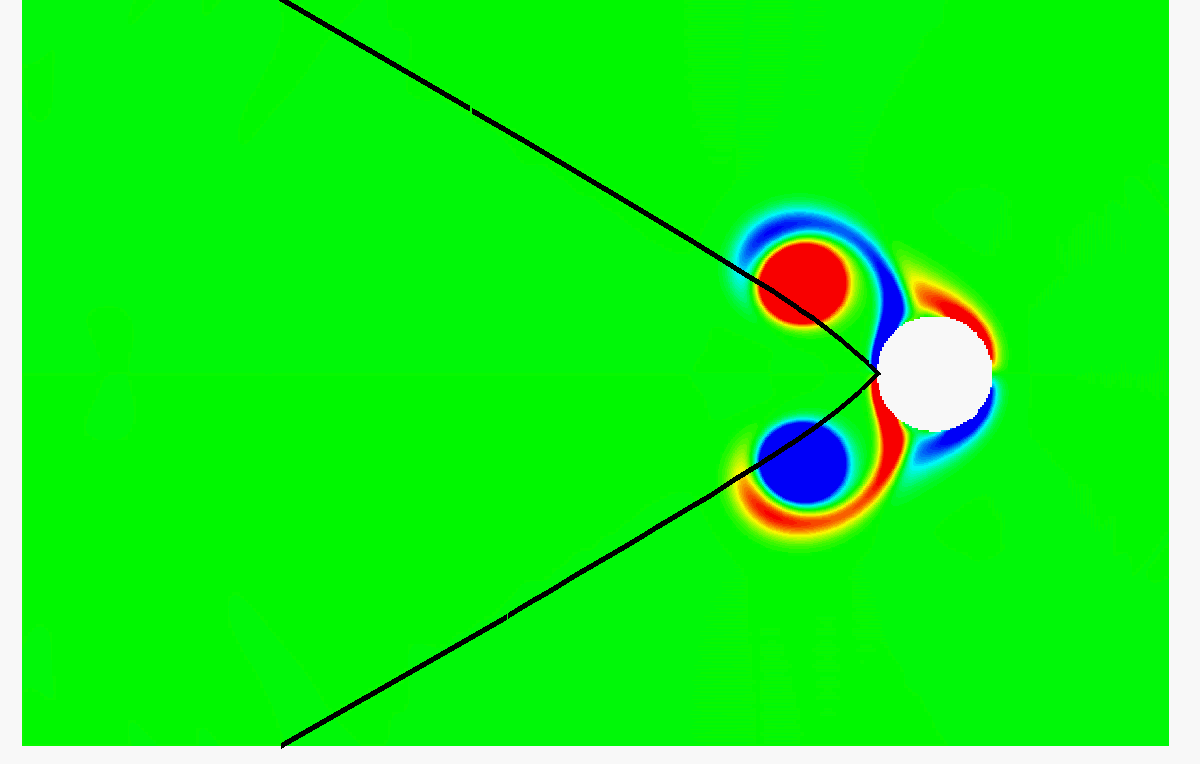}
 {$t=2$}
\end{subfigure}
\begin{subfigure}
\centering \includegraphics[scale=0.12]{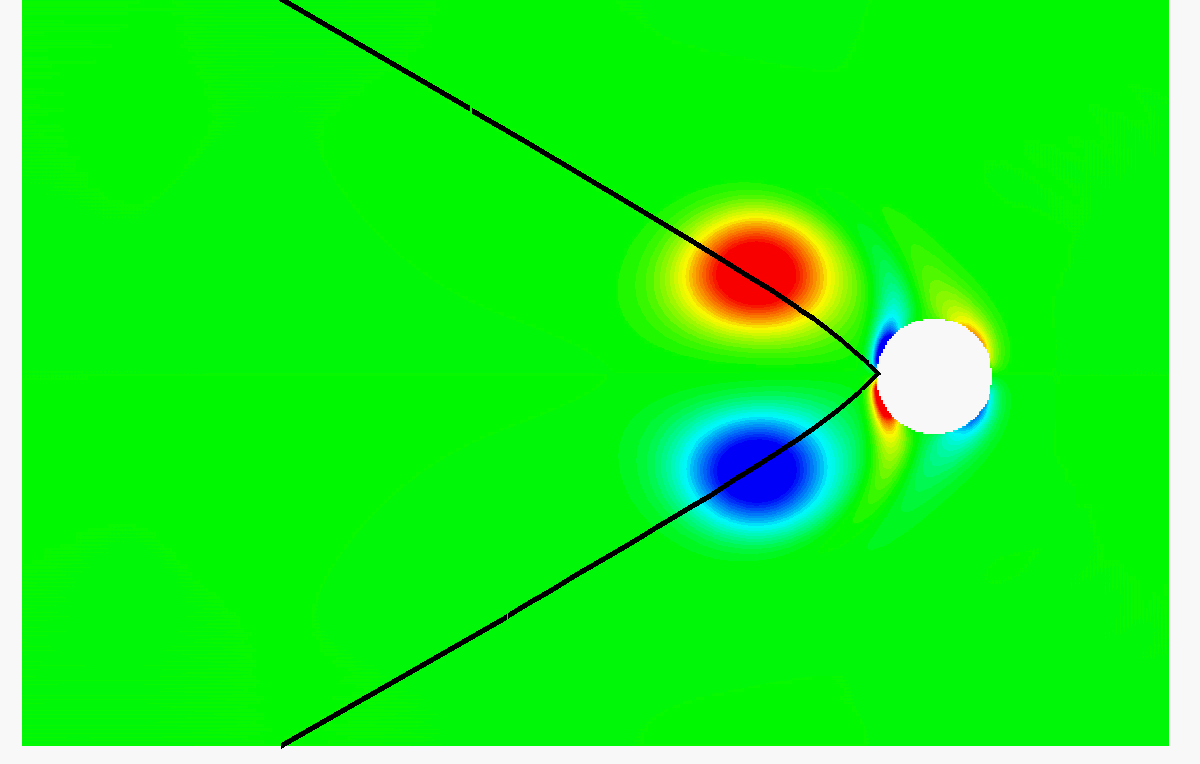}
 {$t=9$}
\end{subfigure}
\caption{Viscous interactions in the body-fixed frame corresponding to starting position L-2-a of Figure \ref{TrFo}. }
\label{L-2-a}
\end{figure}

   Similar qualitative trends are seen for interactions with starting positions corresponding to the initial points L-2-a, L-4-a, L-6-a  and L-8-a  in Figure \ref{TrFo}, i.e. with the vortices starting above the F\"{o}ppl equilibrium curves.  Similar trends are observed in the development of the boundary layers, and one again finds that the instantaneous left equilibrium curves attract the vortices and they are pushed back along these curves as they diffuse. A typical sequence is seen in Figure \ref{L-2-a}.

\begin{figure}
\centering
\includegraphics[scale=0.35]{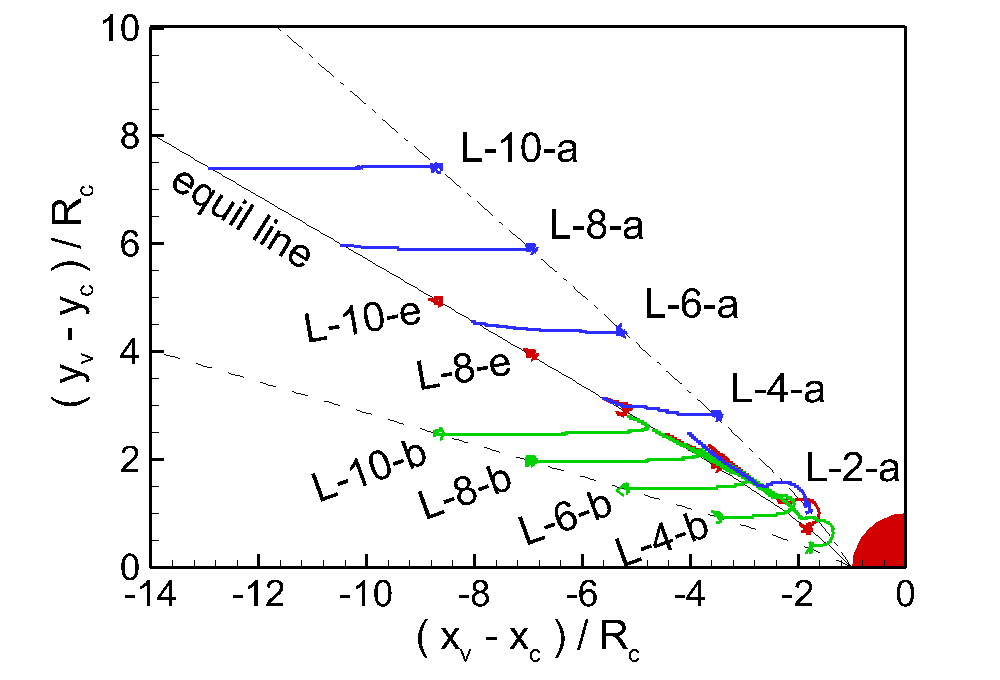}
\caption{Initial positions of the vortices  on the left F\"{o}ppl equilibrium curves and movement of the vorticity maximum point during the evolution in each case. }
\label{MoVMaxTr}
\end{figure}

   To summarize, these interactions generally show that the trailing vortex configuration is stable, even when there is boundary layer separation and entrainment, and that the instantaneous left F\"{o}ppl curves behave like attractors in all cases. 
The movement of the point of vorticity maximum during each evolution is shown in Figure \ref{MoVMaxTr}.

\begin{figure}
\centering
\includegraphics[scale=0.18]{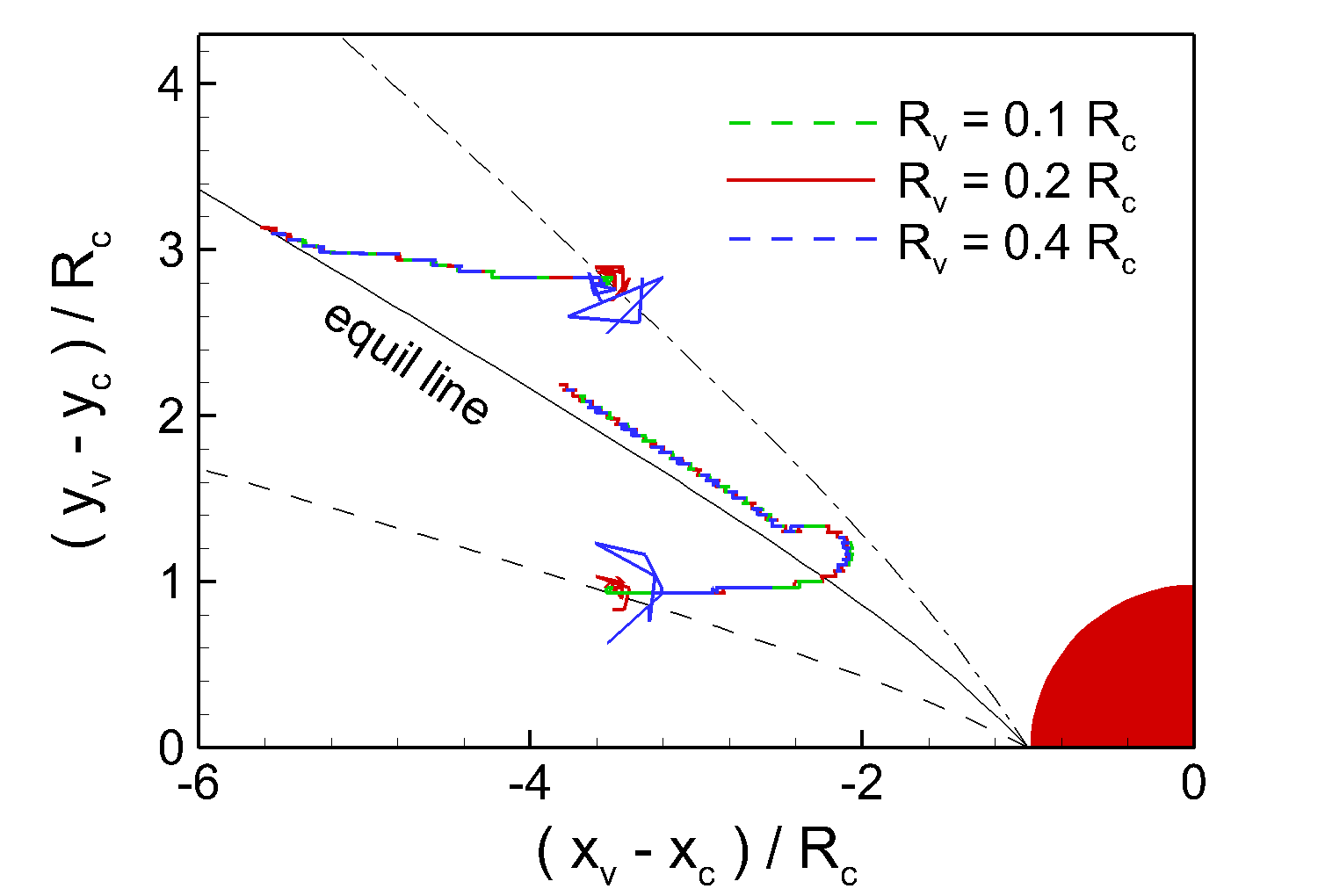}
\includegraphics[scale=0.24]{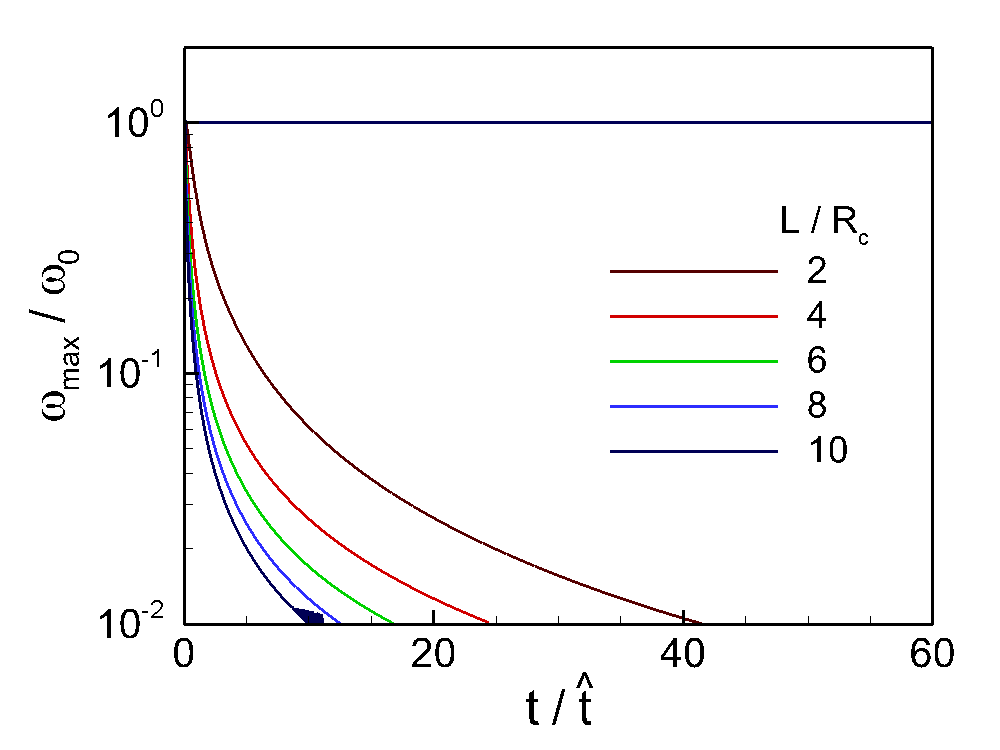}
\caption{Movement of the vorticity maximum point for different cases of initial vortex radius (left ) and decay of vorticity maximum with time (right).}
\label{VorRadiusVorMax}
\end{figure}

As mentioned earlier, a few runs changing the initial vortex core radius while keeping the initial vortex strength the same were also done. The left box of Figure \ref{VorRadiusVorMax} shows the movement of the vorticity maximum for two starting positions for each of three different radii. It is clear that in these runs the radius has very little effect, the trajectories of the vorticity maximum shown in red, blue and green, almost lie on top of each other. This is most likely because decreasing/increasing the initial radius, while keeping the strength same, results in increasing/decreasing, respectively, the initial vorticity in the core. This in turn results in increasing/decreasing the initial vorticity diffusion rate, so that the effect of the change in initial vorticity induced by the change in vortex radius is annulled in just a few time steps. 
A typical decay of the vorticity maximum (logarithmic scale) with time is shown in the right box of Figure \ref{VorRadiusVorMax} for initial positions of the vortex on the equilibrium curve.

\subsubsection{Velocity of cylinder.}

\begin{figure}
\centering
\includegraphics[scale=0.17]{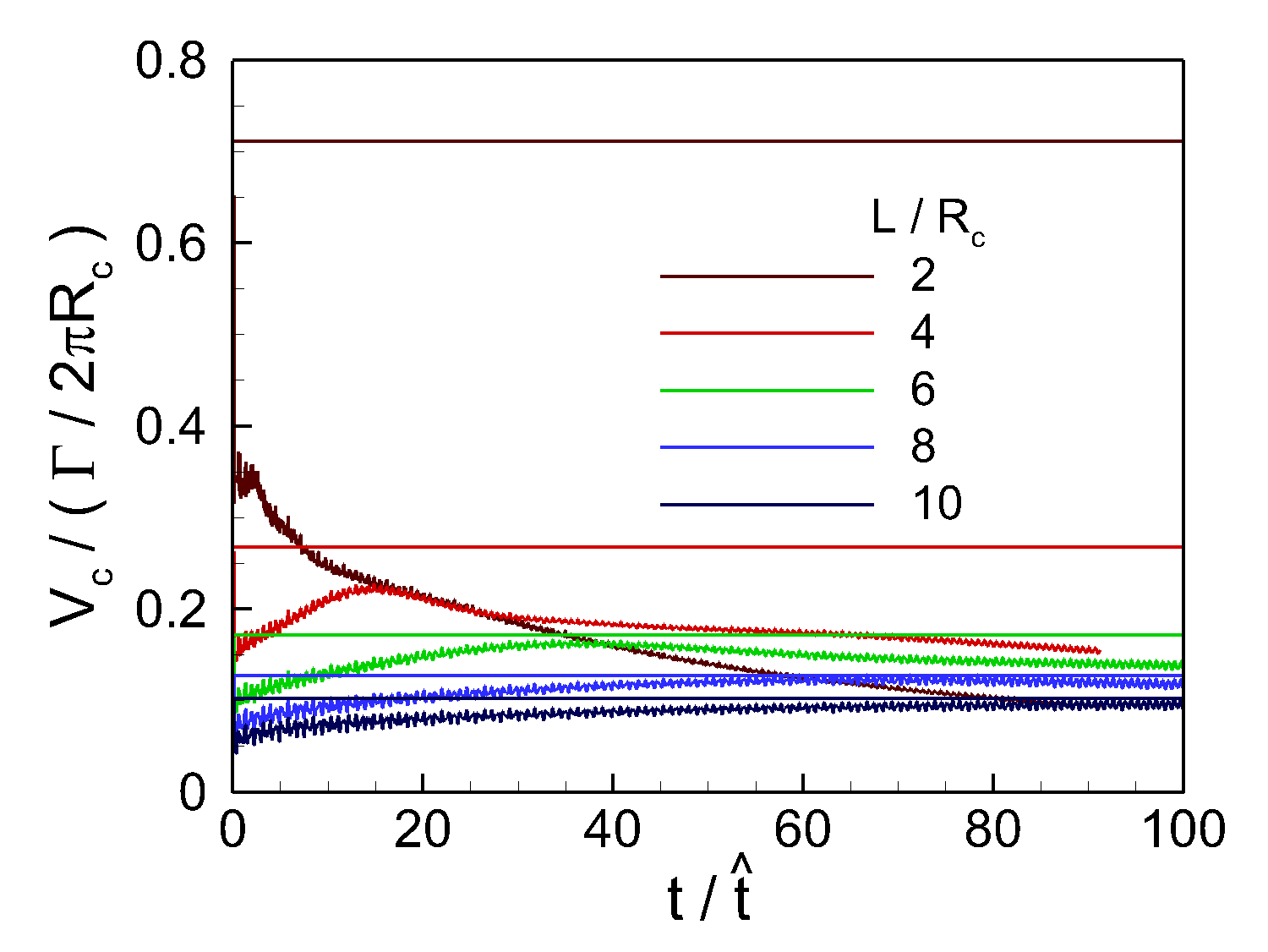} 
\includegraphics[scale=0.25]{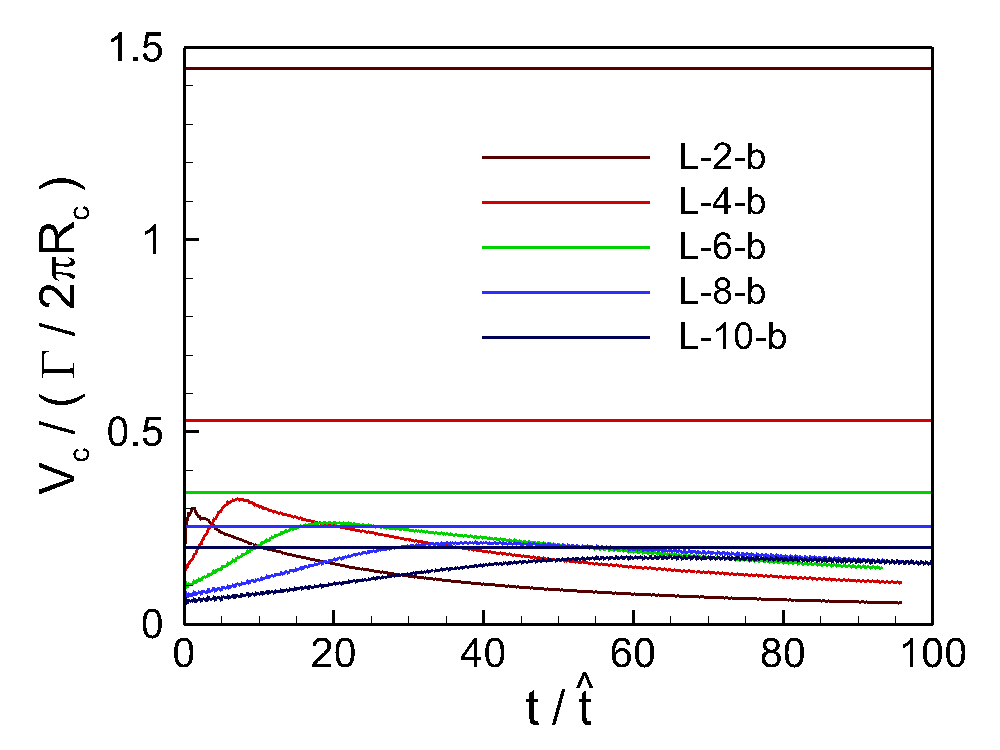}
\includegraphics[scale=0.25]{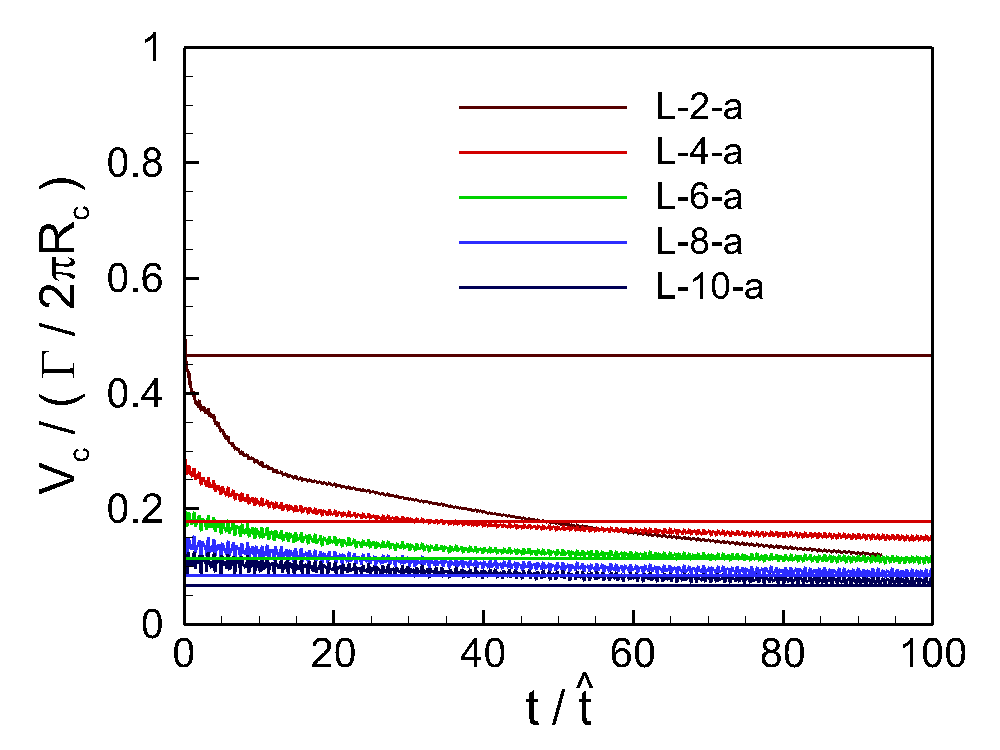}
\caption{Cylinder forward speed versus time corresponding to various starting positions on and near the left equilibrium curves, see Figure \ref{TrFo}. Top left=on the curves, top right=below the curves, bottom=above the curves. The solid straight lines represent the values of $V_{c,0}/(\Gamma / 2 \pi R_c)$for each $L/R_c$ value.}
\label{Vel-time-L}
\end{figure}

  To illustrate the effect of the vortex interactions on the cylinder forward speed, plots of cylinder versus time are shown in Figure \ref{Vel-time-L} for cases on, below and above the equilibrium curves, respectively. 
\begin{figure}
\centering
\includegraphics[scale=0.18]{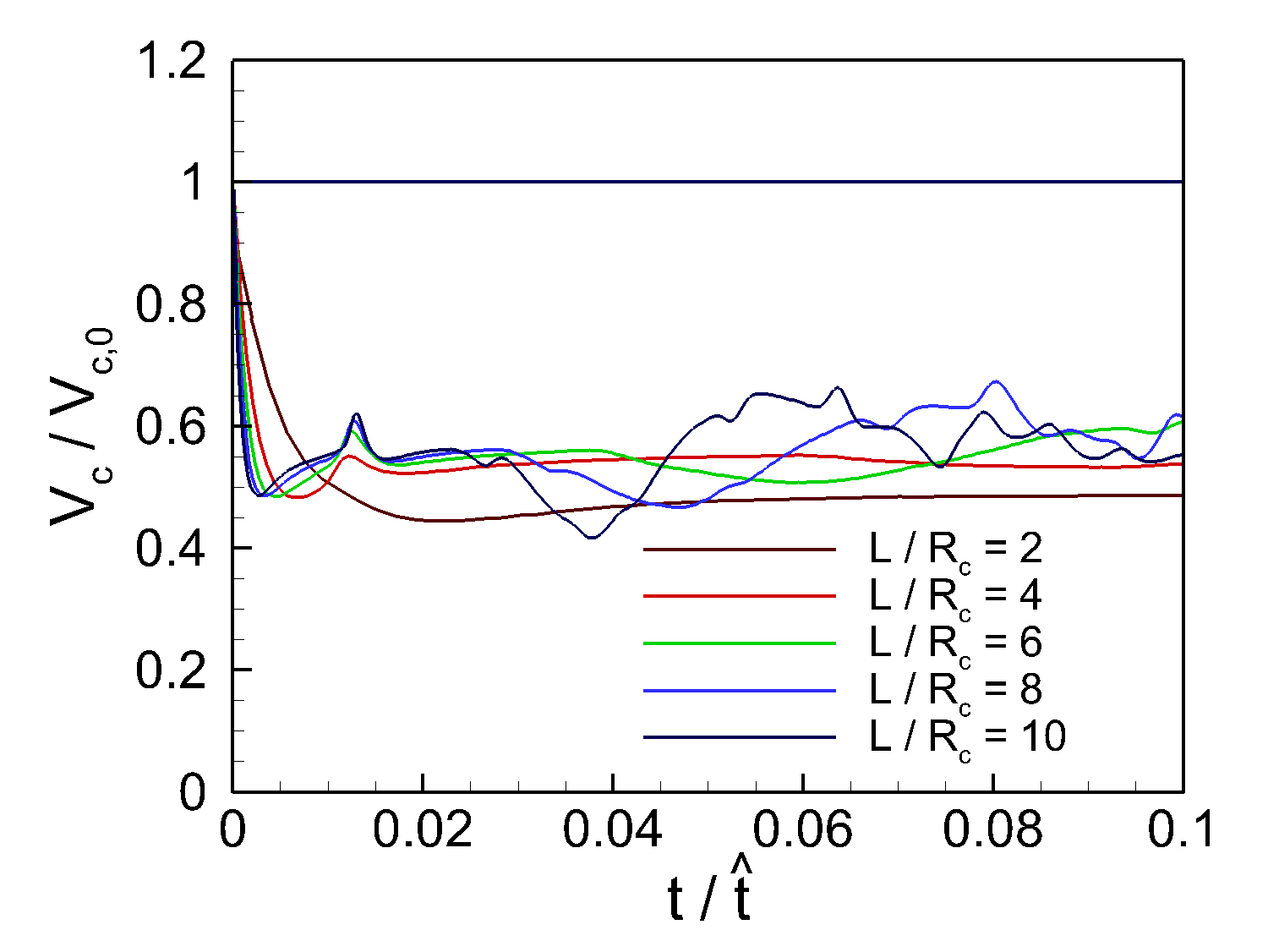} 
\caption{Transient behavior in the cylinder velocity corresponding to starting positions on the left equilibrium curve.}
\label{transients}
\end{figure}
The velocity of the cylinder is normalized by $V_{c,0}$. Time could be normalized by $\hat{t}:=R_c/V_{c,0}$.  This choice of $\hat{t}$ is  natural, but it does introduce a differential compression of the non-dimensionalized time axis since $V_{c,0}$ is a function of $L/R_c$; see equation (\ref{eq:velposition}), $L$ denotes the initial distance of the vortex from the cylinder center. A better sense of the behavior of the cylinder motion is obtained by non-dimensionalizing time by $2 \pi R_c^2/\Gamma$, since both $R_c$ and $\Gamma$ are constants in all the cases. These plots are shown in Figure \ref{Vel-time-L}. It should be noted that for the off-equilibrium starting positions since there is no natural choice for $V_{c,0}$, the latter was computed from equation (\ref{eq:velposition}).

   Two general comments may first be made about these plots. There is an initial transient behavior observed when viscosity is turned on, as previously mentioned, which causes the velocity to quickly fall below or rise above the inviscid equilibrium speed. The transient behavior corresponding to the top left case of Figure \ref{Vel-time-L} is shown on a time interval of $0.1$ seconds in Figure \ref{transients}.  Moreover, some large-scale high frequency oscillations are seen which are, however, an artifact of the Lattice Boltzmann Method caused by the treatment of moving boundaries. When the grid nodes are covered and uncovered in the motion of the cylinder, the momentum exchange between the cylinder and ambient fluid is disturbed. But this disturbance, as clearly seen, does not influence the motion of the cylinder in the time-averaged sense.

Focusing first on the top left box, where the solid lines correspond to the inviscid equilibrium speed, one notices that the transient behavior results in an initial `braking'. For other than the closest starting position, the cylinder then accelerates towards the inviscid equilibrium speed. The approach to this speed is closer and persists for longer times farther away the staring position of the vortices.  After peak speed is reached the cylinder slows down. The deviation from the inviscid solutions becomes larger closer the starting position of the vortex is to the cylinder with the peak speed decreasing and occurring at earlier times. For the closest starting position, as in Figure \ref{L-2-e}, the speed almost monotonically decreases after the transient phase. For starting positions below the curves, one observes an initial acceleration phase as the vortices are  attracted towards the curves and then the cylinder speed starts slowing down more significantly than in the cease starting from the equilibrium curves. For starting positions above the curves, no acceleration phase is observed even as the vortices are  attracted (in the opposite direction) towards the curves and the cylinder speed slows down monotonically from the initial time. The deceleration is more for starting positions closer to the cylinder. 

\begin{figure}
\centering
\begin{subfigure}
 \centering \includegraphics[scale=0.12]{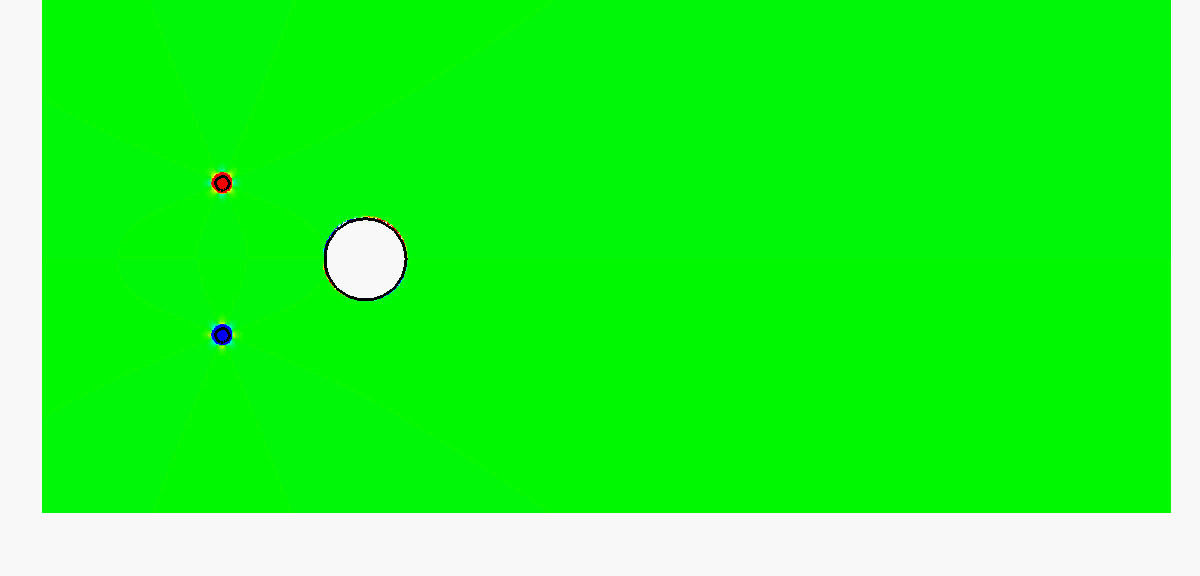} 
 {$t=0$}
\end{subfigure}
\begin{subfigure}
\centering \includegraphics[scale=0.12]{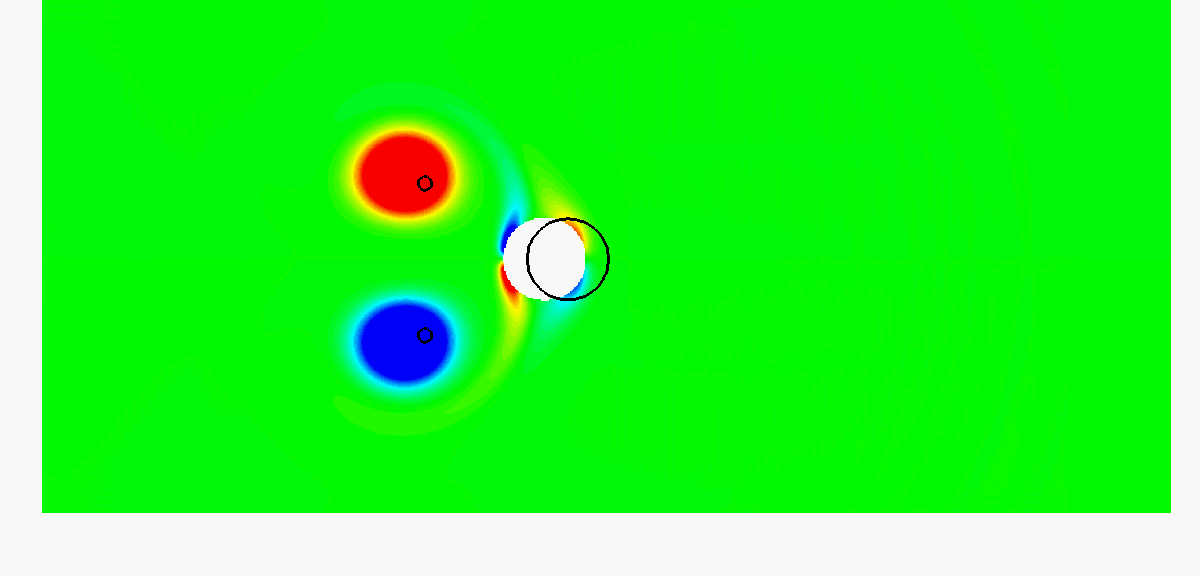}
 {$t=4$}
\end{subfigure}
\begin{subfigure}
\centering \includegraphics[scale=0.12]{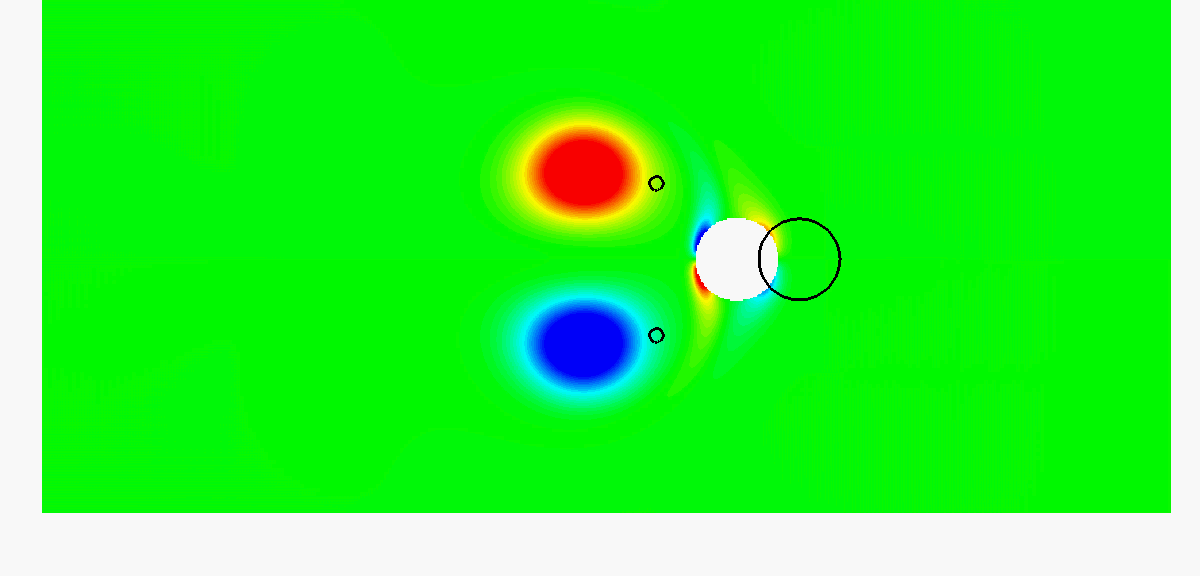}
 {$t=8$}
\end{subfigure}
\begin{subfigure}
\centering \includegraphics[scale=0.12]{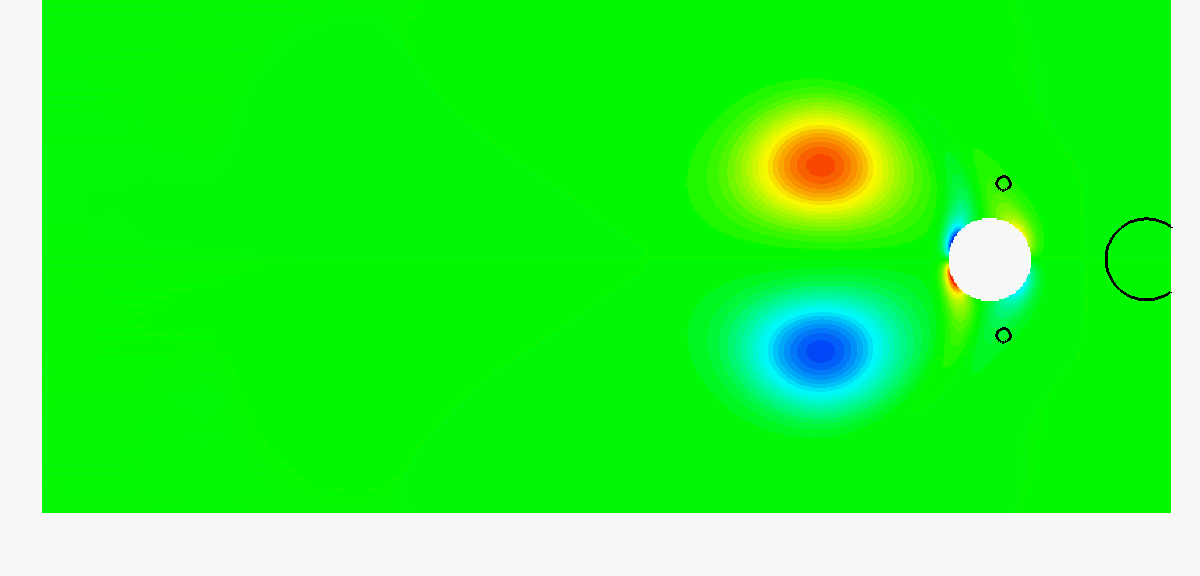}
 {$t=14$}
\end{subfigure}
\caption{Viscous interactions in the body-fixed frame from a starting position on the left equilibrium curve compared with the inviscid evolution of the cylinder and point vortices (black circles) . }
\label{left-b}
\end{figure}

   In Figure \ref{left-b}, snapshots of the interactions are shown in a spatially-fixed frame for positions starting on the left equilibrium curves along with the inviscid evolution of the cylinder and point vortices. 
One clearly sees the slowing down effect of viscous drag. The slowing down is observed to be weaker for the starting position farther away from the cylinder consistent with the plots of Figure \ref{Vel-time-L}.


\subsubsection{Linear momentum exchange during interactions.} 
  In a dynamically coupled interaction, the total fluid+solid total momentum is conserved. 
\begin{figure}
\centering
 \includegraphics[scale=0.17]{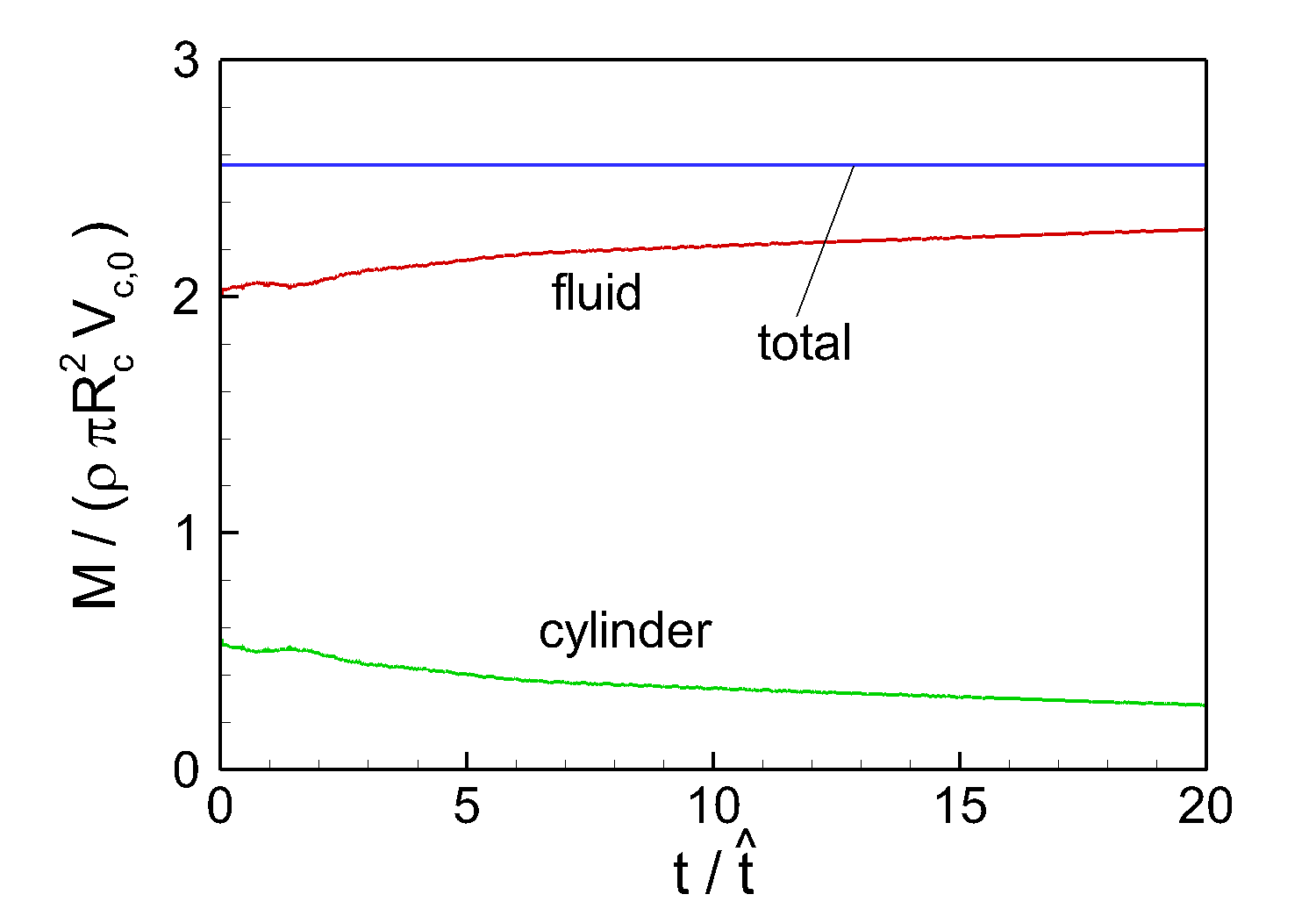} 
 \includegraphics[scale=0.17]{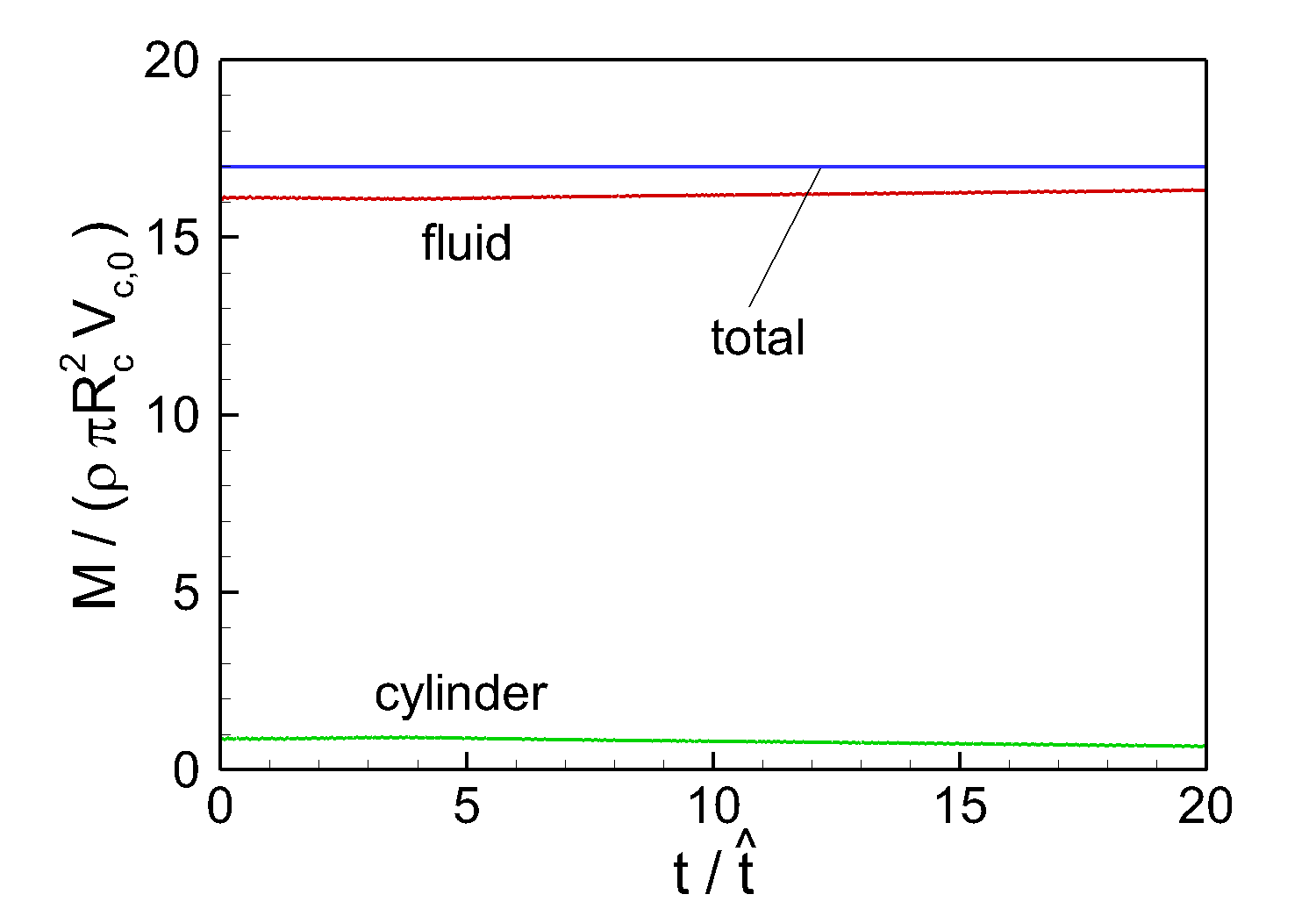}
\caption{Linear momentum exchange for two cases of vortices starting on the F\"{o}ppl left equilibrium curves: left box for starting position L-2-e and right box for starting position L-4-e.}
\label{mom-exchange-L}
\end{figure}
Figure \ref{mom-exchange-L} shows two sets of plots  for two different starting positions of the vortices on the left equilibrium curves that track the linear momentum of the fluid alone, the linear momentum of the  circular cylinder alone and the total linear momentum. The conservation of the total linear momentum is easily seen, even as the fluid and cylinder exchange their linear momenta.


\subsection{Starting configuration: moving F\"{o}ppl equilibria, leading vortices.}

Next, starting positions on and around the F\"{o}ppl equilibrium curves in front of the cylinder are chosen, denoted by the R- positions in Figure \ref{TrFo}. These evolutions show significantly more unstable behavior when compared  with the evolutions associated with the left equilibrium curves. In a few cases, the vortices accelerate to the right  and drift away from the right  equilibrium curves. This drift is faster closer the starting position is to the cylinder. But in a few other cases, interestingly, they pass over the cylinder and drift towards  towards the {\it left} instantaneous F\"{o}ppl curves, and stay on them once they reach them. In such evolutions the cylinder, viewed in a stationary frame, overtakes the vortices by threading through them. Some of these drifts towards the left curves, when the starting position of the vortices is far from the cylinder, seem to occur after the vortices close to their starting positions for a significant amount of time.

\begin{figure}
\centering
\begin{subfigure}
 \centering \includegraphics[scale=0.12]{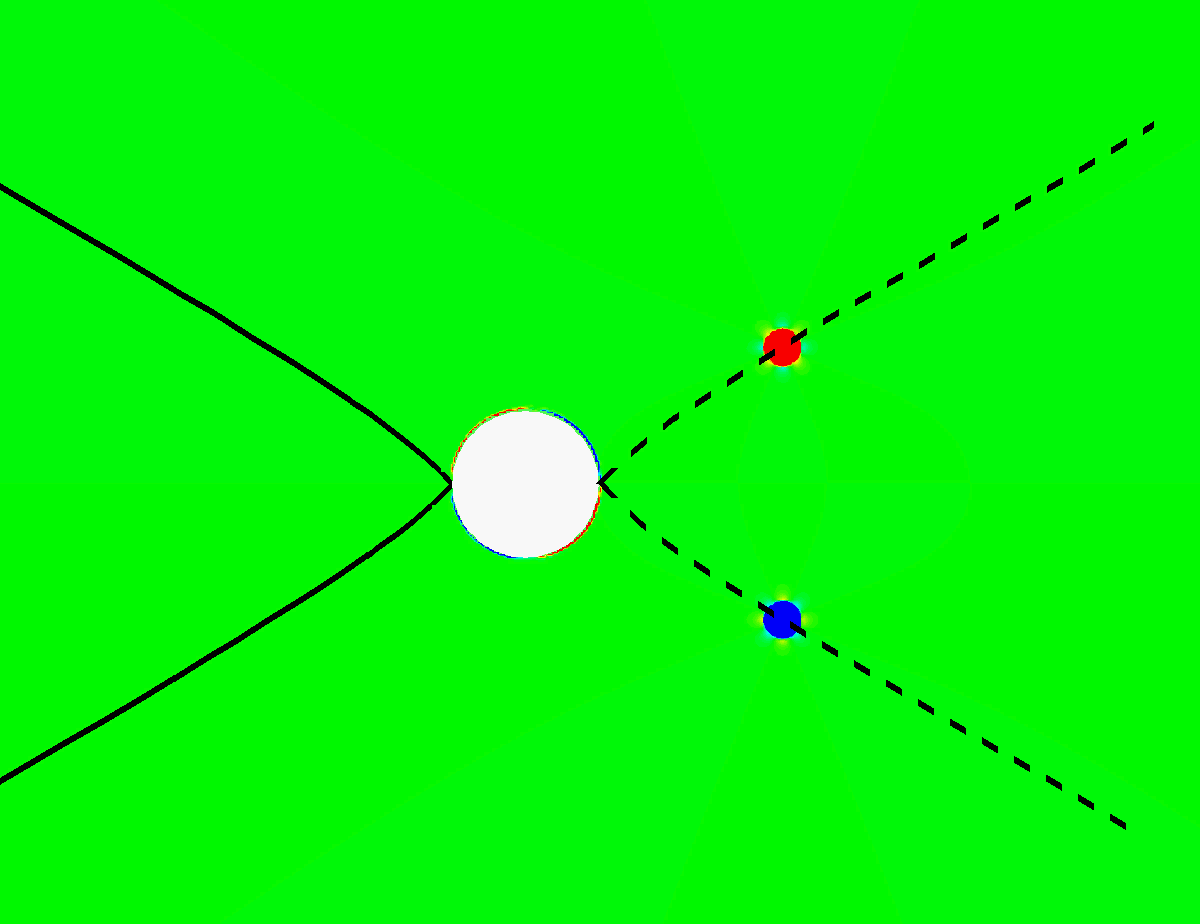} 
 {$t=0$}
\end{subfigure}
\begin{subfigure}
\centering \includegraphics[scale=0.12]{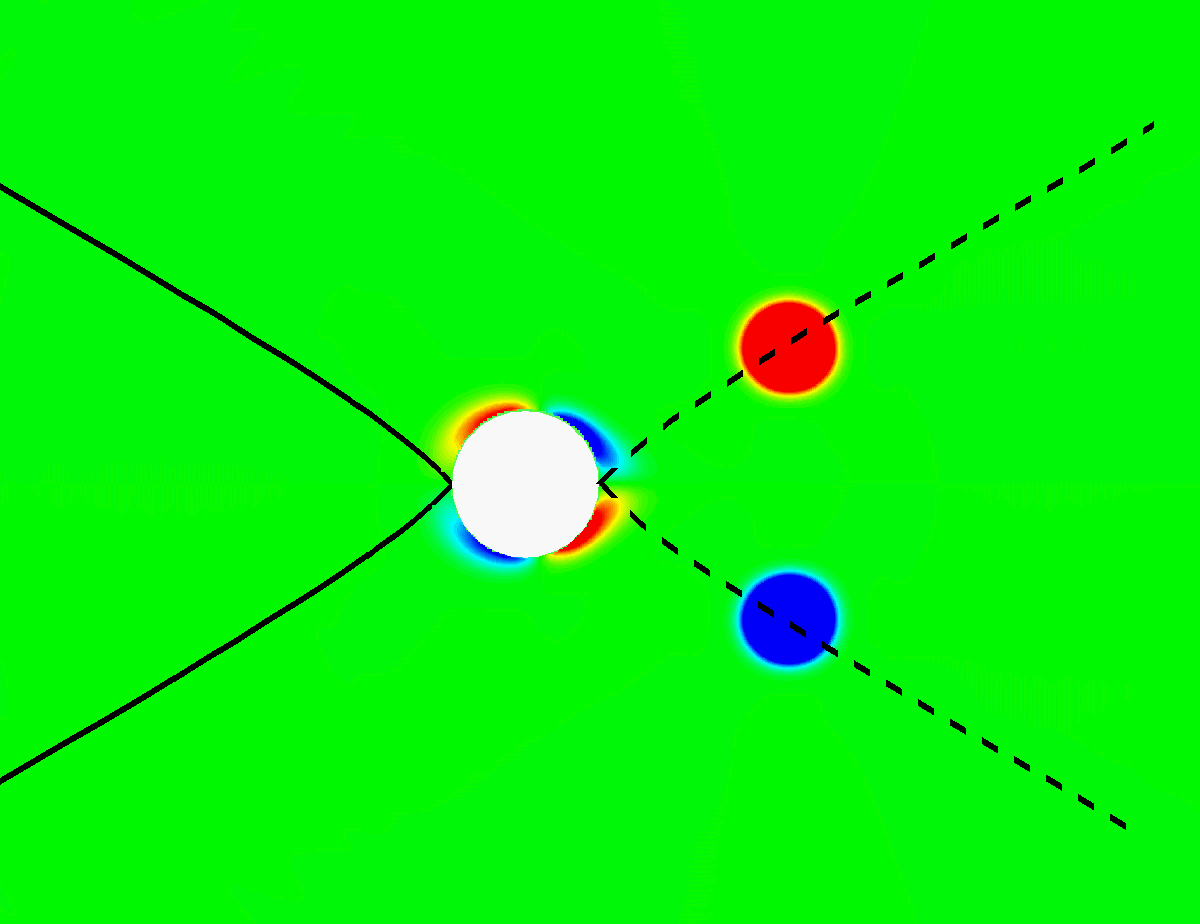}
 {$t=1$}
\end{subfigure}
\begin{subfigure}
\centering \includegraphics[scale=0.12]{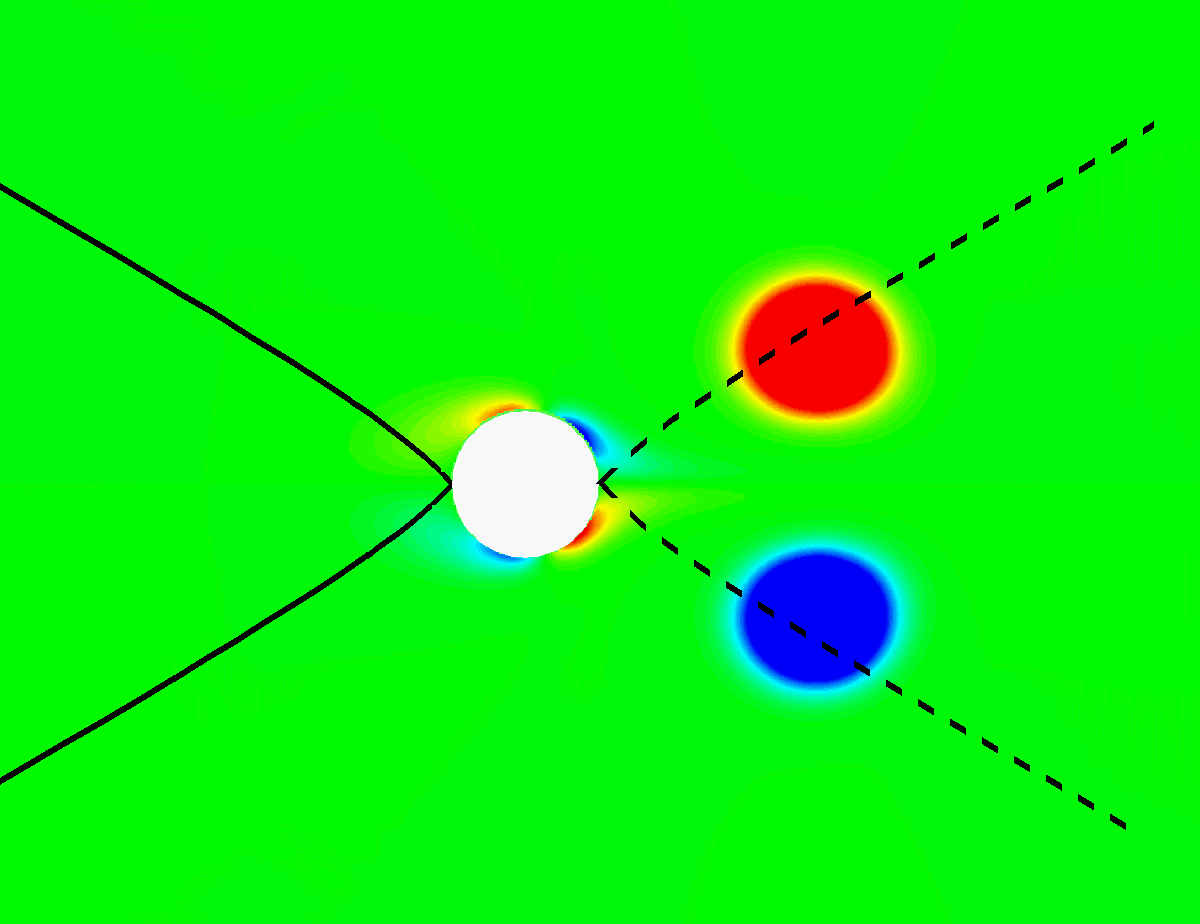}
 {$t=3$}
\end{subfigure}
\begin{subfigure}
\centering \includegraphics[scale=0.12]{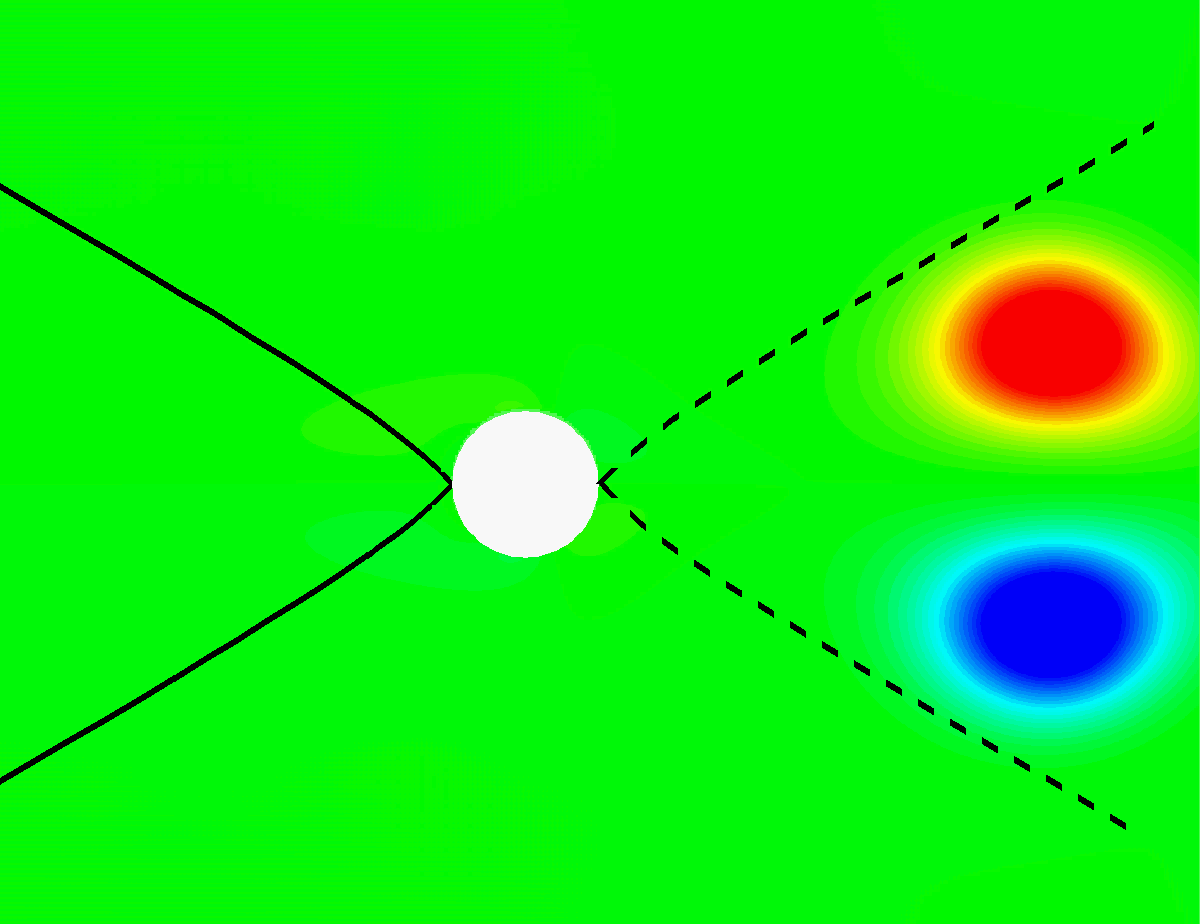}
 {$t=8$}
\end{subfigure}
\caption{Viscous interactions in the body-fixed frame corresponding to starting position R-4-e of Figure \ref{TrFo}. }
\label{R-4-e}
\end{figure}

   In Figure \ref{R-4-e}, snapshots are shown of the interaction in which the vortices start from positions R-4-e (refer Figure \ref{TrFo} for approximate location). The drift to the right is clearly observed.
 For starting position R-2-e the drift to the right is faster since the vortices are closer to each other. There is some boundary layer development in both cases in the initial stages but no significant entrainment is observed. For starting positions R-8-e  and R-6-e on the right equilibrium curves, the vortices stay on the curves for a significant amount of time. The initiation of the drift towards the left instantaneous F\"{o}ppl curves was observed in the movies but due to the significant diffusion that occurs in this time interval the complete movement towards the left curves was not observed.

    In another set of interactions, starting positions below the instantaneous right equilibrium curves were considered. For all these cases, the  vortex pair accelerates to the right and the interactions are very similar to those seen in the above two figures. For the farthest starting position, R-8-b, the vortices linger for longer in the vicinity of the right instantaneous F\"{o}ppl curve before beginning to drift.

  For the set of interactions with starting positions above  the instantaneous right equilibrium curves, the  acceleration to the right occurs for starting position R-2-a and is again qualitatively similar to the previous cases.  For all other starting positions, the vortex pair is attracted to the left instantaneous F\"{o}ppl curve and the cylinder overtakes the pair.  

\begin{figure}
\centering
\begin{subfigure}
 \centering \includegraphics[scale=0.10]{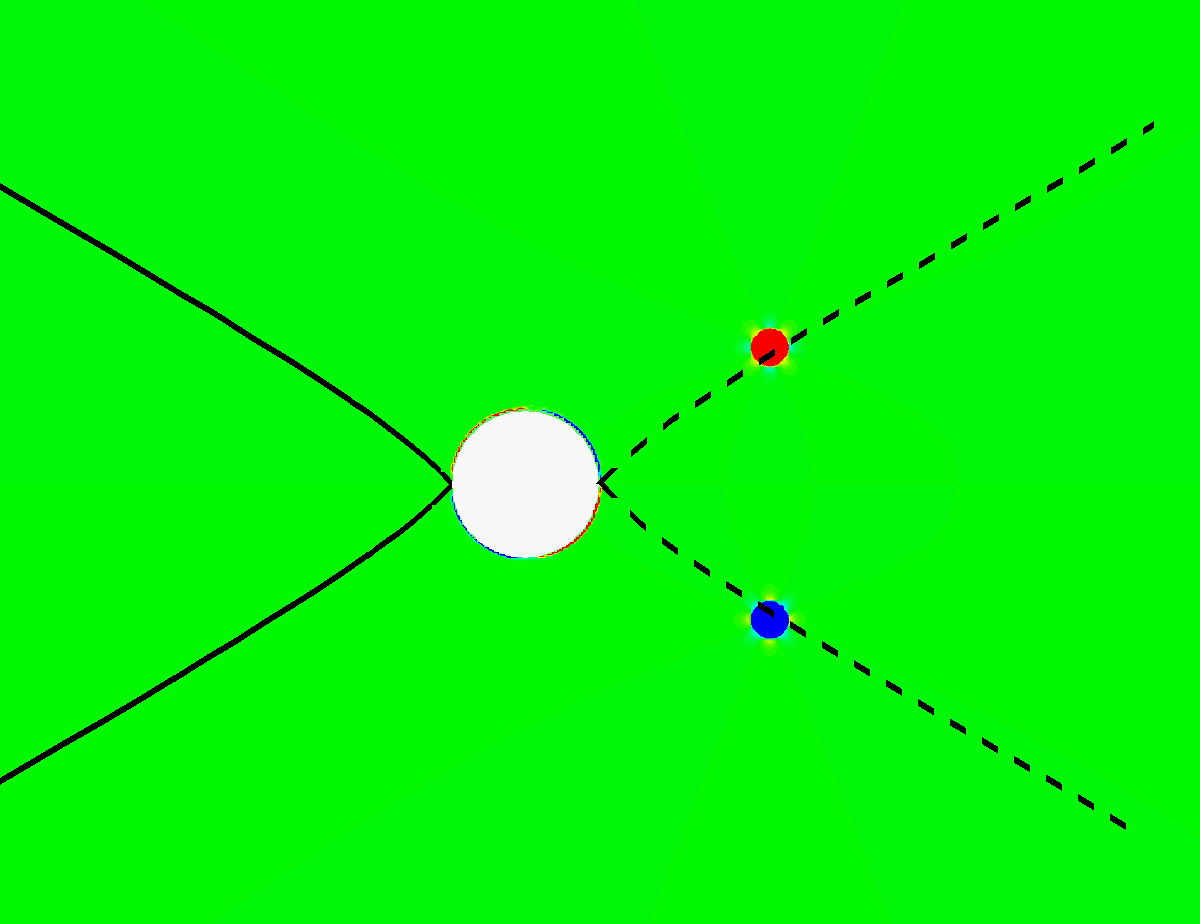} 
 {$t=0$}
\end{subfigure}
\begin{subfigure}
\centering \includegraphics[scale=0.10]{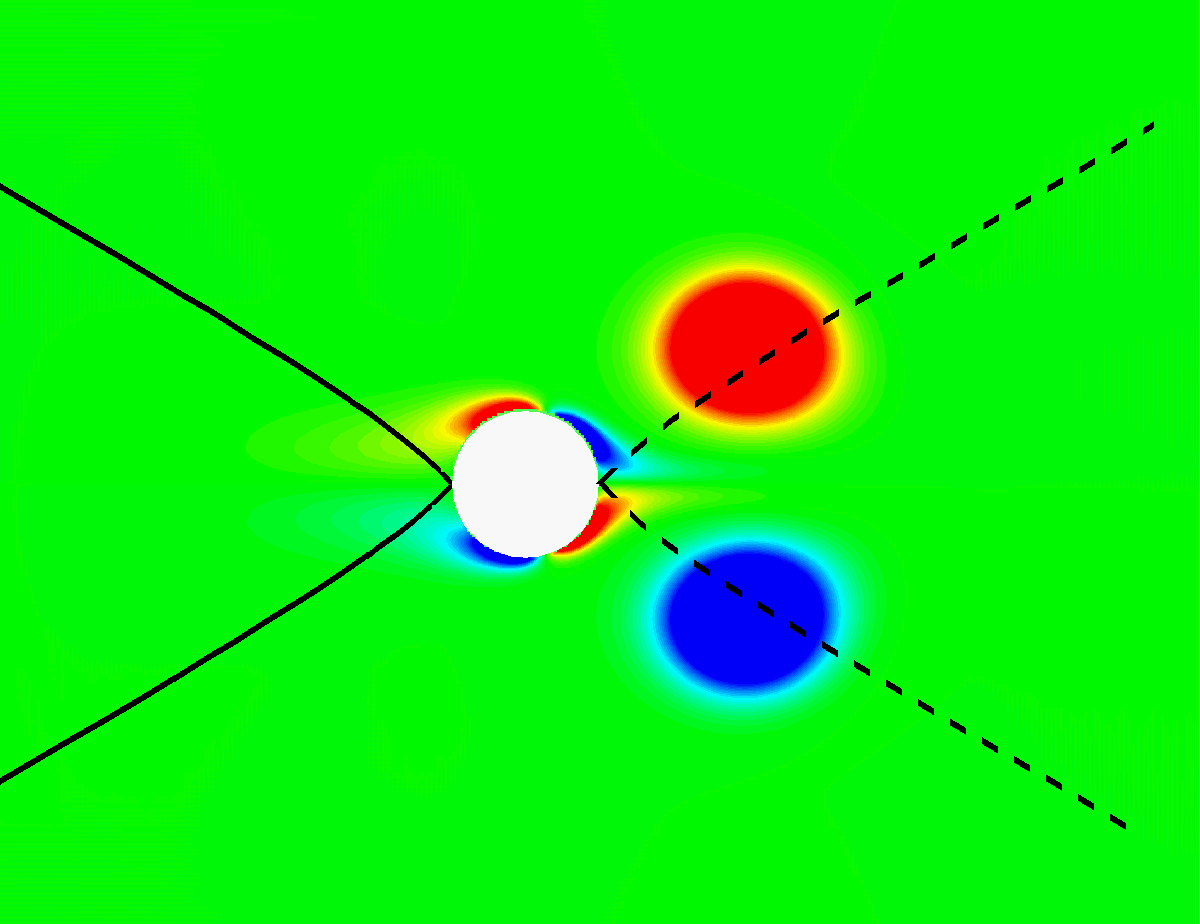}
 {$t=4$}
\end{subfigure}
\begin{subfigure}
\centering \includegraphics[scale=0.10]{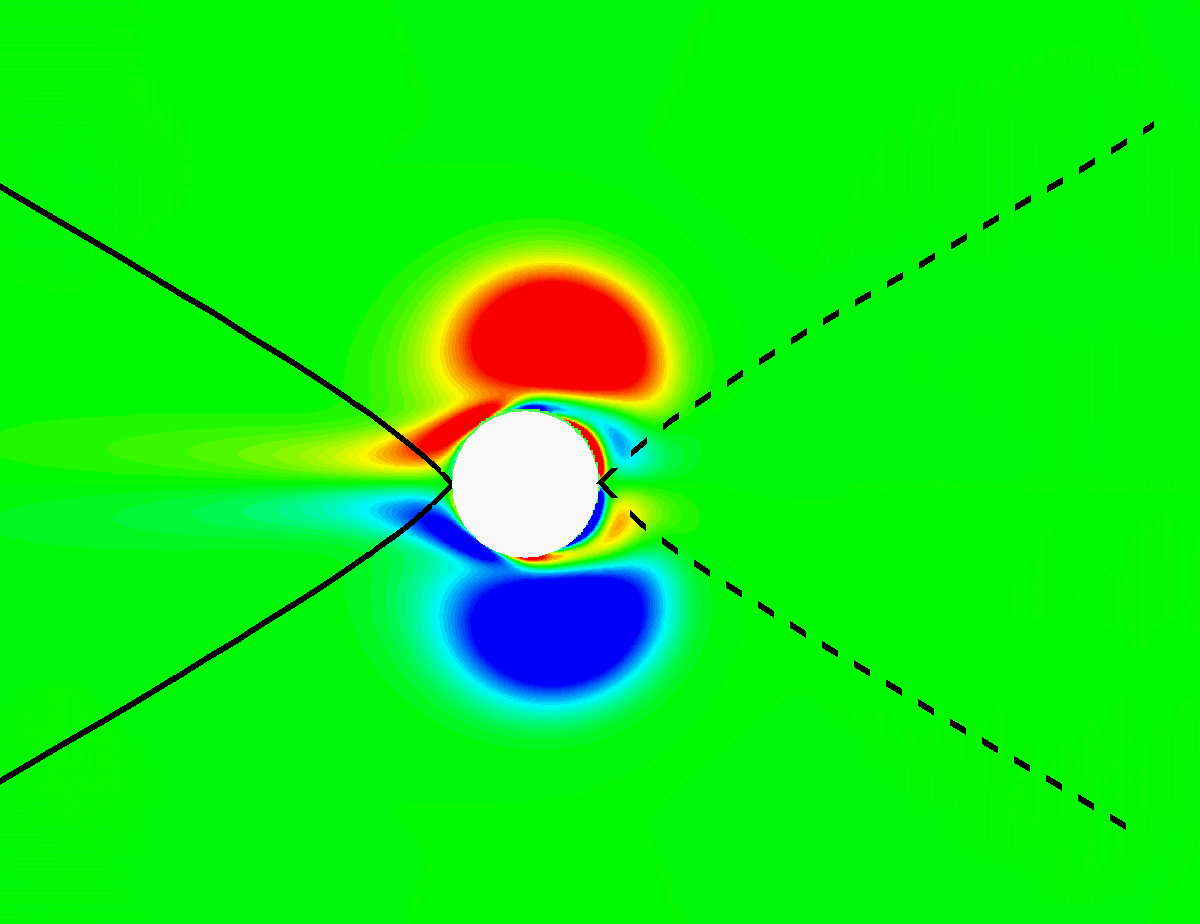}
 {$t=6$}
\end{subfigure}
\begin{subfigure}
\centering \includegraphics[scale=0.09]{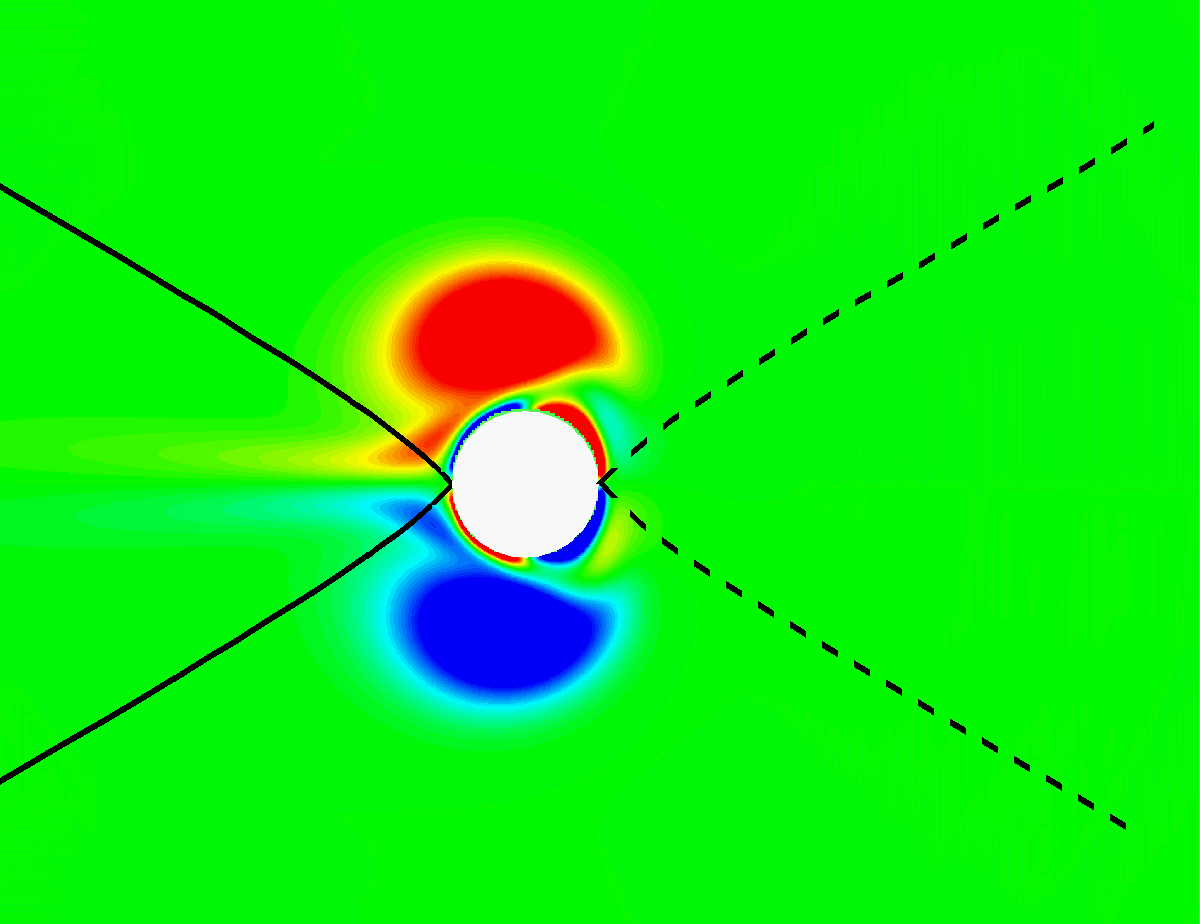}
 {$t=6.25$}
\end{subfigure}
\begin{subfigure}
\centering \includegraphics[scale=0.09]{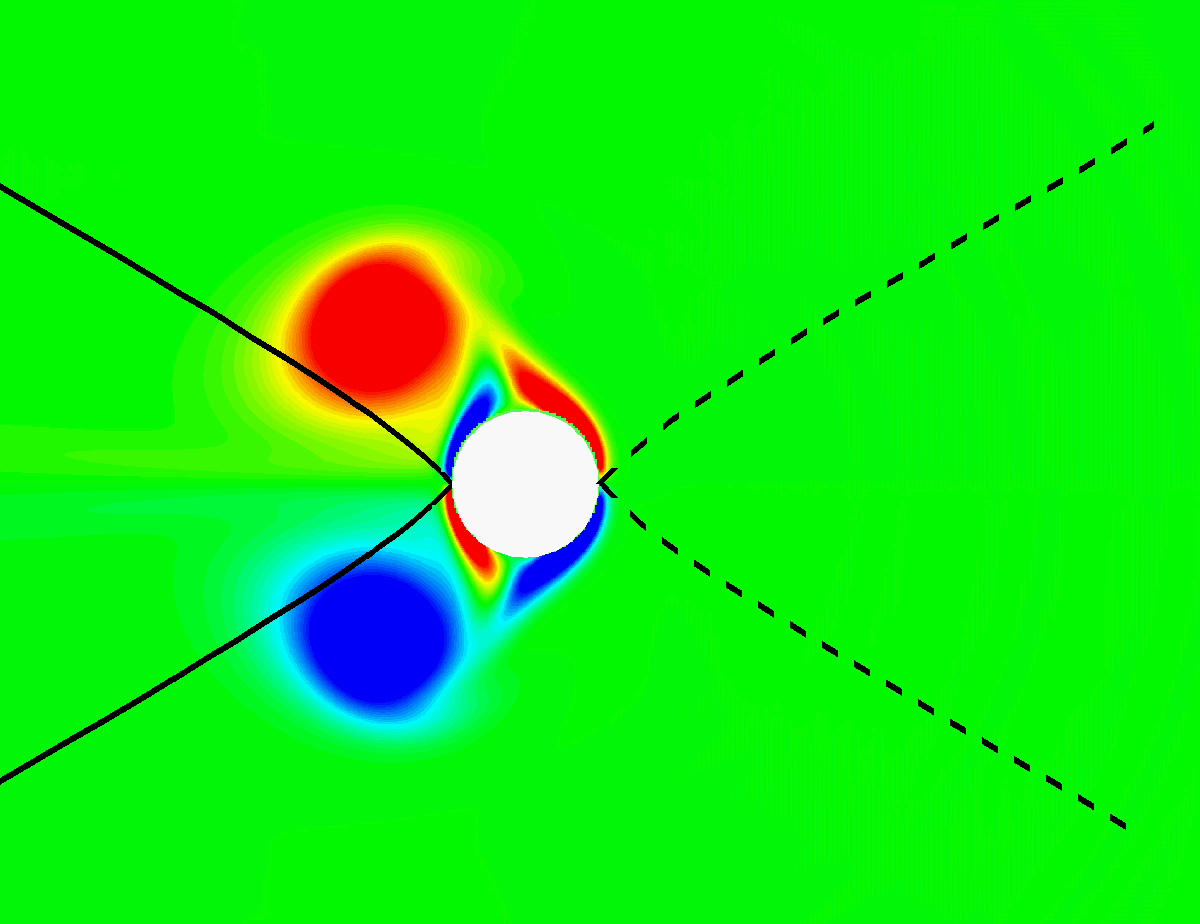}
 {$t=7$}
\end{subfigure}
\begin{subfigure}
\centering \includegraphics[scale=0.09]{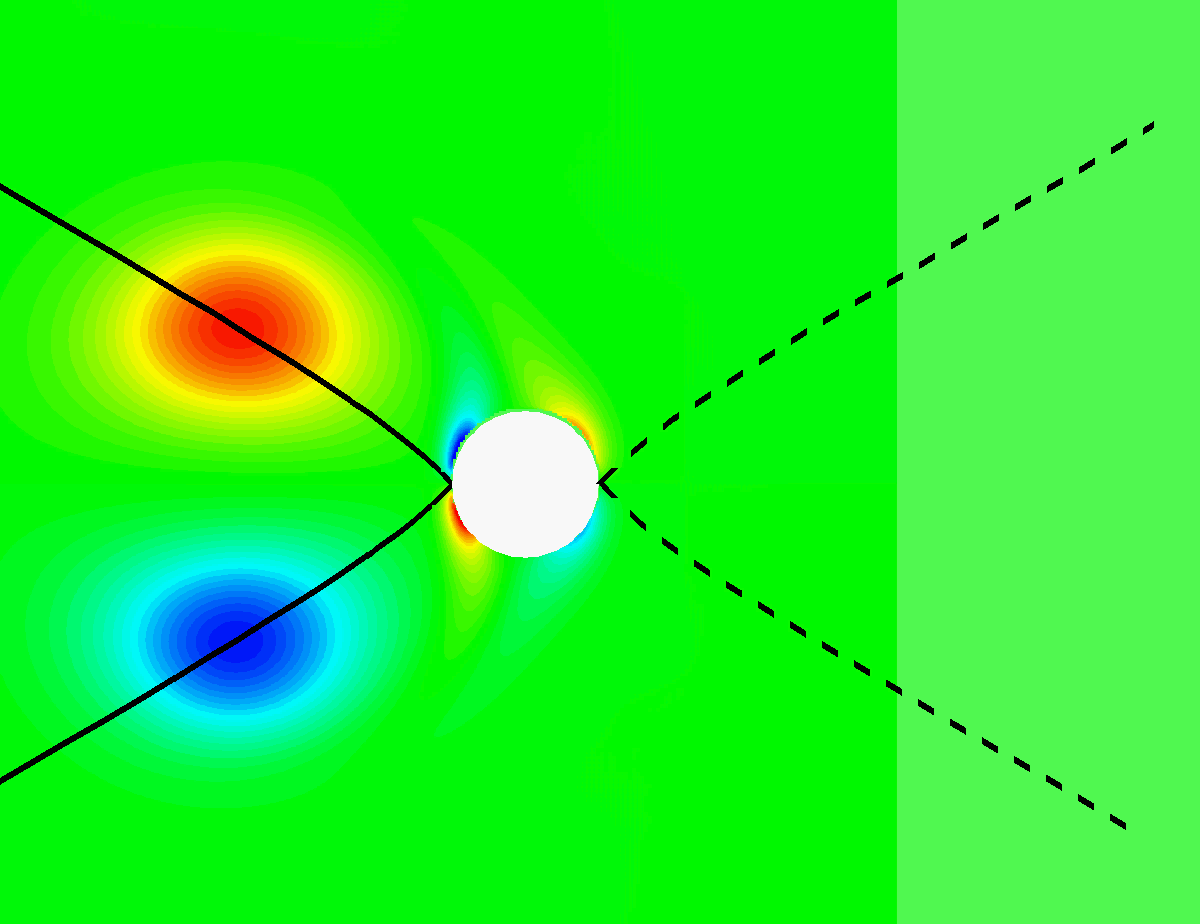}
 {$t=12$}
\end{subfigure}
\caption{Viscous interactions in the body-fixed frame corresponding to starting position R-4-a of Figure \ref{TrFo}. }
\label{R-4-a}
\end{figure}

Snapshots are shown in Figure \ref{R-4-a} for starting position  R-4-a of Figure \ref{TrFo}. It is interesting to note the changes in the viscous boundary layers as the cylinder threads through the pair, in particular, the switch in the sign of the vorticity.  One also observes some entrainment of vorticity from the boundary layers as the vortices pass close to the cylinder. 

\begin{figure}
\centering
\includegraphics[scale=0.35]{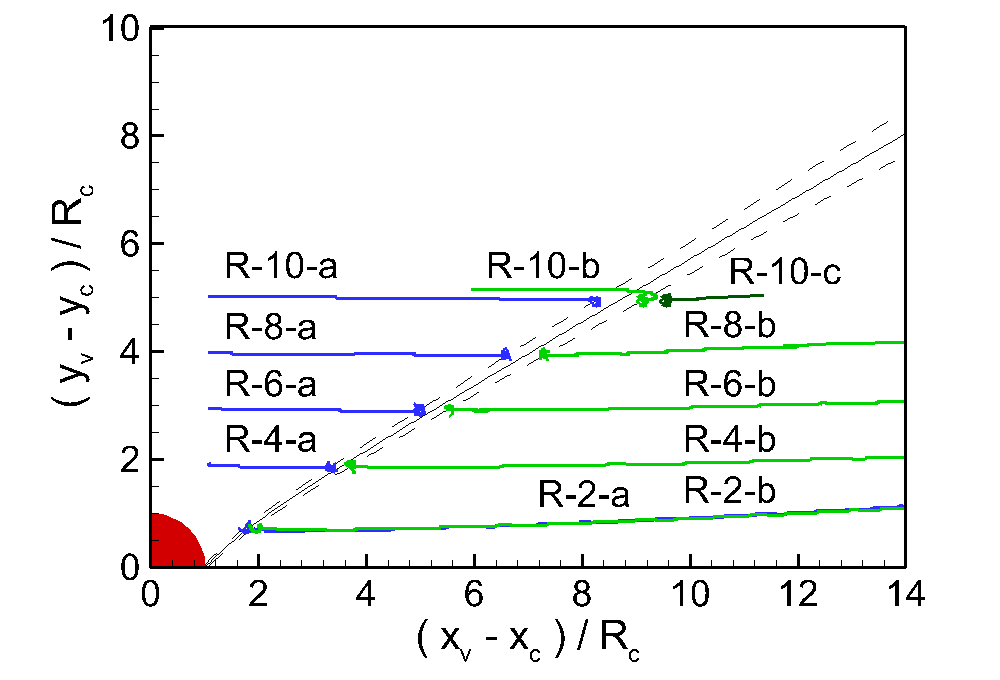}
\caption{Initial positions of the vortices  on the right F\"{o}ppl equilibrium curves and movement of the vorticity maximum point during the evolution in each case. }
\label{MoVMaxLe}
\end{figure}

 The motions are summarized by again keeping track of the vorticity maximum and are shown in Figure \ref{MoVMaxLe}.

\subsubsection{Velocity of cylinder.}

\begin{figure}
\centering
\includegraphics[scale=0.17]{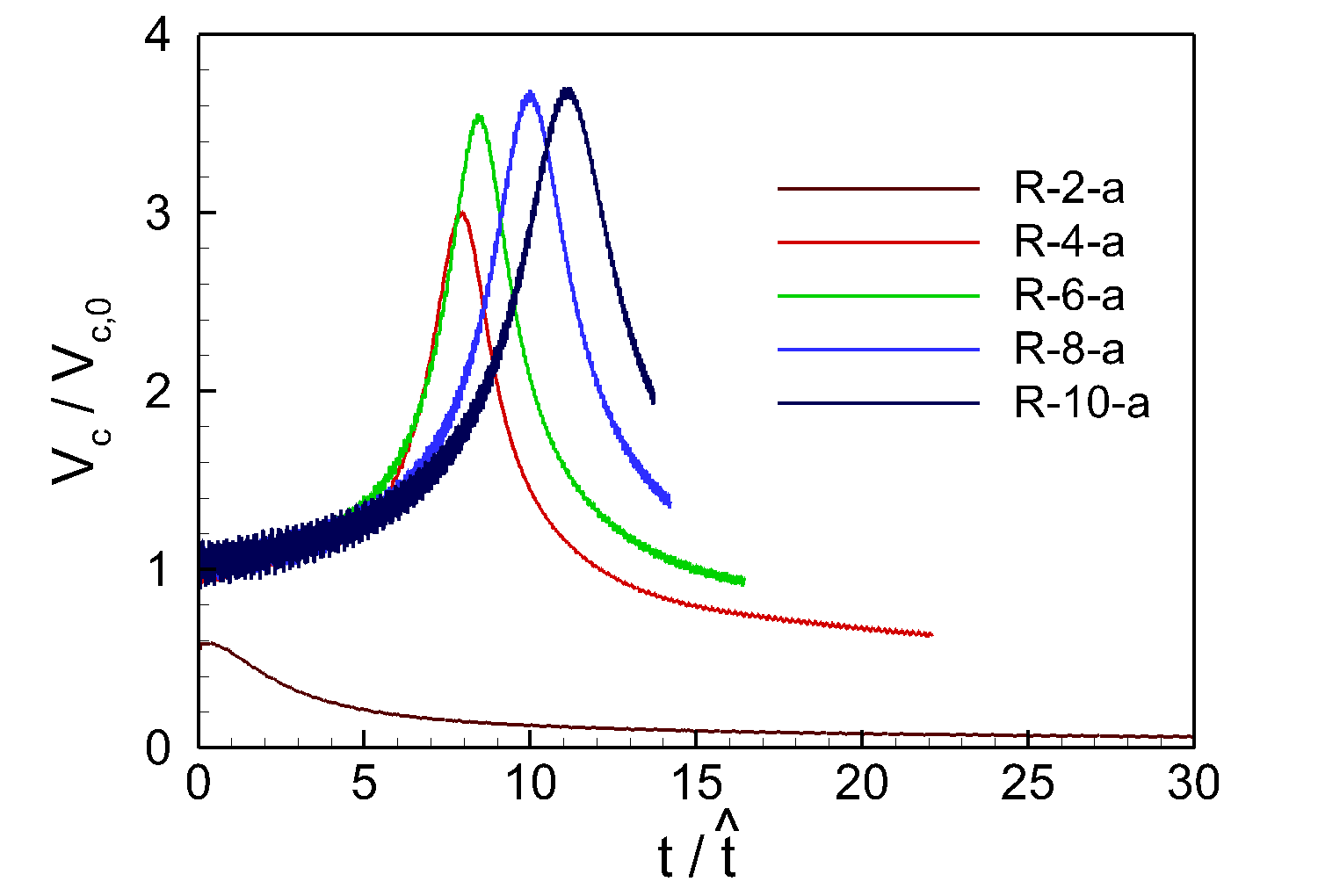}
\includegraphics[scale=0.17]{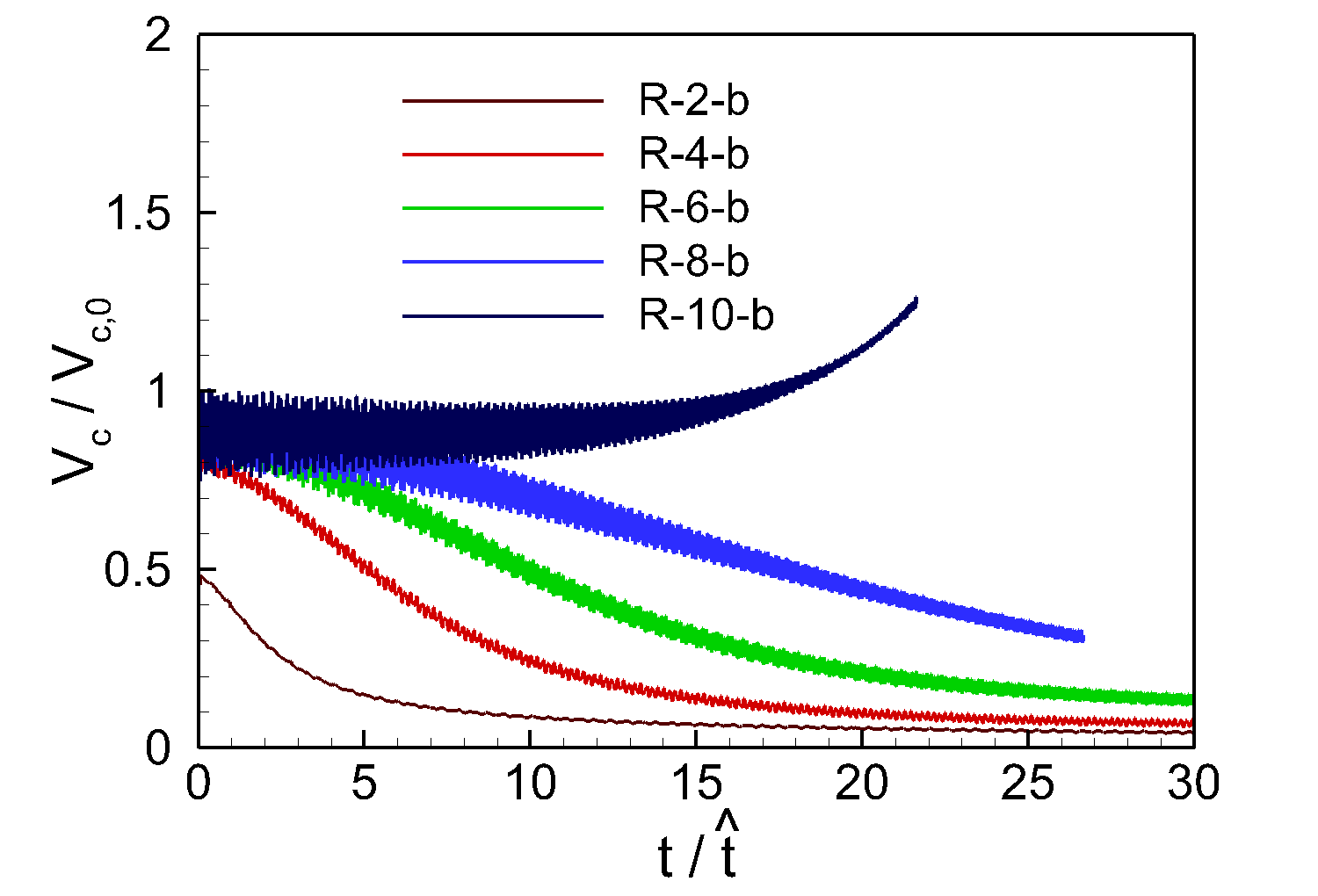}
\caption{Cylinder forward speed versus time corresponding to various starting positions above and below right equilibrium curves, see Figure \ref{TrFo}. Left=above the curves, right=below the curves.}
\label{Vel-time-R}
\end{figure}

  As for the left equilibrium configurations, to illustrate the effect of the vortex interactions on the cylinder forward speed, plots of cylinder versus time are shown in Figure \ref{Vel-time-R} for cases above and below the right equilibrium curves, respectively. Recall from the simulations that in most of the former cases the cylinder threads through the vortices which are attracted to the left equilibrium curves, and in most of the latter cases the vortices accelerate away from the cylinder towards the right. 
Focusing on the left box, one clearly sees that the threading-through phenomenon of the cylinder results in a strong acceleration phase followed by a strong deceleration phase. In fact, the velocity peaks occur when the vortices pass the cylinder. For starting position R-2-a, for which the vortices accelerate to the right, the cylinder decelarates. For starting positions below the curves, the cylinder decelerates for all the cases in which the vortices accelerate to the right, except for starting position R-10-b for which the threading-through phenomenon is again seen and the vortices drift towards the left (see Figure \ref{MoVMaxLe}). Attention may also be drawn to starting position R-10-c, a little to the right of R-10-b, for which the vortices again drift away to the right.

\subsubsection{Linear momentum exchange during interactions.} 
  
\begin{figure}
\centering
 \includegraphics[scale=0.3]{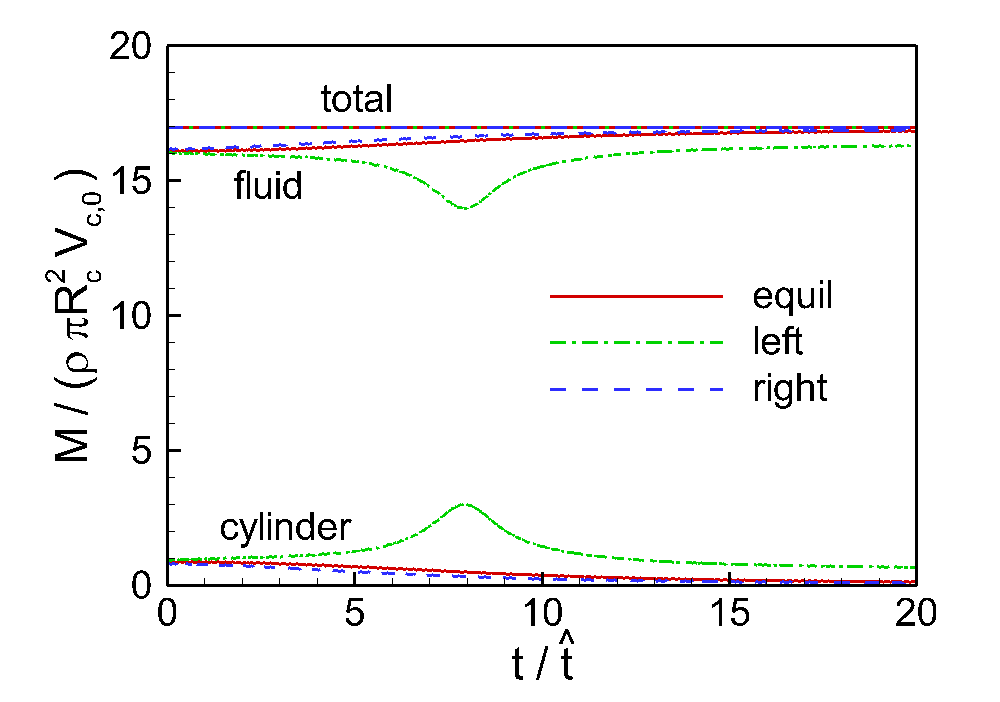} 
\caption{Linear momentum exchange for cases starting from R-4-e, R-4-a and R-4-b, i.e. on, above and below the F\"{o}ppl right equilibrium curves.}
\label{mom-exchange-R}
\end{figure}
Figure \ref{mom-exchange-R} shows plots of linear momentum evolution for three different starting positions of the vortices on the right equilibrium curves: R-4-e, R-4-a and R-4-b. The conservation of the total linear momentum is easily seen in all three case with the straight lines lying on top of each other. It may be recalled from the discussions above that for the cases R-4-e and R-4-b, the vortices leave the cylinder behind and accelerate to the right. For case R-4-a, the cylinder threads through the vortices--and it is seen that the exchange of linear momentum between fluid and cylinder is the maximum for this case (green dashed curves).

\subsection{Starting configuration: moving vertical line equilibria .}

 A few starting vortex configurations on the vertical line passing through the center of the cylinder were also examined. 
\begin{figure}
\centering
\begin{subfigure}
 \centering \includegraphics[scale=0.06]{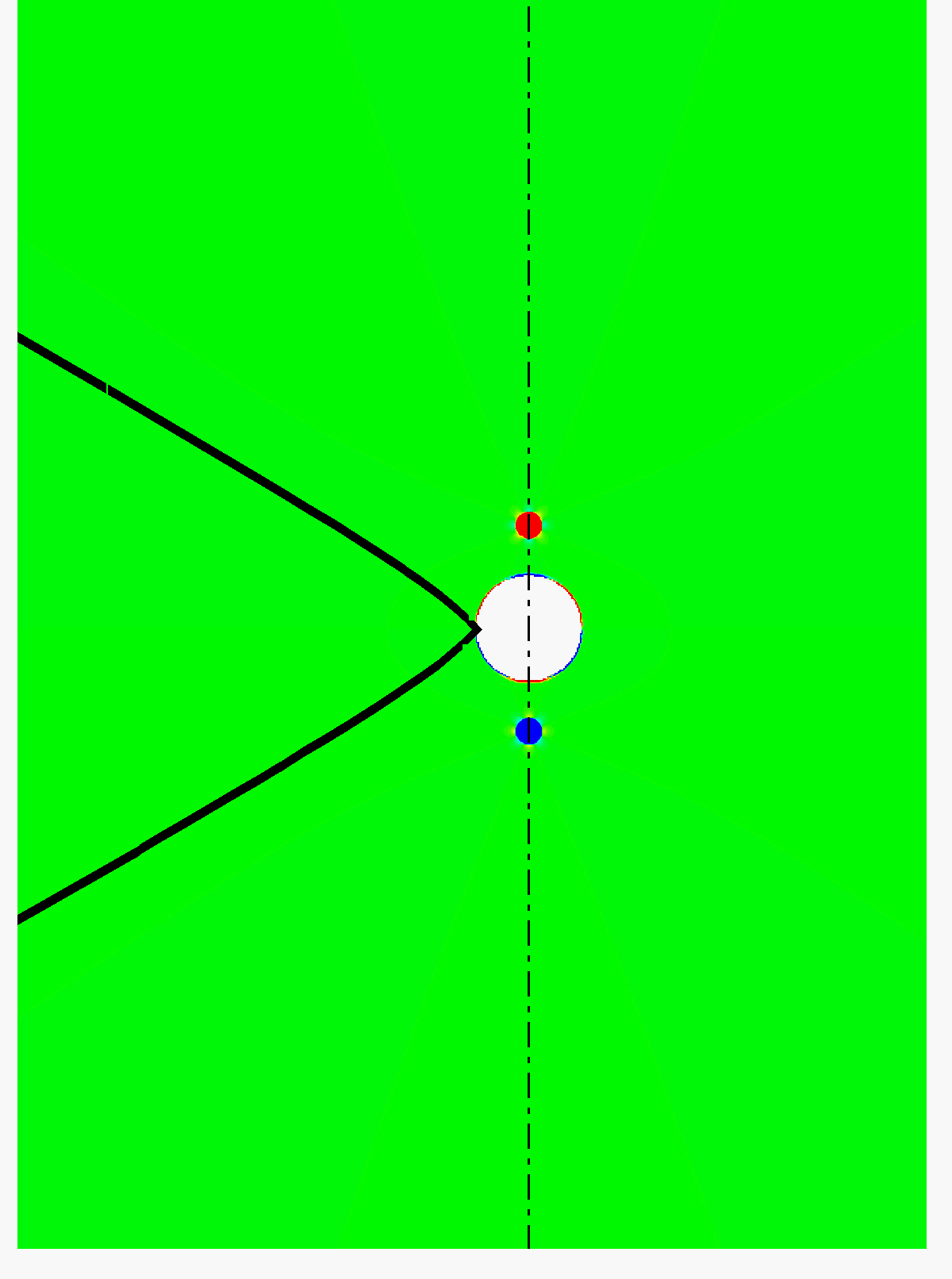} 
 {$t=0$}
\end{subfigure}
\begin{subfigure}
\centering \includegraphics[scale=0.06]{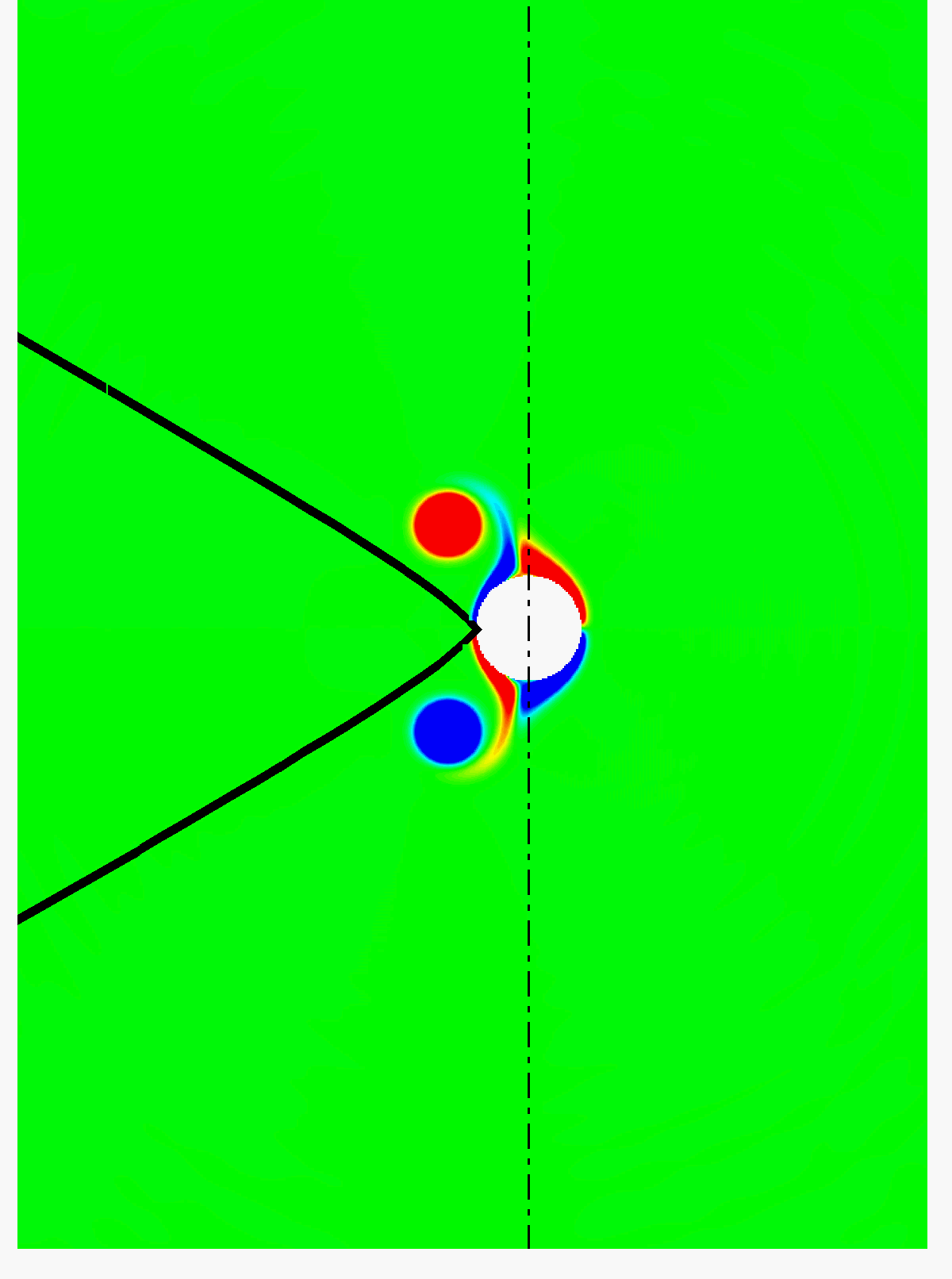}
{$t=1$}
\end{subfigure}
\begin{subfigure}
\centering \includegraphics[scale=0.06]{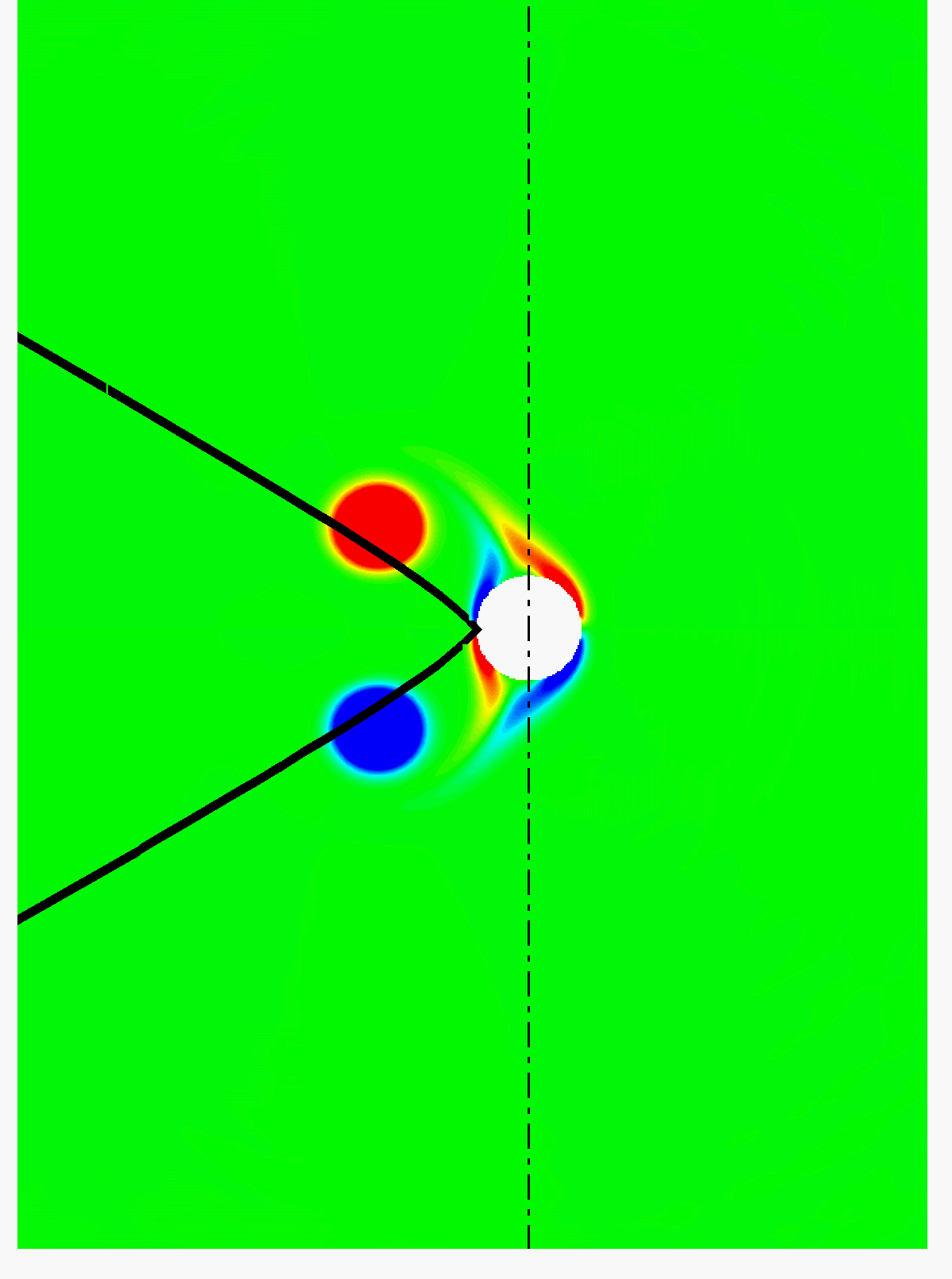}
 {$t=2$}
\end{subfigure}
\begin{subfigure}
\centering \includegraphics[scale=0.06]{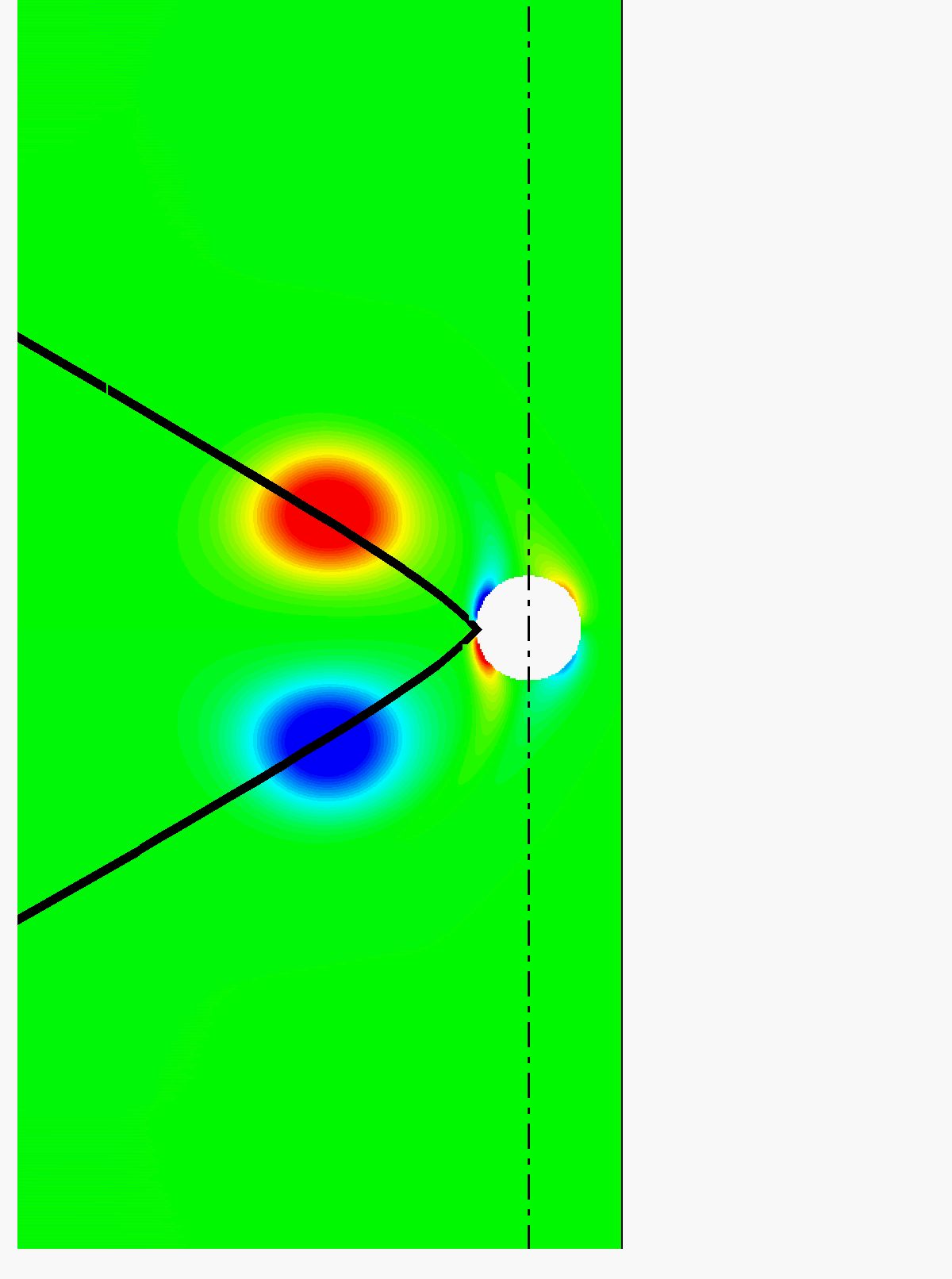}
 {$t=10$}
\end{subfigure}
\caption{Viscous interactions in the body-fixed frame corresponding to a starting position V-2-e on the vertical line in Figure \ref{TrFo}. }
\label{vertical-a}
\end{figure}
None of these evolutions show any stability. In all cases, the vortices drift away from the vertical line. But again, interestingly, they move towards the left instantaneous F\"{o}ppl curves, and stay on them once they reach them. The drift appears to be faster closer the starting position is to the cylinder. Boundary layer development,  and some separation and entrainment is also observed. A typical sequence is shown in Figure \ref{vertical-a} for starting position V-2-e of Figure \ref{TrFo}. 

\section{Viscous simulations: elliptic cylinder.}

   Simulations were also carried out for translating configurations of vortex pairs and elliptic cylinders of various aspect ratios. The equilibirum curves for the inviscid model in such cases are given by Hill's equations. Ellipses of aspect ratios 2 and 4 were chosen, and both cases of major axis parallel to $x$-axis and major axis parallel to $y$-axis were considered. The qualitative behavior of the interaction is very similar to the cases with a circular cylinder and some representative (smaller-sized) snapshots are presented in this section. 

\subsection{Starting configuration: moving Hill equilibria, trailing vortices and ellipse.}
Starting positions on the left Hill curves again display stable behavior---with the vortices either remaining on them, for farther starting positions, or returning to them if starting from nearer positions or slightly displaced positions. Since no real dynamical features other than core diffusion are observed  in the former set of evolutions (as in the case of the circular cylinder), only the evolutions from nearer positions or displaced positions are shown. Moreover, there is no noticeable qualitative difference between configurations in which the ellipse has it major axis parallel to the $x$-axis  and in which the ellipse has its major axis  parallel to the $y$-axis 

\begin{figure}
\centering
\begin{subfigure}
\centering  \includegraphics[scale=0.35]{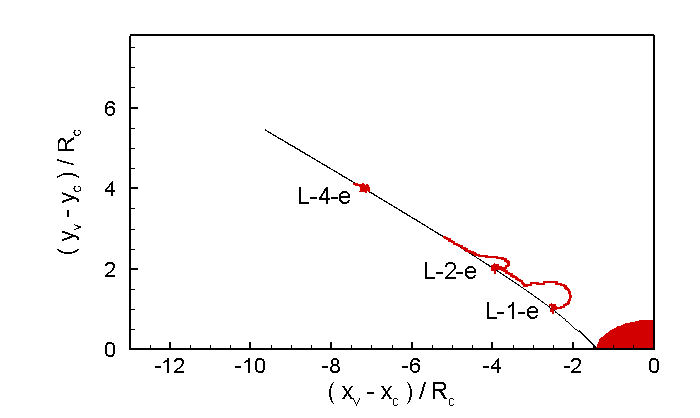} 
 \end{subfigure}
\begin{subfigure}
\centering \includegraphics[scale=0.35]{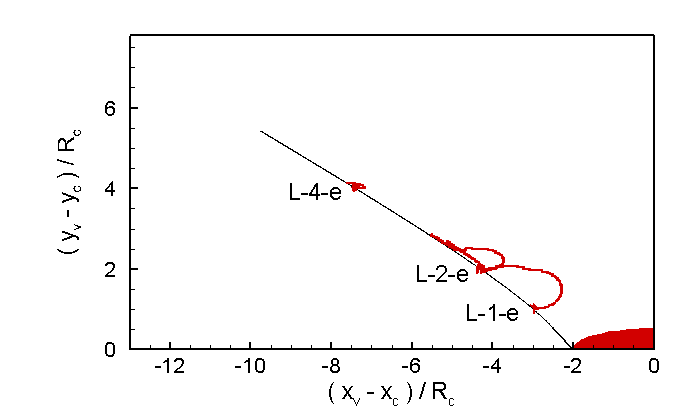}
\end{subfigure}
\begin{subfigure}
\centering  \includegraphics[scale=0.35]{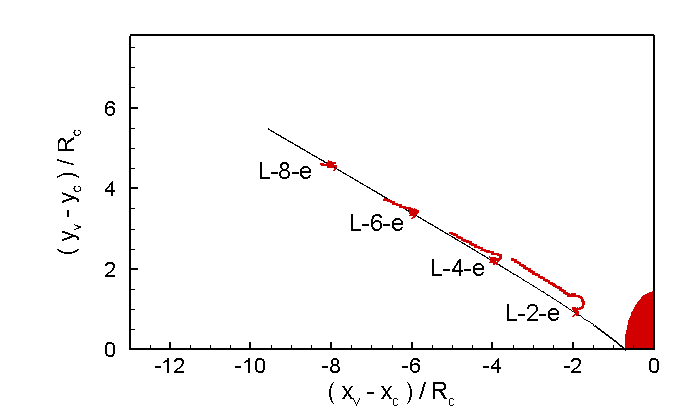} 
 \end{subfigure}
\begin{subfigure}
\centering \includegraphics[scale=0.35]{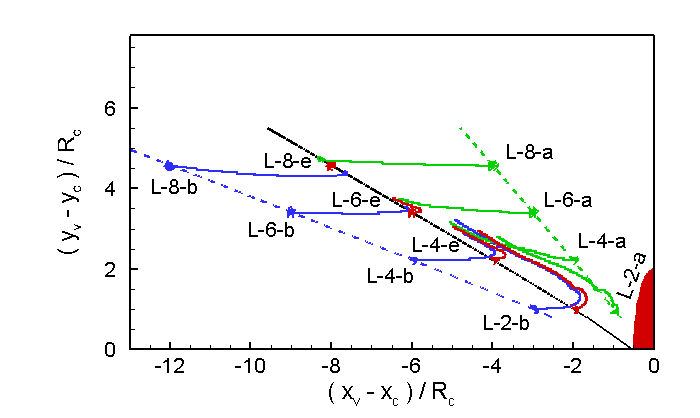}
\end{subfigure}
\caption{Initial positions of the vortices on the left Hill equilibrium curves and the drift of the vorticity maximum point for interactions involving elliptic cylinders of aspect ratios 2 and 4. Configurations in which the ellipse has its major axis parallel to the $x$-axis  and parallel to the $y$-axis, respectively, are both shown.}
\label{IPosEllL}
\end{figure}

\begin{figure}
\centering
\begin{subfigure}
 \centering \includegraphics[scale=0.08]{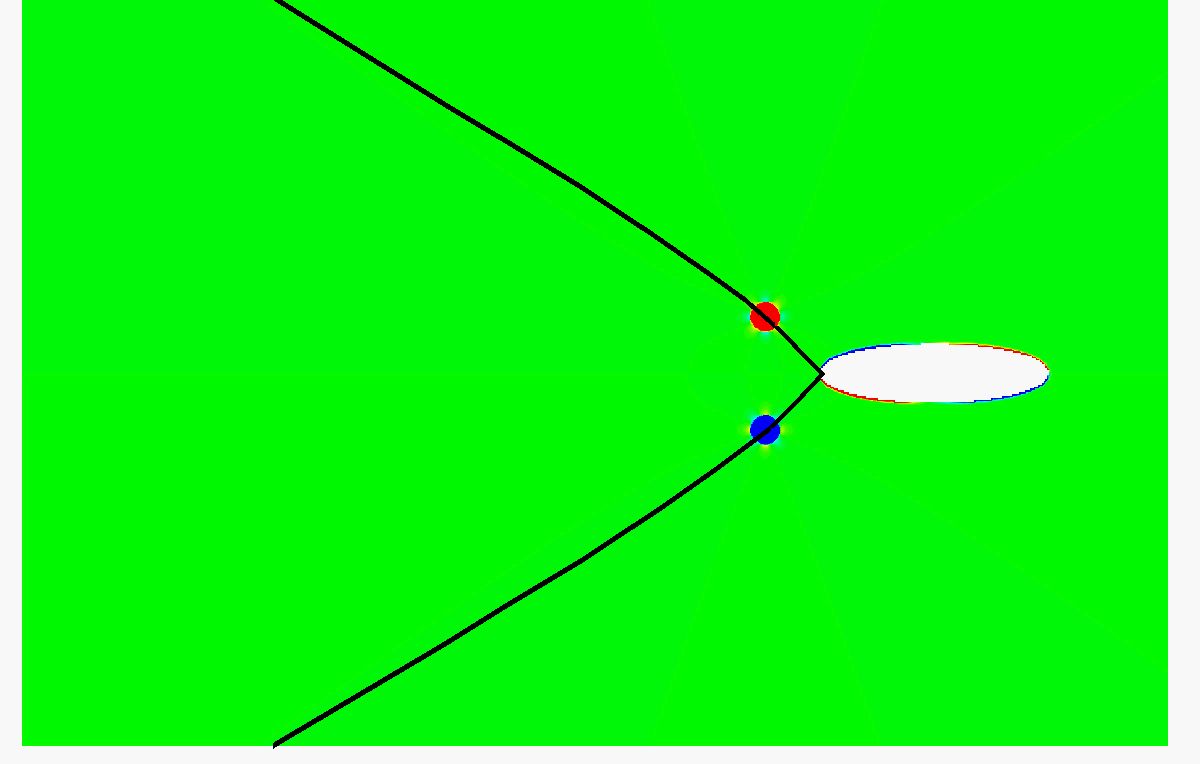} 
 {$t=0$}
\end{subfigure}
\begin{subfigure}
\centering \includegraphics[scale=0.08]{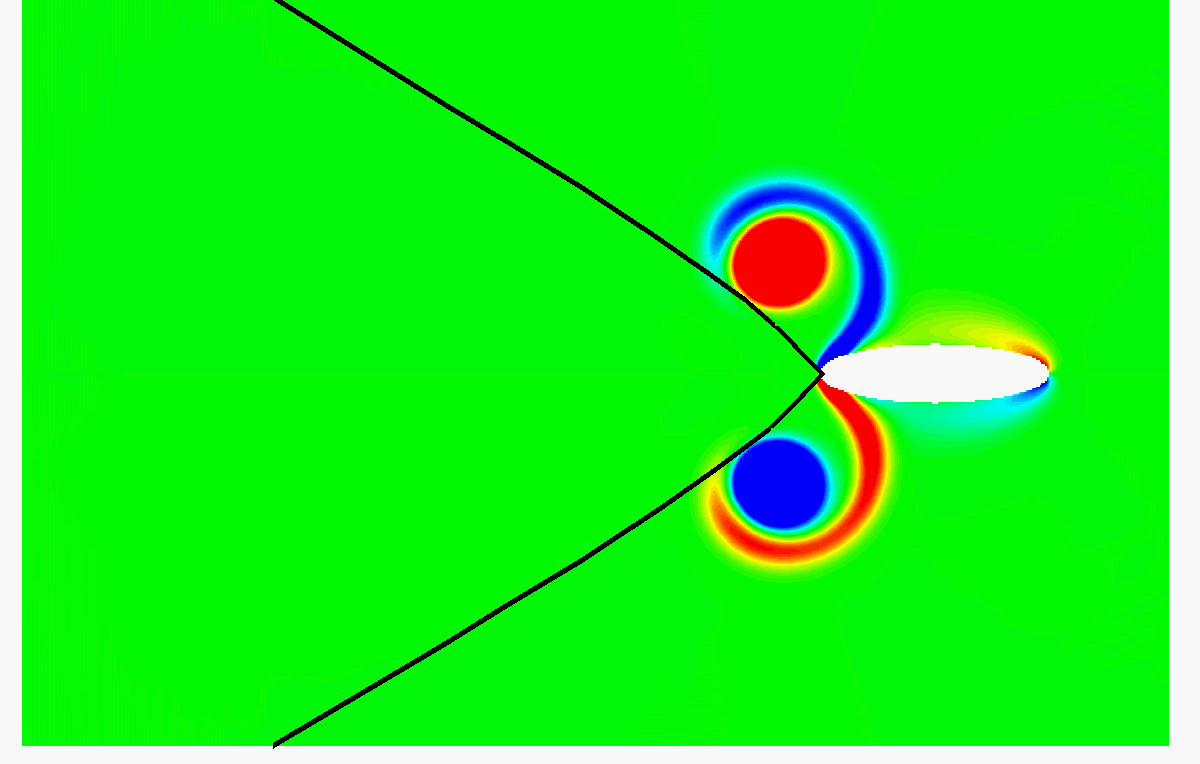}
 {$t=2$}
\end{subfigure}
\begin{subfigure}
\centering \includegraphics[scale=0.08]{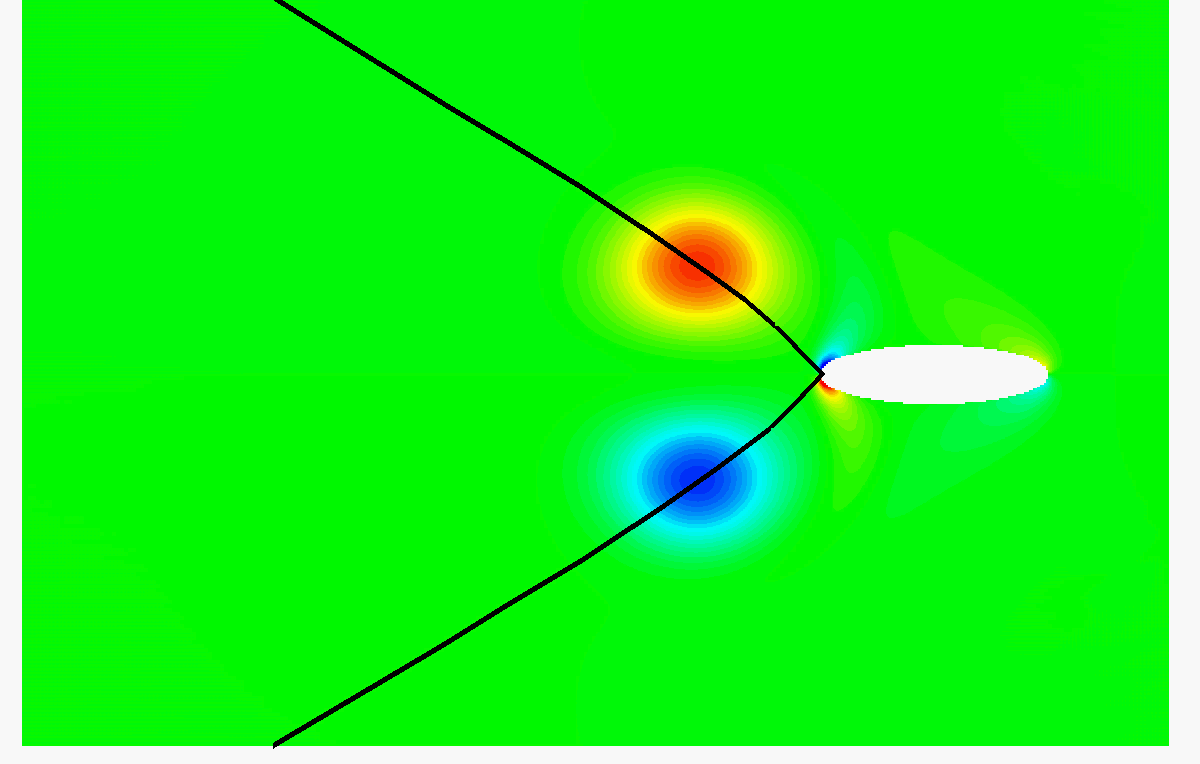}
 {$t=10$}
\end{subfigure}
\caption{Viscous interactions with an elliptic cylinder of aspect ratio 4, major axis along the $x$-axis, in the body-fixed frame from starting position L-1-e on the left Hill equilibrium curves }
\label{a4-hor-L-1-e}
\end{figure}

    The initial positions of the vortices and the drift of the vorticity maximum for the two configurations for ellipses of apsect ratios 2 and 4, respectively, are shown in Figure \ref{IPosEllL}.
  Representative snapshots at different times for a starting position on the equilibrium curve, closest to the ellipse of aspect ratio 4, are shown in Figure \ref{a4-hor-L-1-e}.
As for the circular cylinder, the vortices are initially displaced from the curves while entraining boundary layer vorticity of the opposite sign and are then pushed back to the curves and drift slowly rearward along the curves. 

\begin{figure}
\centering
\begin{subfigure}
 \centering \includegraphics[scale=0.08]{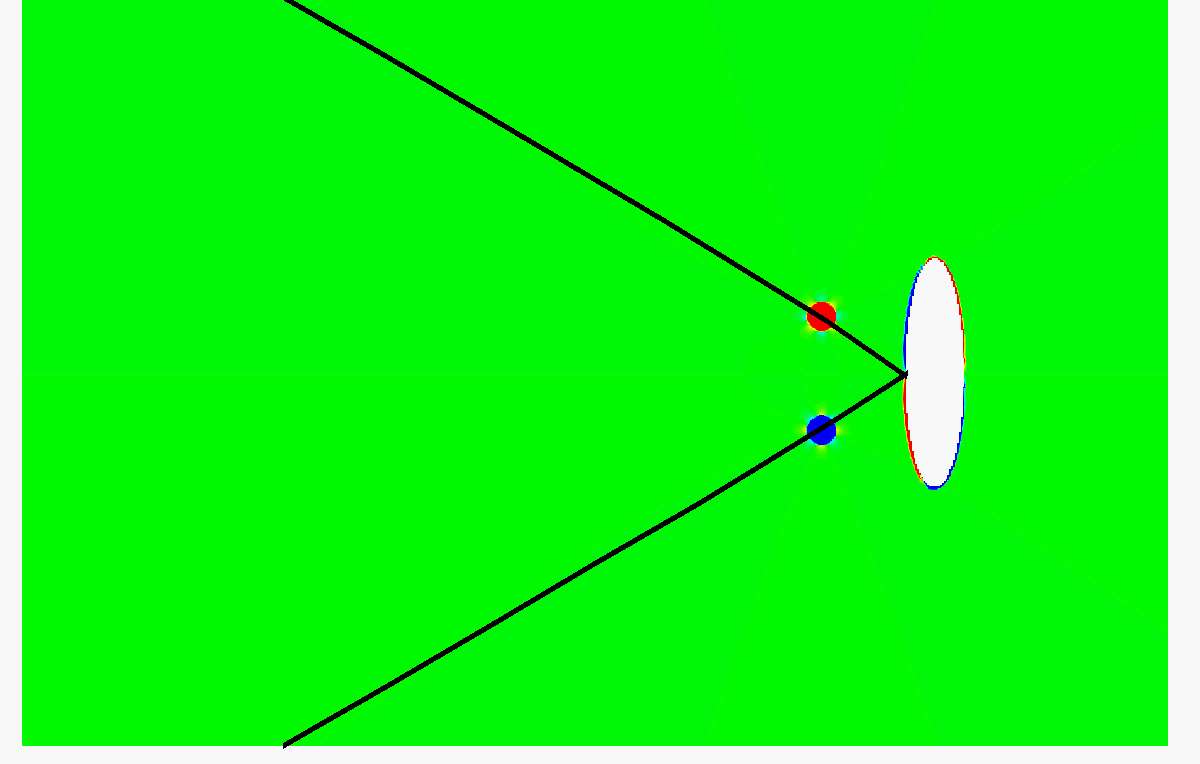} 
 {$t=0$}
\end{subfigure}
\begin{subfigure}
\centering \includegraphics[scale=0.08]{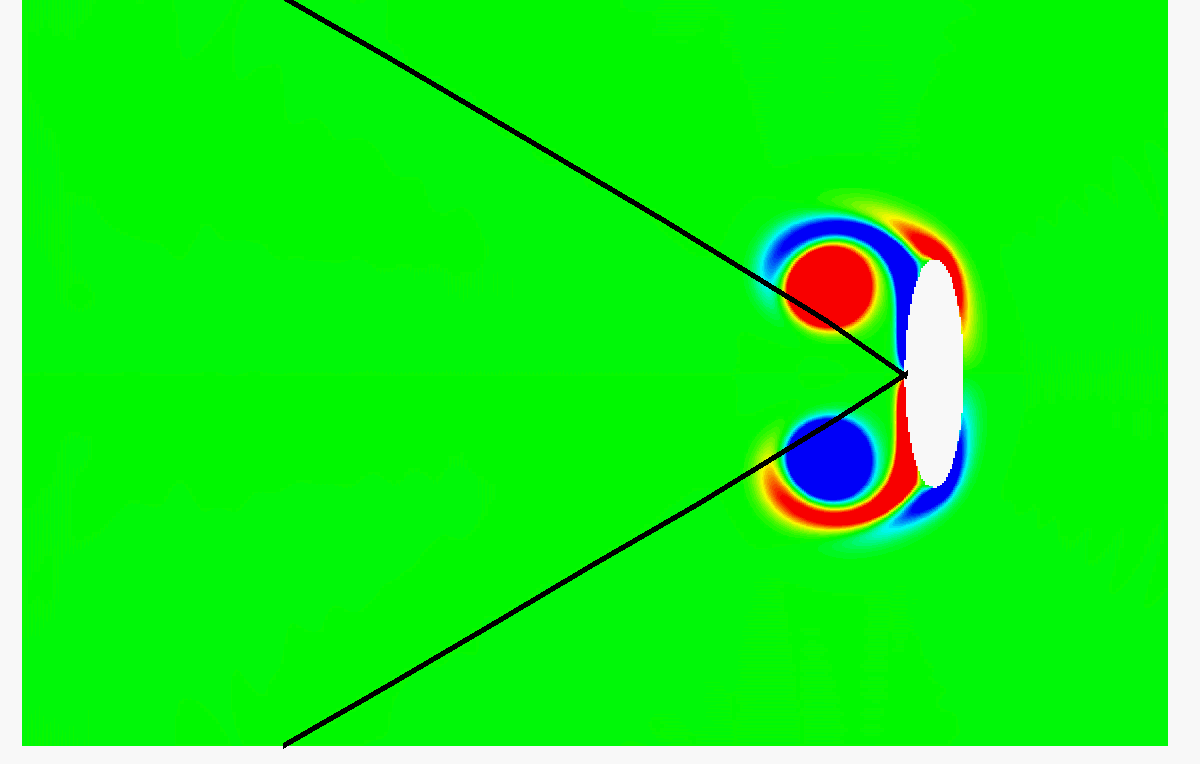}
 {$t=1.5$}
\end{subfigure}
\begin{subfigure}
\centering \includegraphics[scale=0.08]{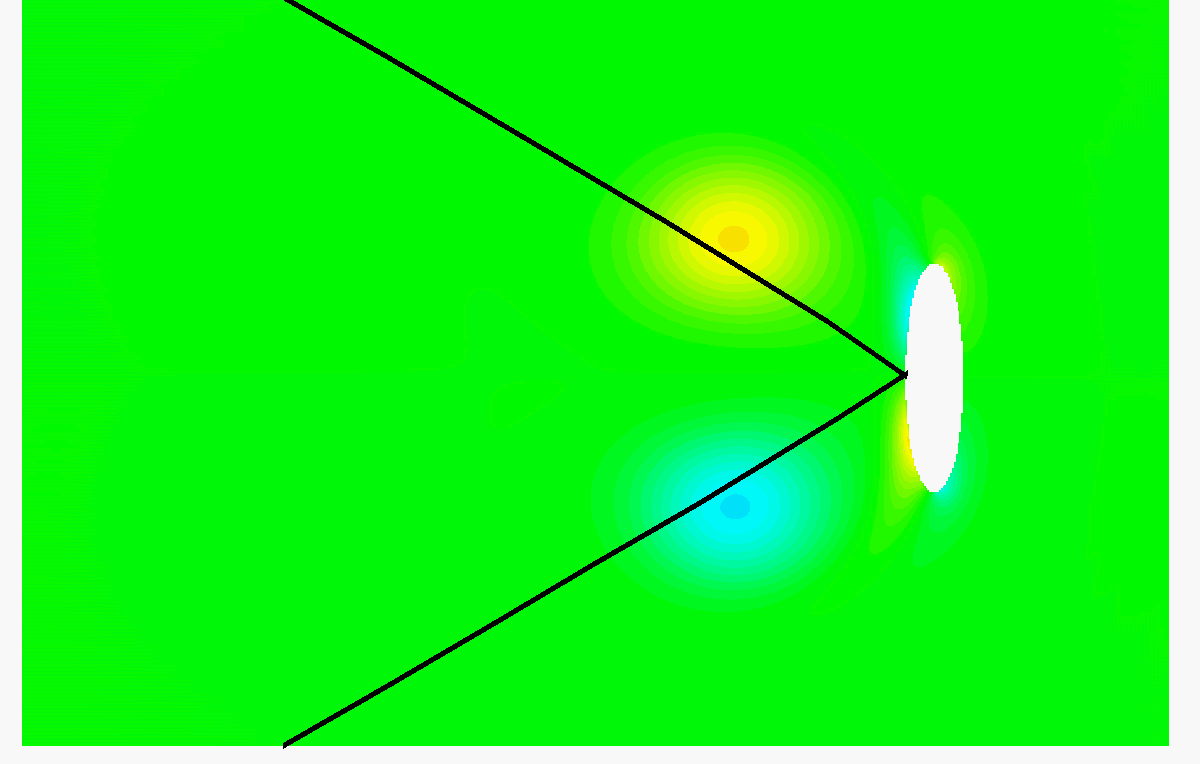}
 {$t=15$}
\end{subfigure}
\caption{Viscous interactions with an elliptic cylinder of aspect ratio 4, major axis along the $y$-axis, in the body-fixed frame from starting position L-2-e on the left Hill equilibrium curves. }
\label{a4-ver-L-2-e}
\end{figure}

For the same ellipse positioned vertically and with the vortices starting in a similar position as in the previous case, the evolution is captured in the snapshots shown in Figure \ref{a4-ver-L-2-e}.

\subsection{Starting configuration: moving Hill equilibria, leading vortices and ellipse with major axis along $x$-axis.}
\begin{figure}
\centering
\begin{subfigure}
\centering  \includegraphics[scale=0.35]{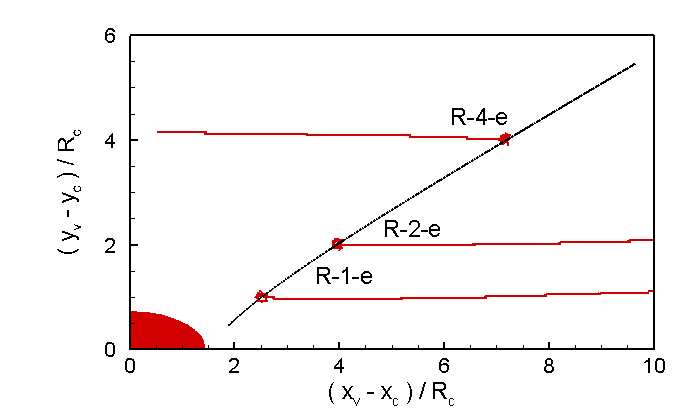} 
 \end{subfigure}
\begin{subfigure}
\centering \includegraphics[scale=0.35]{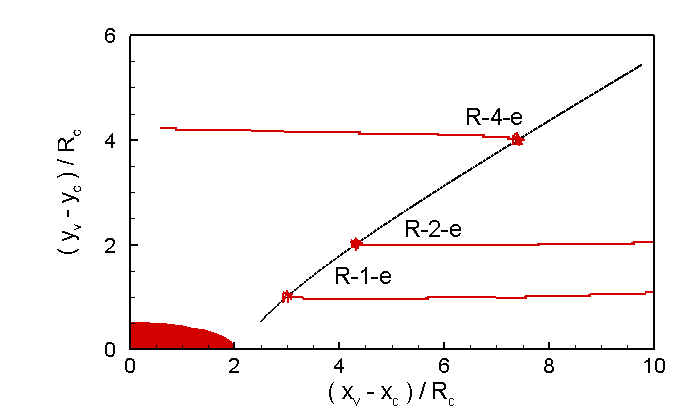}
\end{subfigure}
\caption{Initial positions of the vortices on the right Hill equilibrium curves and the drift of the vorticity maximum point for interactions involving elliptic cylinders with major axis along the $x$-axis of aspect ratios 2 and 4, respectively.}
\label{IPosHorEllR}
\end{figure}

\begin{figure}
\centering
\begin{subfigure}
 \centering \includegraphics[scale=0.08]{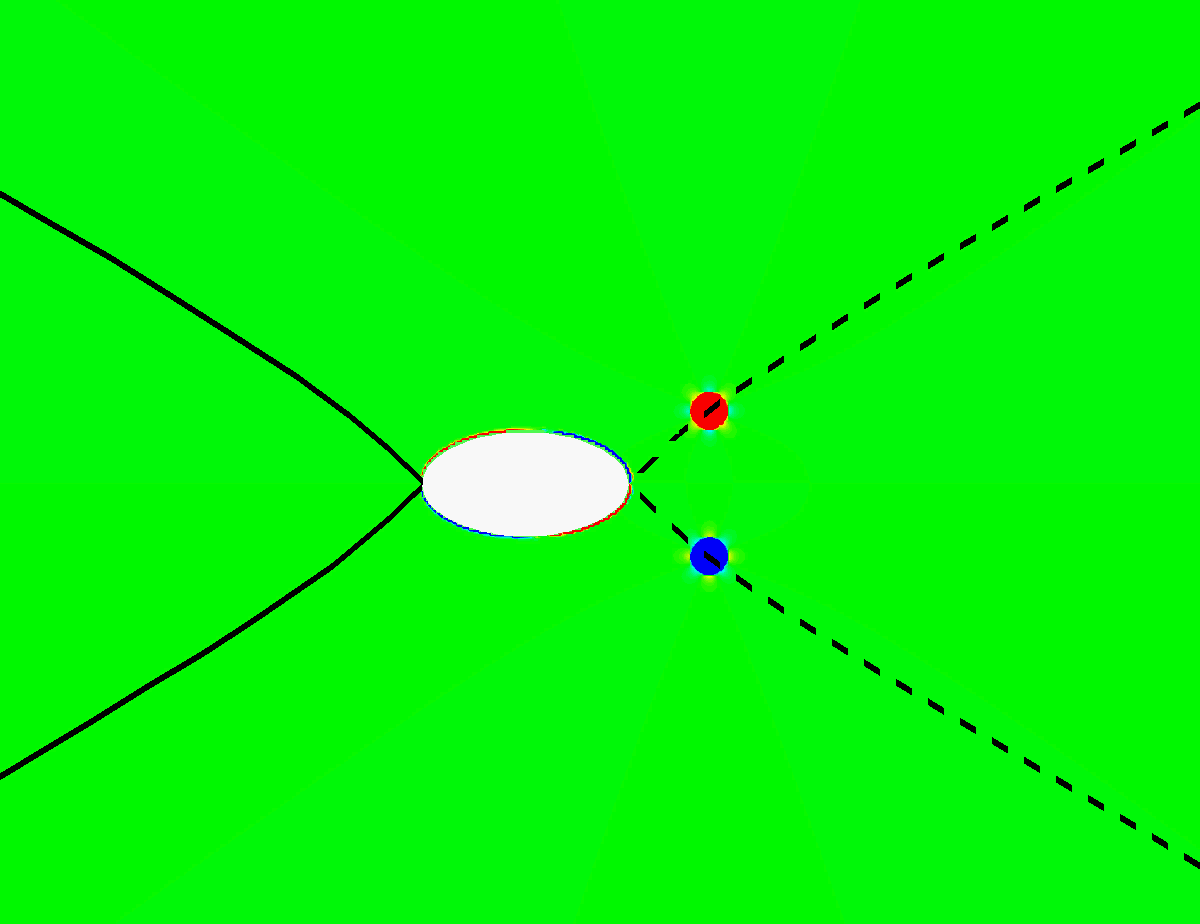} 
 {$t=0$}
\end{subfigure}
\begin{subfigure}
\centering \includegraphics[scale=0.08]{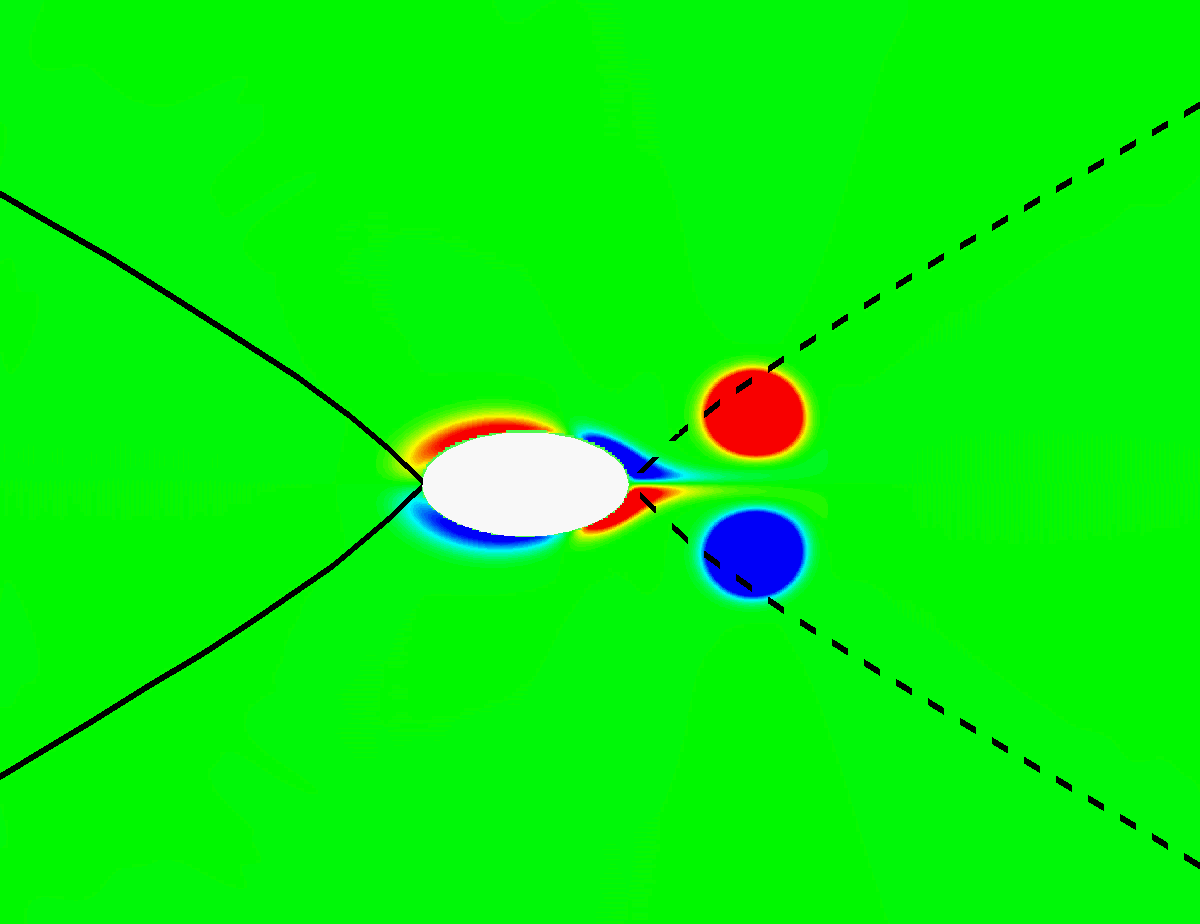}
 {$t=1$}
\end{subfigure}
\begin{subfigure}
\centering \includegraphics[scale=0.08]{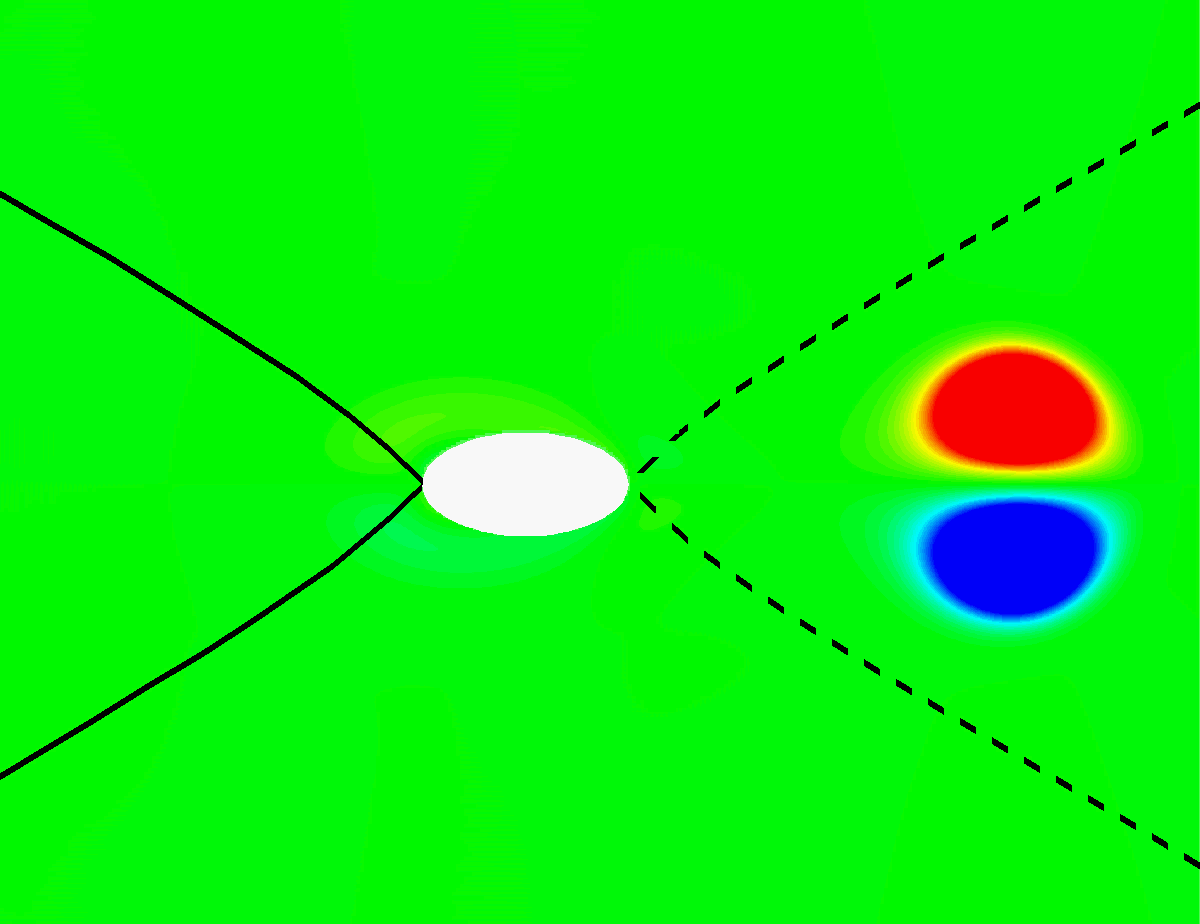}
 {$t=3$}
\end{subfigure}
\caption{Viscous interactions with an elliptic cylinder of aspect ratio 2, major axis along the $x$-axis, in the body-fixed frame from a starting position R-1-e on the right Hill equilibrium curves }
\label{a2-hor-R-y1}
\end{figure}

\begin{figure}
\centering
\begin{subfigure}
 \centering \includegraphics[scale=0.08]{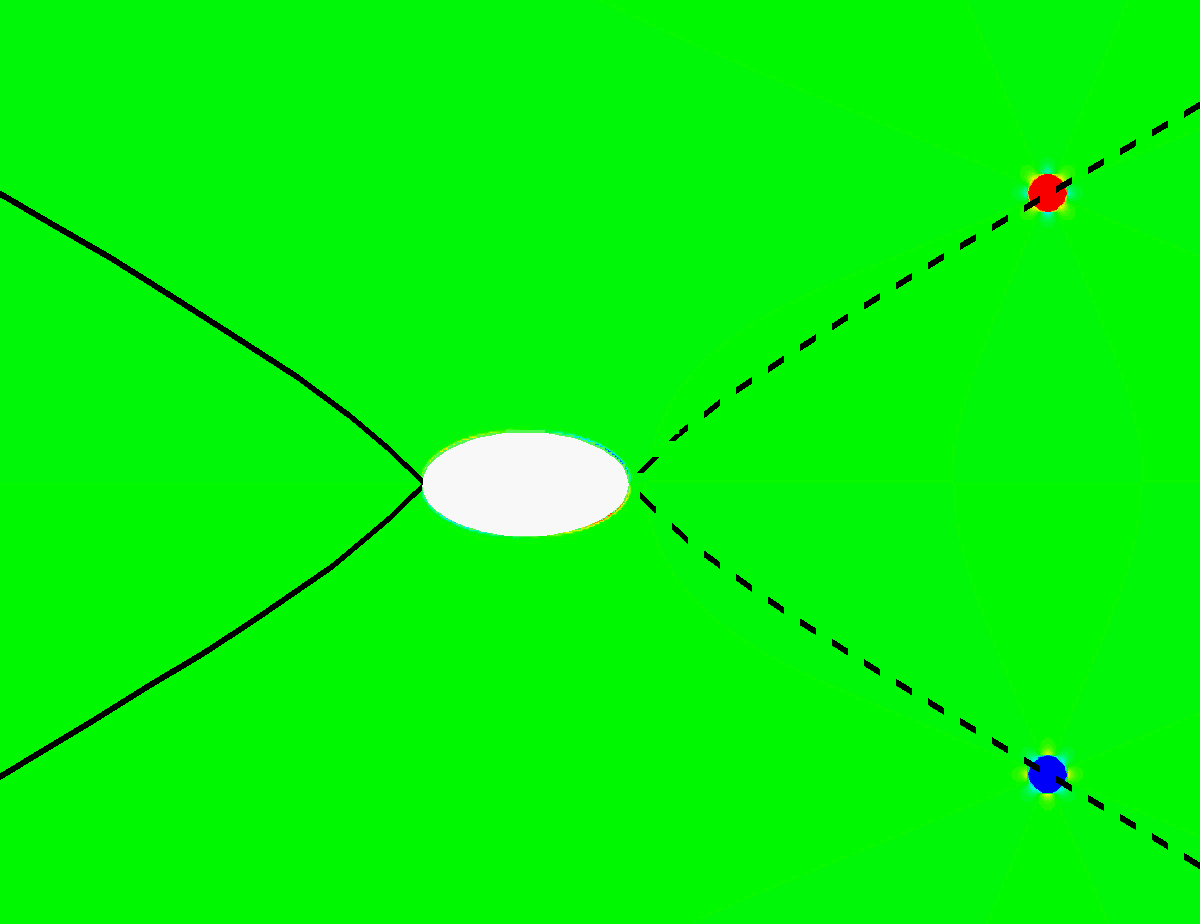} 
 {$t=0$}
\end{subfigure}
\begin{subfigure}
\centering \includegraphics[scale=0.08]{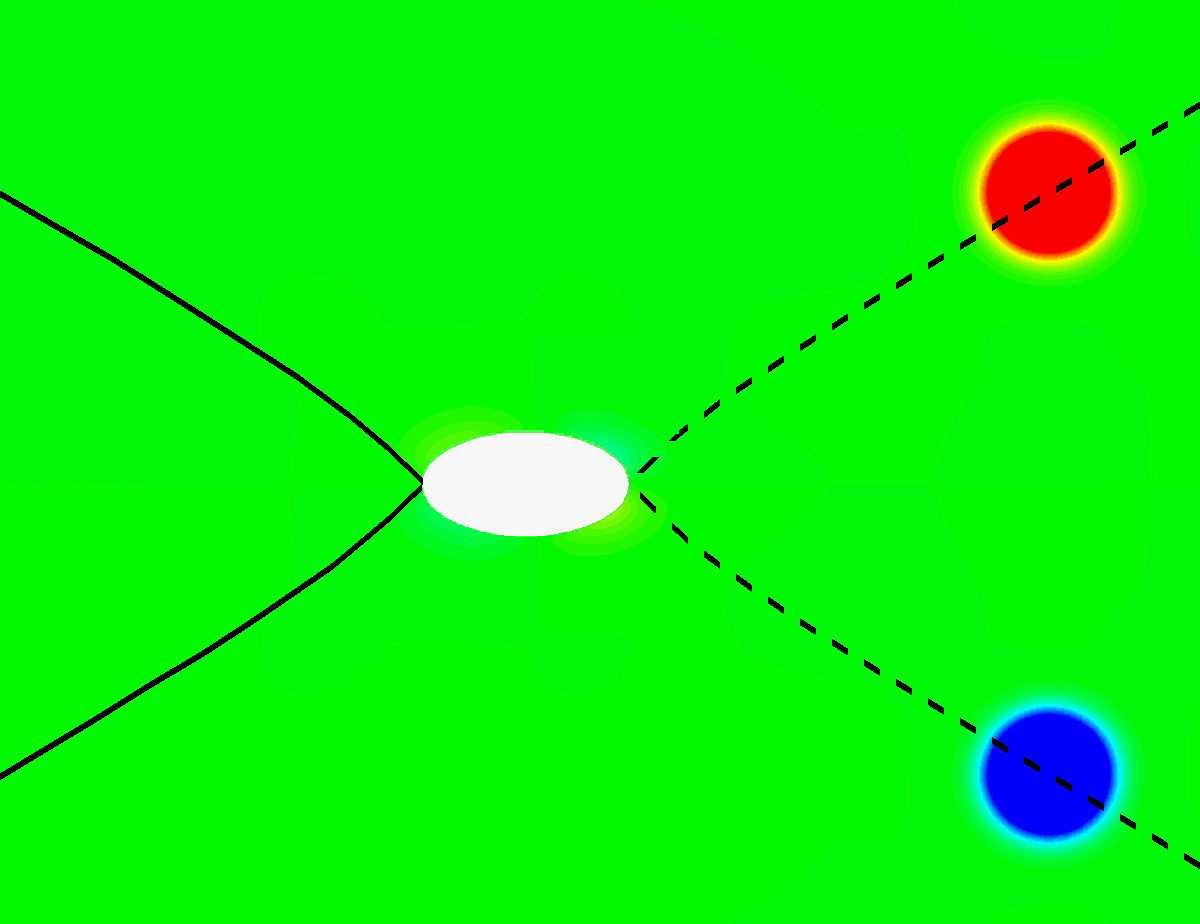}
 {$t=2$}
\end{subfigure}
\newline
\begin{subfigure}
\centering \includegraphics[scale=0.08]{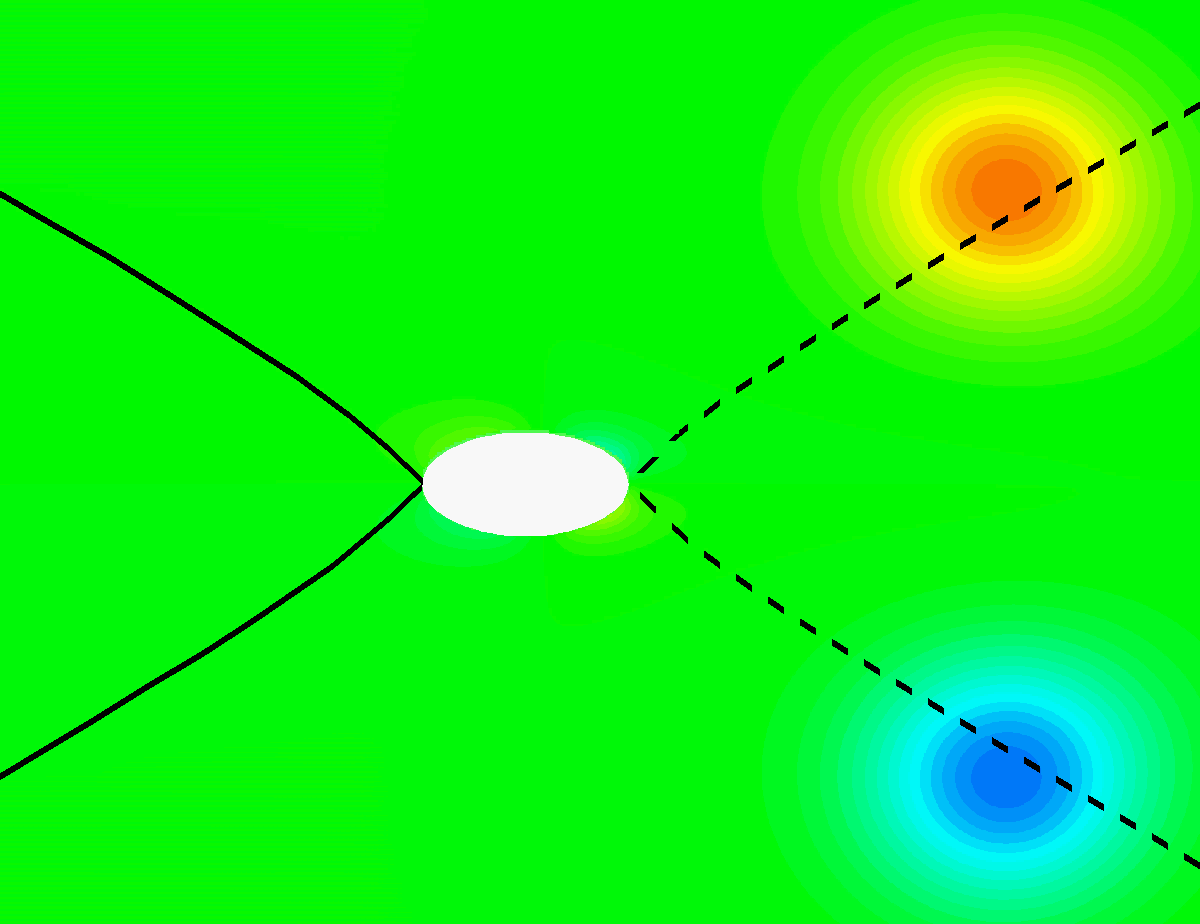}
 {$t=16$}
\end{subfigure}
\begin{subfigure}
\centering \includegraphics[scale=0.08]{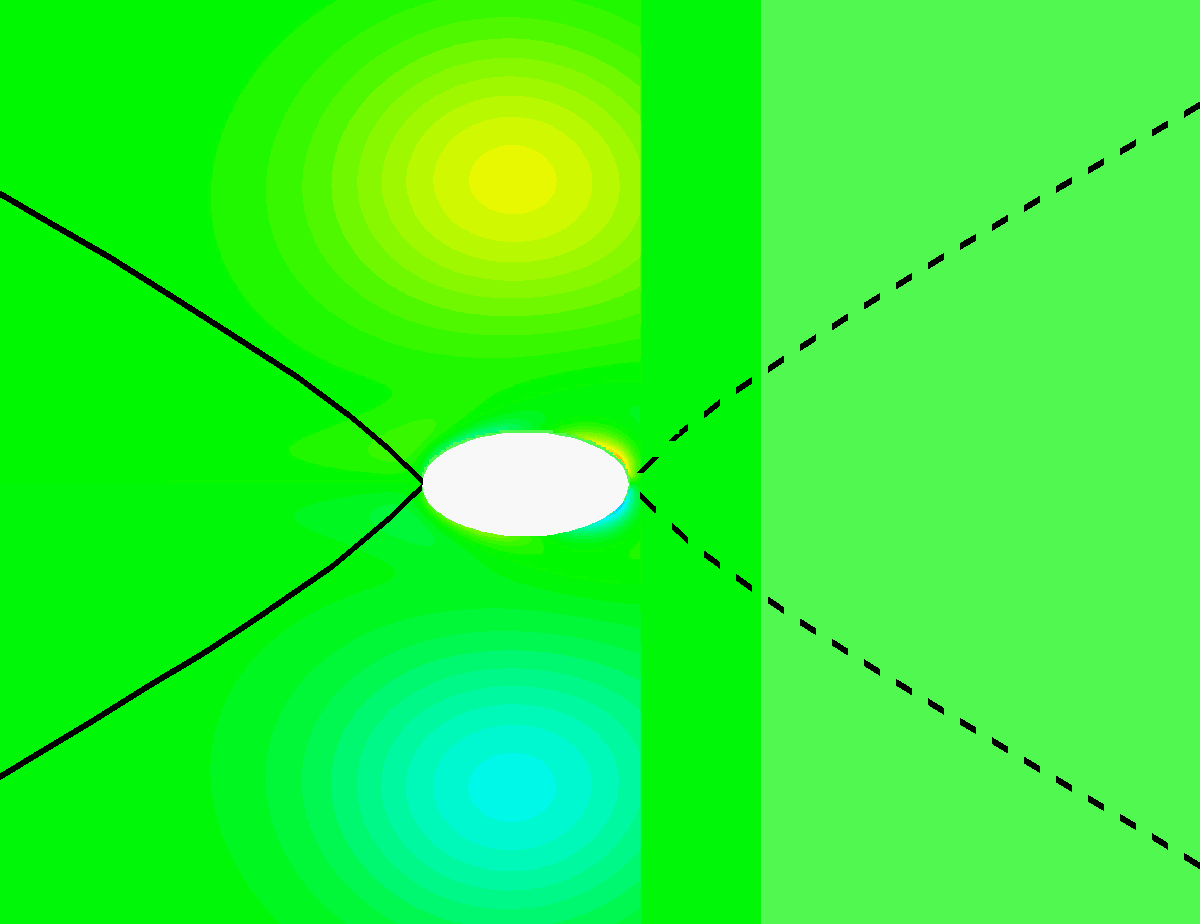}
 {$t=26$}
\end{subfigure}
\caption{Viscous interactions with an elliptic cylinder of aspect ratio 2, major axis along the $x$-axis, in the body-fixed frame corresponding to a starting position R-4-e on the right Hill equilibrium curves}
\label{a2-hor-R-y4}
\end{figure}

First, the configurations in which the major axis of the ellipse is along the $x$-axis were considered and for starting vortex positions on the equilibrium curves. The intial positions of the vortices and the drift of the vorticity maximum for the two ellipses are shown in Figure \ref{IPosHorEllR}.
Snapshots of the interactions for the ellipse of aspect ratio 2 are shown in Figures \ref{a2-hor-R-y1} amd \ref{a2-hor-R-y4} for two of the starting positions on the equilibrium curves. From the first position the vortices drift to the right and from the second position they eventually drift towards the left equilibrium curves i.e. the elliptic cylinder threads through the vortices 
The interactions for the ellipse with aspect ratio 4 are not shown since they are qualitatively similar.
\subsection{Starting configuration: moving Hill equilibria, leading vortices and ellipse with major axis along $y$-axis.}
\begin{figure}
\centering
\begin{subfigure}
\centering  \includegraphics[scale=0.35]{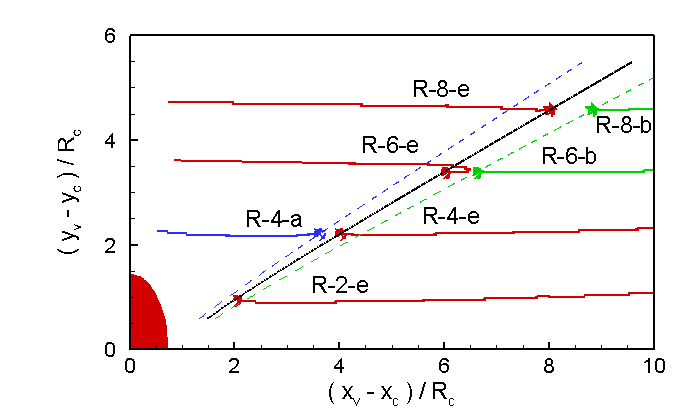} 
 \end{subfigure}
\begin{subfigure}
\centering \includegraphics[scale=0.35]{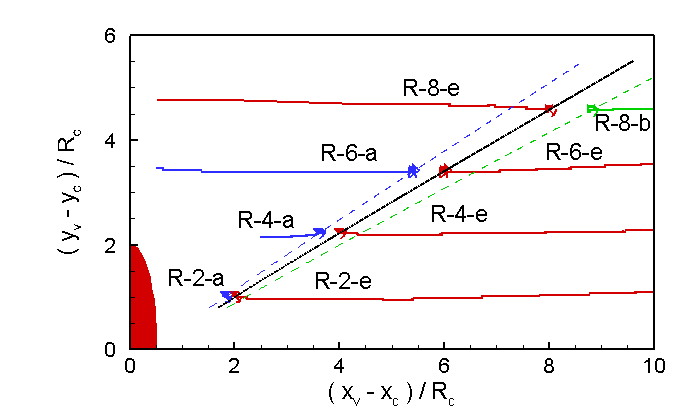}
\end{subfigure}
\caption{Initial positions of the vortices on and close to the right Hill equilibrium curves and the drift of the vorticity maximum point for interactions involving elliptic cylinders with major axis along the $y$-axis of aspect ratios 2 and 4, respectively.}
\label{IPosVerEll}
\end{figure}

Next, cases of the same ellipses with their major axis along the $y$-axis were considered and similar qualitative behavior is observed. The starting configurations and the drifts of the points of vorticity maximum are shown in Figure \ref{IPosVerEll}.

\begin{figure}
\centering
\begin{subfigure}
 \centering \includegraphics[scale=0.08]{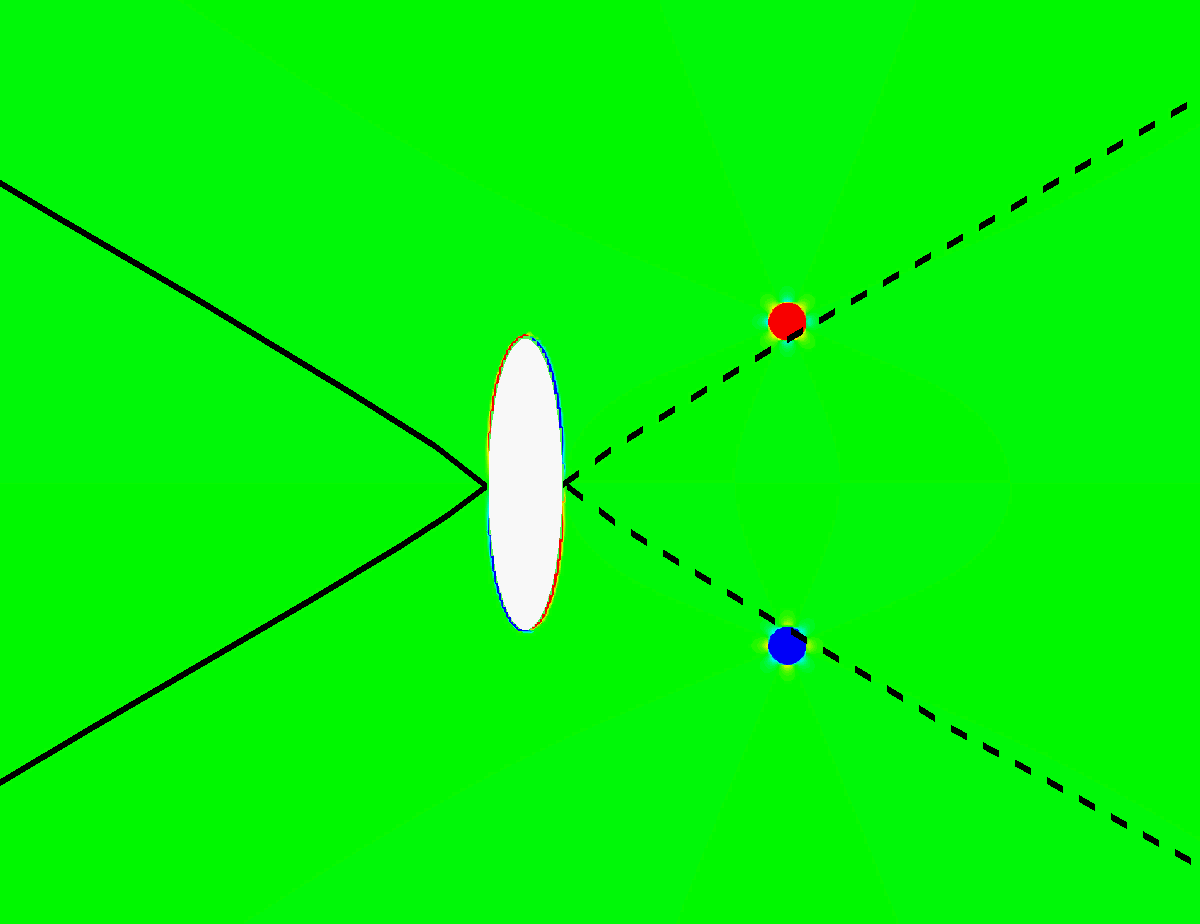} 
 {$t=0$}
\end{subfigure}
\begin{subfigure}
\centering \includegraphics[scale=0.08]{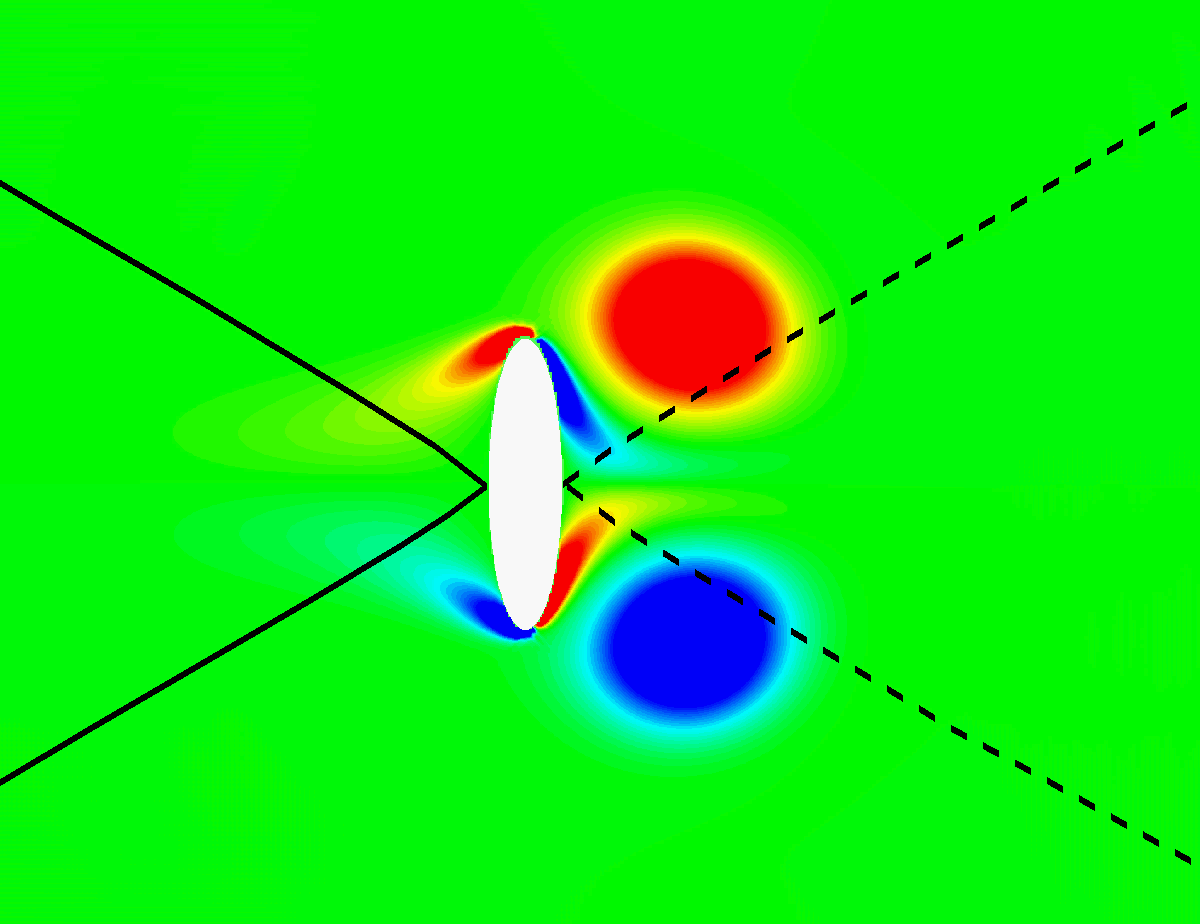}
 {$t=6$}
\end{subfigure}
\begin{subfigure}
\centering \includegraphics[scale=0.08]{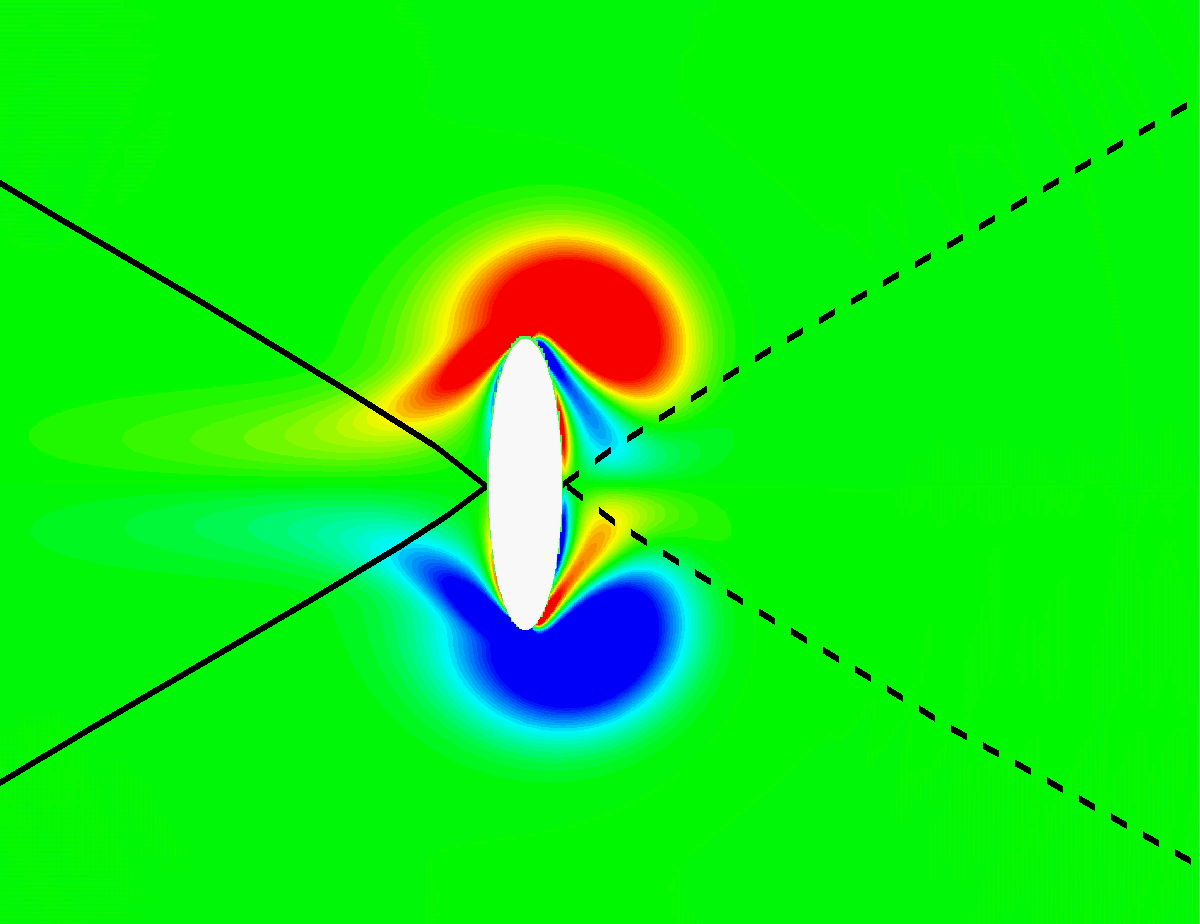}
 {$t=7$}
\end{subfigure}
\begin{subfigure}
\centering \includegraphics[scale=0.08]{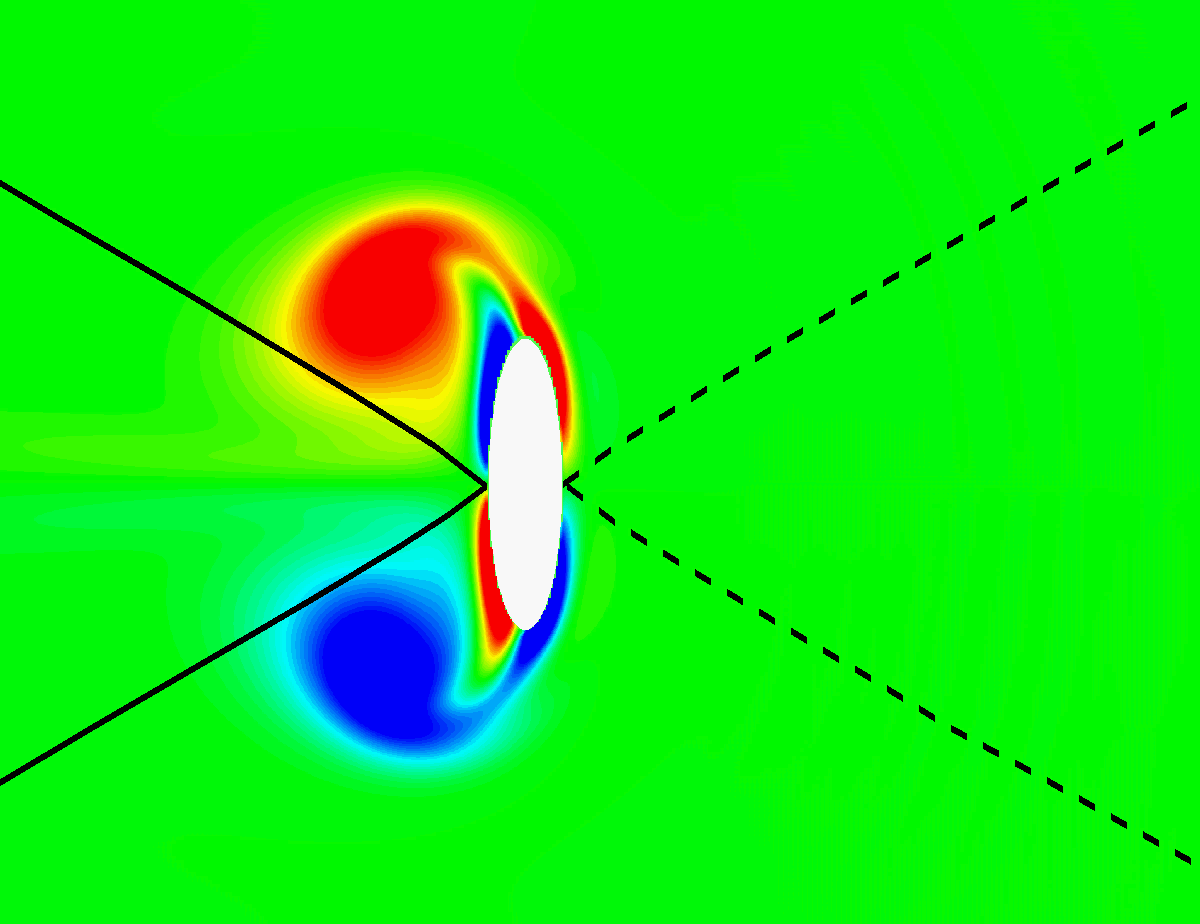}
 {$t=8$}
\end{subfigure}
\begin{subfigure}
\centering \includegraphics[scale=0.08]{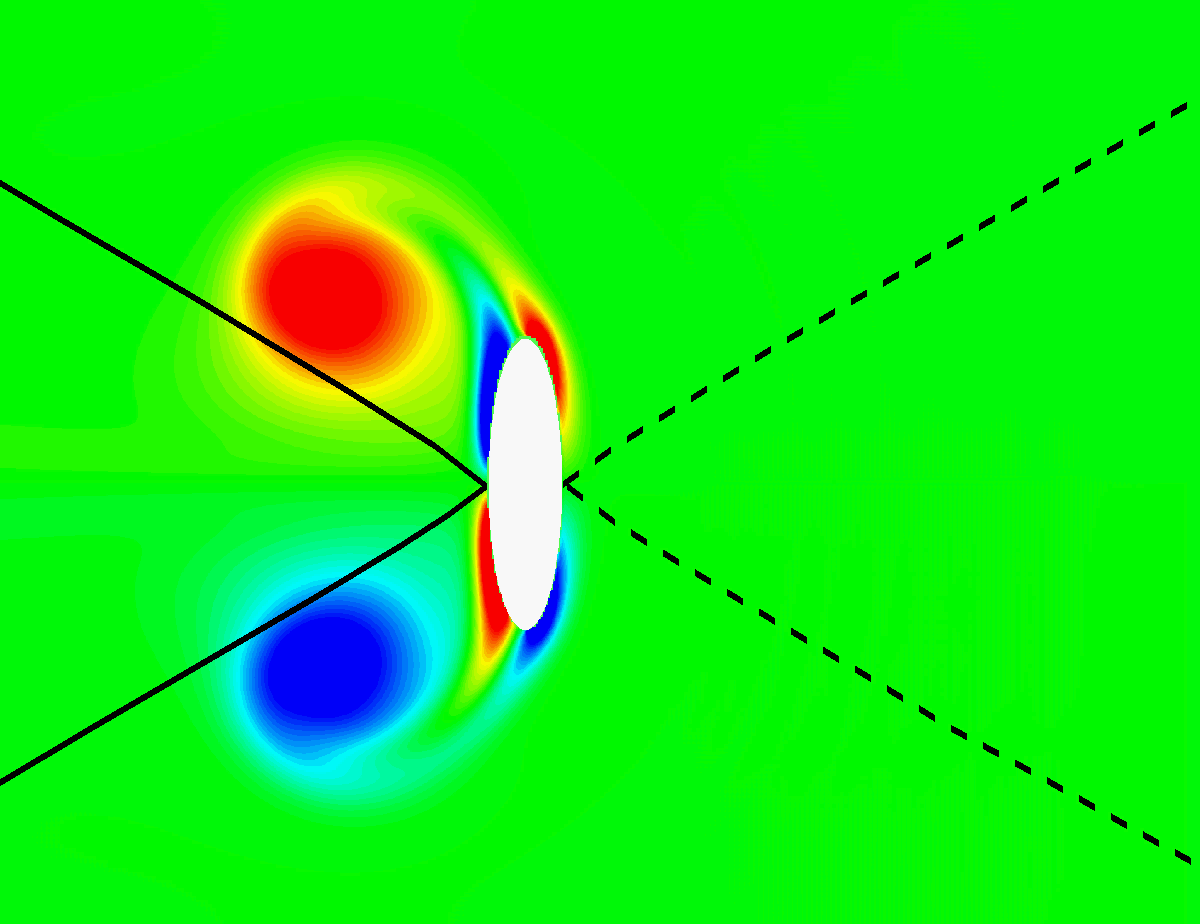}
 {$t=9$}
\end{subfigure}
\begin{subfigure}
\centering \includegraphics[scale=0.08]{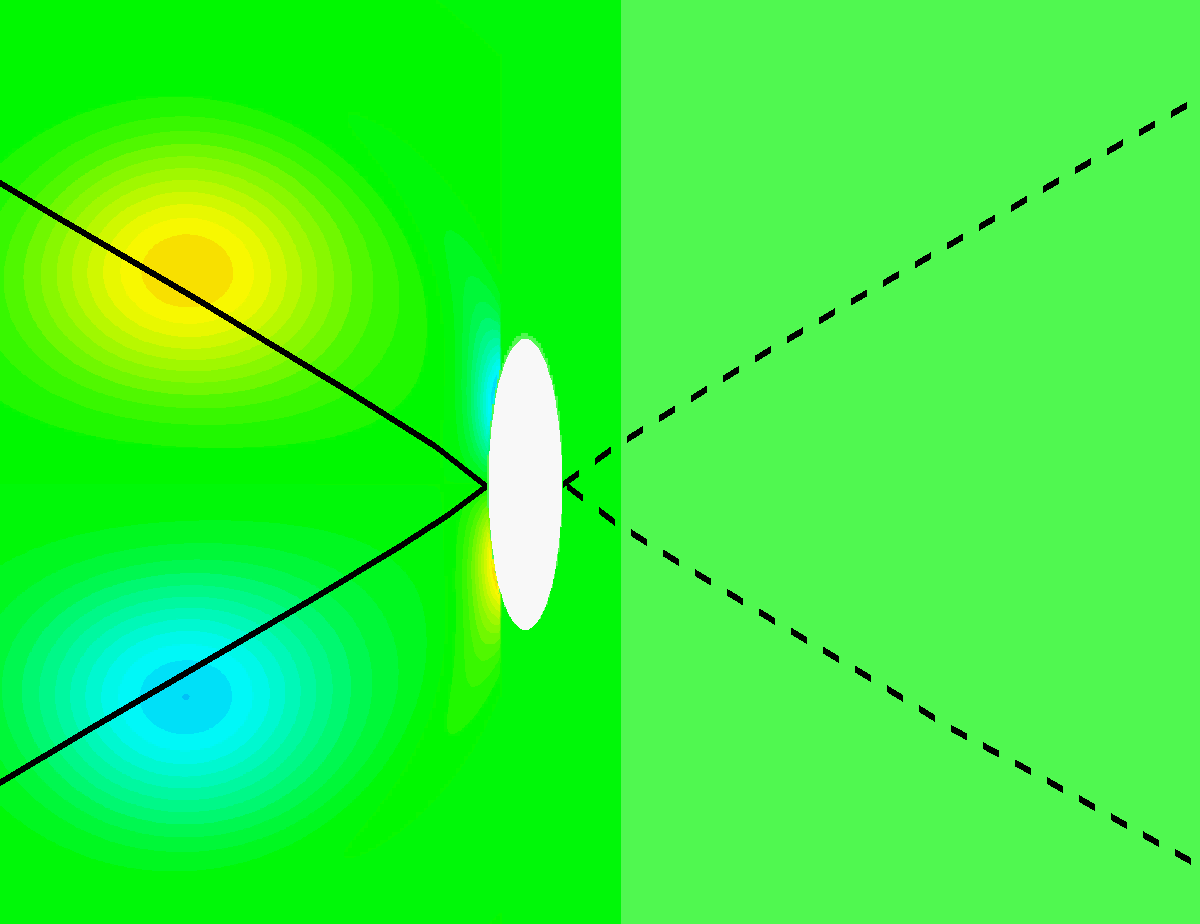}
 {$t=19$}
\end{subfigure}
\caption{Viscous interactions with an elliptic cylinder of aspect ratio 4, major axis along the $y$-axis, in the body-fixed frame from starting position R-4-a slightly above the right equilibrium curve.}
\label{a4-ver-off-R-x4}
\end{figure}

  Again, equilibrium-curve and off-equilibrium-curve starting positions were considered. Similar to previous cases of leading vortices, there is a stronger tendency for the cylinder to thread through the vortices and for the vortices to be attracted to the left equilibrium curves for starting positions that are slightly above the right equilibrium curves. For starting positions that are slightly below the right equilibrium curves there is a strong tendency to drift away to the right away from the ellipse. In Figure  \ref{a4-ver-off-R-x4},  the attraction towards the left equilibrium curves is shown for case of the vertical  ellipse of aspect ratio 4 from a starting position close to the ellipse. 

\section{Discussions and future directions.}

     Viscous simulations, in a dynamically coupled setting, of symmetric configurations of circular and elliptic cylinders interacting with a counter-rotating vortex pair are presented in this paper. The starting configuration of each simulation corresponds exactly to or is close to a moving F\"{o}ppl or moving Hill equilibrium configuration. A few vertical line configurations were also considered. The main objective of the simulations was to observe how much turning on viscosity at $t >0$ affects the evolution of the configuration and makes it deviate from the translating equilibrium configuration.

  The following principal features were observed in the simulations for both circular and elliptical cylinders (of aspect ratios 2 and 4 and with major axes parallel to and perpendicular to the $x$-axis.): 
\begin{itemize}
\item[1.] In all runs, the vorticity diffuses in time, as expected. Boundary layer development is observed, including separation and entrainment of boundary layer vorticity by the primary vortices, and is more significant for starting positions close to the cylinder.  None of the simulations showed any clear pinch-off of the separated shear layer and formation of a separate secondary vortex. 

\item[2.] The trailing configurations of vortices displays remarkably stable behavior. The stability is greater if the starting position of the vortex is farther away from the cylinder. For positions starting exactly on the left equilibrium curves and sufficiently far from the cylinder the configuration translates close to the equilibrium velocity for a significant amount of time. The deviation from the equilibrium velocity is greater for positions on the left  equilibrium curves closer to the cylinder. For positions starting close to, but not exactly on, the left equilibrium curves, the vortices are first attracted back to the left equilibrium curves and the ensuing dynamics is then similar to the cases where the vortices  start exactly on the equilibrium curves. 

\item[3.] The leading configurations of vortices are generally not stable. The vortices either drift away to the right leaving the cylinder behind or, conversely, the vortices drift to the left as the cylinder threads through them. In these latter cases, the drift of the vortices ends on the left equilibrium curves and so these curves again act like attractors. The drift to the left starts after a significant amount of time for starting positions farther from the cylinder. During the threading through event the cylinder experiences a strong acceleration followed by a strong deceleration.

\end{itemize}

       Dwelling on the last point, it is clear from these simulations that viscosity breaks the dynamical symmetry about the $y$-axis. In the dynamically coupled inviscid models, this symmetry exists not only in the translational equilibria but also in the (symmetrically) perturbed dynamics \cite{Sh2006, KaOs2008}. The drift to the right is easily explained since the vortices in such cases are close enough to each other so that the flow of the other vortex dominates. The drift the left is harder to explain. From a fluid dynamics perspective, it is conjectured that this has to do with viscous shear stresses and the location of the stagnation points. 

\begin{figure}
\centering
\begin{subfigure}
\centering  \includegraphics[scale=0.4]{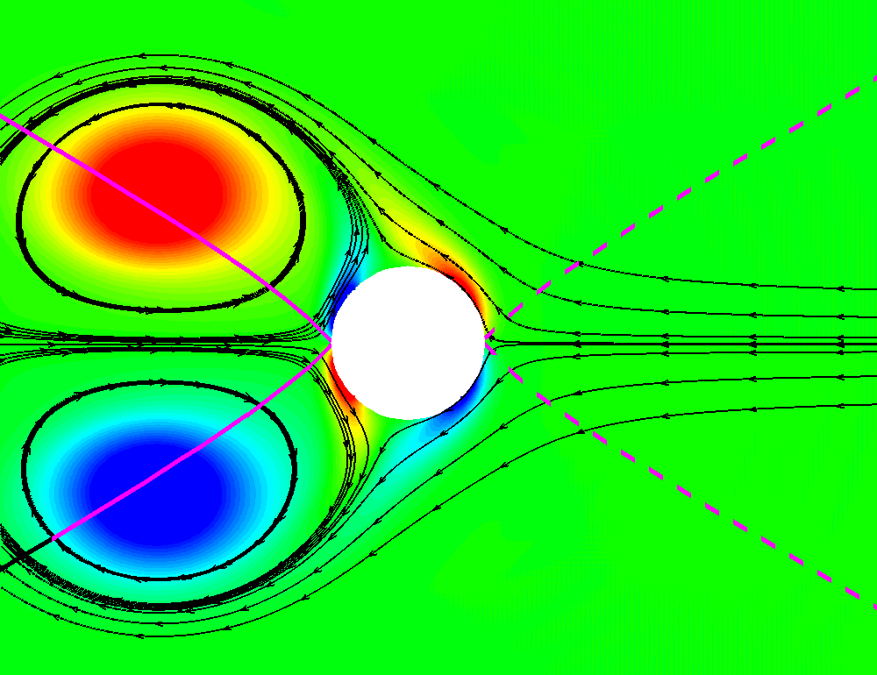} 
\end{subfigure}
\begin{subfigure}
 \centering \includegraphics[scale=0.4]{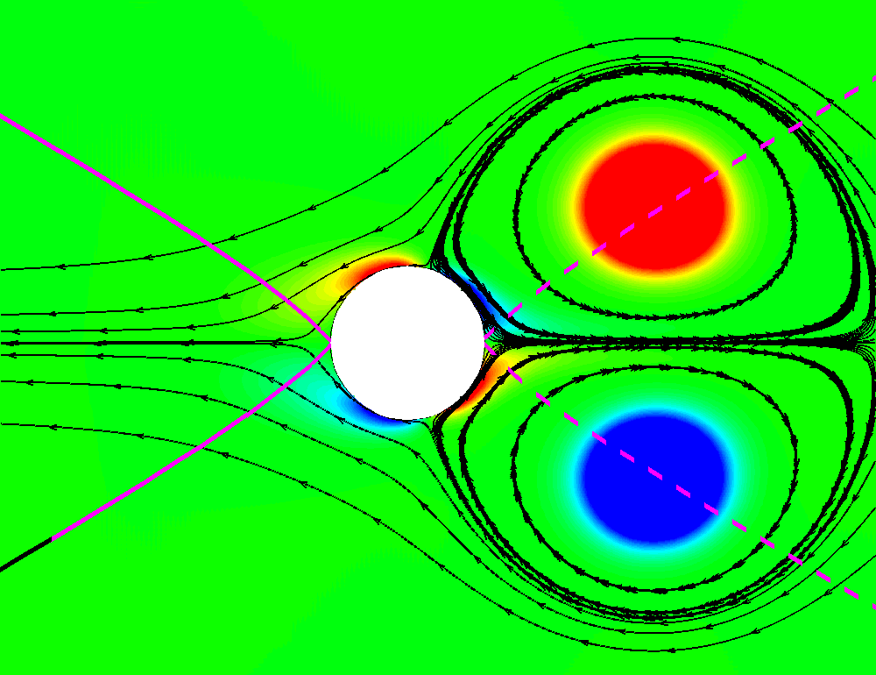} 
\end{subfigure}
\caption{Streamlines in a translating frame and vorticity contours when the vortices are situated on the left and right equilibrium curves for the case of a circular cylinder.}
\label{stagpoints}
\end{figure}

To elaborate, consider Figure \ref{stagpoints} for the circular cylinder case.
The pictures show streamlines of the flow in a body-fixed translating frame, along with the vorticity contours and the F\"{o}ppl equilibrium curves for both the trailing pair and the leading  pair configurations. In each configuration there are three stagnation points. The number of these surface stagnation points is the same as for the inviscid flow and their locations closely match as well.  Let $\theta$ be measured counter-clockwise from the positive  $x$-axis. 
In the trailing configuration, one stagnation point is caused by the vortex flow approaching the cylinder and the other two by the vortex flow leaving the cylinder. Label these $l_O,l^+$ and $l_-$, with angles $\theta=180^\circ, 180^\circ > \theta > 90^\circ$and $-180^\circ < \theta < -90^\circ$, respectively. In the leading configuration, the flow associated with each stagnation point is the opposite. Label these $r_O,r^+$ and $r_-$, with angles $\theta=180^\circ, 90^\circ > \theta > 0^\circ$and $-90^\circ < \theta < 0^\circ$, respectively. Strong wall shear stresses are formed in the neighborhood of $l_O, r^+$ and $r_-$ as the boundary layer develops on both sides of these stagnation points. Noting the location of the stagnation points,  it is plausible that the strong wall shear stresses in the neighborhood of $r^+$ and $r_-$ lead to an integrated force in the positive $x$-direction which, albeit only slightly greater than for the trailing configuration, is sufficient to trigger the instability of threading through. As the cylinder starts threading through this effect is enhanced as $r^+$ and $r_-$ approach and pass the $y$-axis. To prove or disprove the conjecture above would require an accurate computation of both the wall shear stress {\it and} the pressure distribution on the surface, which is work planned for the future. 

 The pushing of the cylinder by the trailing vortices for significant distances and the  threading of the cylinder through the leading vortices are both interesting phenomena and have potential  applications to the transport of free-to-move solid objects by vortices. It is tempting to also relate this result to the swimming of fish mentioned in the Introduction. However, one must remember that the counter-rotating vortices shed by tail fins of swimming fish typically have opposite circulations than in the configurations considered in this paper.

     These results also demonstrate that inviscid models do not become irrelevant when viscosity is added, and that though viscosity causes significant changes it is not always in the form of new vortices shed by the body. Point vortex models of these interactions could presumably be extended to include these effects, once clearly  understood, by methods which do not necessarily entail the addition of new point vortices. Clearly, the left branches of the F\"{o}ppl and Hill equilibria curves are still relevant in the viscous setting. 

  The cases presented in this paper are only a preliminary set of cases of fundamental interactions of vortices with a neutrally buoyant solid. Several other cases could be investigated. To name a few, breaking the reflection symmetry about the $x$-axis in the initial configuration and considering asymmetric interactions, or considering multiple pairs of symmetric counter-rotating vortices interacting with the cylinder. It may be recalled that there exists a translating equilibrium corresponding to two pairs of counter-rotating vortices and a circular cylinder  in the inviscid models \cite{Sh2006}.


\newpage

\section*{Appendix: Review of results for the inviscid models} 

 First, recall all the translating equilibria in the inviscid models.  

\subsection*{Moving F\"{o}ppl Equilibria.} 

\begin{figure}
\begin{center}
\includegraphics[scale=0.7,angle=0]{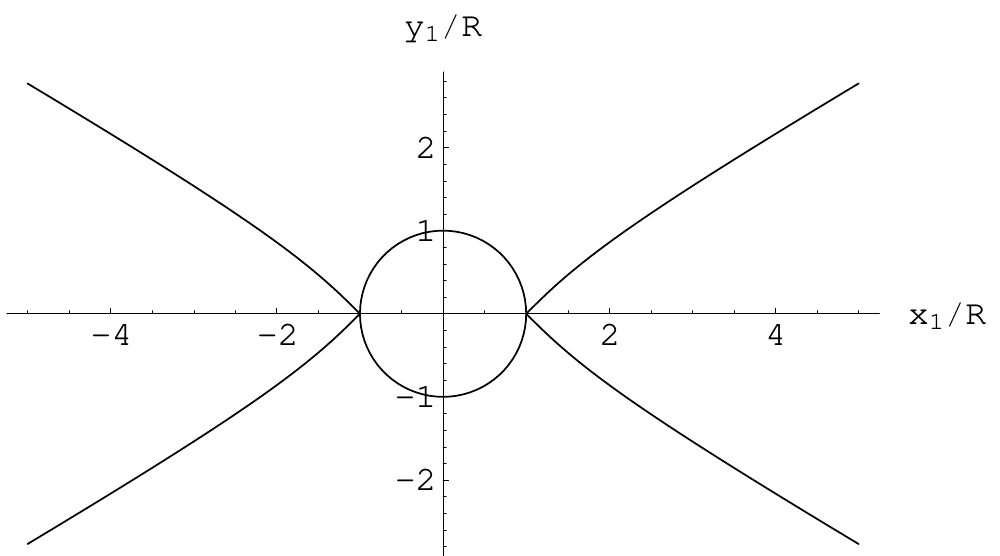}
\caption{The non-dimensionalized F\"{o}ppl equilibrium curves.}
\label{foppl}
\end{center}
\end{figure}

In the moving F\"{o}ppl equilibrium, the point vortices lie on the following curves:
\begin{align}
l^2-R^2&=\pm 2 l y, \label{eq:fopplcurve} 
\end{align}
where $R$ is the radius of the cylinder and $l^2=x^2+y^2$. The four branches are shown in Figure~\ref{foppl}, plotted in non-dimensioanlized coordinates. 

Corresponding to the location of each pair $(x_1,y_1)$ and $(x_1,-y_1)$, with strengths $\pm \Gamma > 0$, the cylinder forward velocity $V$ is given by 
\begin{align}
V&=\frac{\Gamma}{4 \pi } \frac{l_1^4}{y_1(l_1^4-R^4)}, \label{eq:velposition} 
\end{align}
Each vortex in the translating pair has to satisfy both equations~(\ref{eq:fopplcurve}) and~(\ref{eq:velposition}). \\

\textit{\underline{Vertical Line Equilibria.}} \\

  In this equilibrium, the vortices are located on the $y$-axis at $(0,y_1)$and $(0,-y_1)$. The corresponding forward speed of the cylinder is given by the relation
\begin{align}
&-\left(V+\frac{ \Gamma_1}{\pi} \left( \frac{ y_1}{R^2}-\frac{1}{y_1}\right) \right) \left(1+ \frac{R^2}{y_1^2} \right) + \frac{\Gamma_1}{\pi R^2} y_1 - \frac{R^2 \Gamma_1}{\pi y_1^3}  \nonumber \\
&  \hspace{0.5in} + \Gamma_1 \left( \frac{1}{2\pi} \frac{y_1}{y_1^2 - R^2} - \frac{1}{2\pi} \frac{(y_1^2+R^2)y_1}{y_1^4 +R^4+2R^2 y_1^2}+ \frac{1}{4 \pi y_1}\right)&=0 \label{eq:velpositionN}
\end{align}


\paragraph{Remarks.}
\begin{itemize}
\item[1.] Equations~(\ref{eq:fopplcurve}) and~(\ref{eq:velposition}) are invariant under the transformation $x_1 \rightarrow -x_1$, implying that for each pair position there is another position along the $x$-axis, reflected about the $y$-axis, which also satisfies the two equations (for the same  $V$ and $\Gamma$).  
\item[2.] Since the left side of equation~(\ref{eq:fopplcurve}) is always positive, one chooses the $+$ sign if $y_1 >0$ and the $-$ sign if $y_1 <0$. 
\end{itemize}

\subsection*{Moving Hill equilibria.}
The ellipse is in the $z$-plane with semi-major axis $a$, \underline{parallel to the direction of motion}, and semi-minor axis $b$, and  with
\[c^2=a^2-b^2. \]Coordinates $(x,y)$ in the $z$-plane are related to the coordinates $(\zeta_x,\zeta_y)$ in the $\zeta$-plane by: 
\begin{align}
x+iy&=(\zeta_x + i \zeta_y)+ \frac{c^2}{4(\zeta_x + i \zeta_y)}, \nonumber \\
&=(\zeta_x + i \zeta_y)+ \frac{c^2(\zeta_x - i \zeta_y)}{4\mid \zeta \mid^2}, \nonumber \\
\Rightarrow x&=\left( 1 + \frac{c^2}{ 4\mid \zeta \mid^2} \right) \zeta_x, \quad y=\left( 1 - \frac{c^2}{ 4\mid \zeta \mid^2} \right) \zeta_y \label{eq:zzeta}
\end{align}
The equilibrium curve is given in the \underline{$\zeta$-plane} by 
\begin{align}
 \frac{4  \zeta_y^2}{\zeta_x^2 + \zeta_y^2}\left(k^6 (\zeta_x^2 + \zeta_y^2)^3- \lambda \right)&=\left(k^2( \zeta_x^2 + \zeta_y^2)-1\right)^2\left(k^2 (\zeta_x^2 + \zeta_y^2)-\lambda\right), \label{eq:hillcurve}
\end{align}
where 
\[k=\frac{2}{a+b}, \quad \lambda=\frac{k^2(a^2-b^2)}{4}\]
For a circle of radius $R$, $k=1/R$, $\lambda=0$ and (\ref{eq:hillcurve}) reduces to (\ref{eq:fopplcurve}). 

  To get the equilibrium curves in the $z$-plane, where the ellipse lies, one takes the above curve and maps it to the $z$-plane, using (\ref{eq:zzeta}). It is difficult to obtain an  analytical expression for this curve in terms of $x$ and $y$, the curve in the $z$-plane is computed numerically. 

  The analog of expression (\ref{eq:velposition}) which gives the ellipse forward speed for a vortex pair of strengths $\pm \Gamma >0$ located on the equilibrium curves at $(x_1,y_1)$ and $(x_1,-y_1)$ is: 
\begin{align}
\frac{k \Gamma}{2  \pi V}&=- \frac{(k^2 \mid \zeta_1 \mid^2-1)^2(k^2 \mid \zeta_1 \mid^2+1)(\lambda - k^4 \mid \zeta_1 \mid^4)^2}{k \mid\zeta_1
 \mid \left( \lambda + \lambda k^8 \mid \zeta_1 \mid^8 -2 \lambda k^6 \mid \zeta_1 \mid^6-2 \lambda k^4 \mid \zeta_1 \mid^4 + \lambda k^2 \mid \zeta_1 \mid^2 + k^{10} \mid \zeta_1 \mid^{10} \right)} \nonumber \\
& \hspace{3in} \sqrt{\frac{k^2 \mid \zeta_1 \mid^2-\lambda}{k^6 \mid \zeta_1 \mid^6- \lambda}} \label{eq:velpositionell}
\end{align}
where $\mid \zeta_1 \mid^2=\zeta_{x1}^2+\zeta_{y1}^2$, with $(\zeta_{x1}, \zeta_{y1})$ being the coordinates in the $\zeta$-plane corresponding to the point vortex located at $(x_1,y_1)$. 

\begin{figure}
\centering
\includegraphics[scale=0.26]{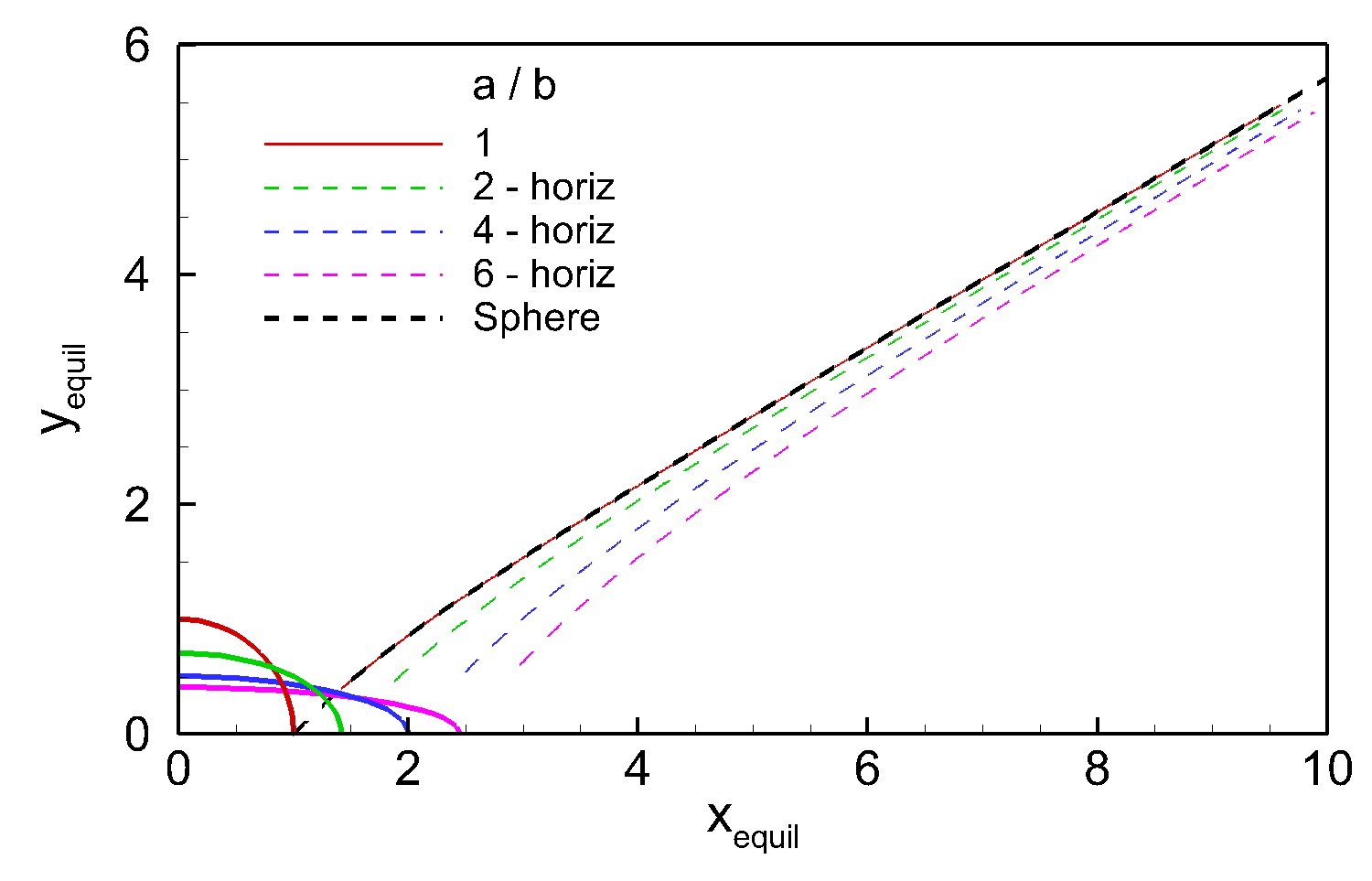}
\includegraphics[scale=0.26]{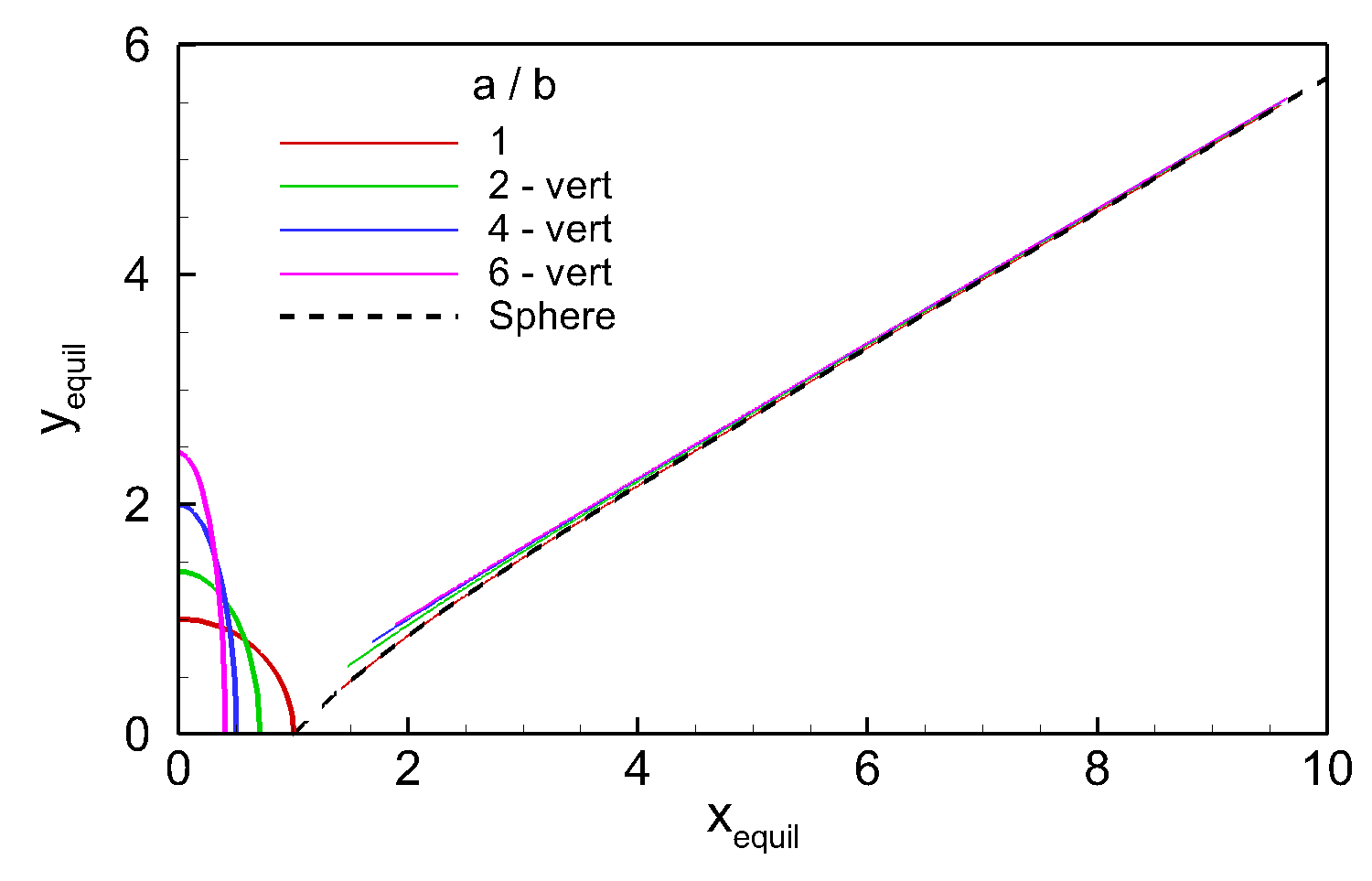}
\caption{The Hill equilibrium curves.}
\label{Hill}
\end{figure}

Note that if the semi-major axis is \underline{perpendicular to the direction of motion}, one replaces (\ref{eq:zzeta}) by 
\begin{align}
x&=\left( 1 - \frac{c^2}{ 4\mid \zeta \mid^2} \right) \zeta_x, \quad y=\left( 1 + \frac{c^2}{ 4\mid \zeta \mid^2} \right) \zeta_y \label{eq:zzeta2}
\end{align}
The equilibrium curves are plotted for elliptic cylinders of various aspect ratios for both the horizontal and vertical configurations in Figure \ref{Hill}. Note that these curves, like the F\"{o}ppl curves, also have four branches and only the ones in the first quadrant are shown. 

\pagestyle{empty}
\bibliographystyle{unsrt}

\end{document}